\documentclass[showpacs,amssymb,aps,twocolumn]{revtex4}
\usepackage{amsmath}
\usepackage{graphicx}
\usepackage[latin1]{inputenc}
\begin{document}

\title{Forward scattering amplitudes and the thermal operator representation}

\author{F. T. Brandt$^{a}$, Ashok Das$^{b,c}$,
  J. Frenkel$^{a}$ and Silvana Perez$^{d}$}
\affiliation{$^{a}$ Instituto de Física, Universidade de São
Paulo, São Paulo, BRAZIL}
\affiliation{$^{b}$ Department of Physics and Astronomy,
University of Rochester,
Rochester, New York 14627-0171, USA}
\affiliation{$^{c}$ Saha Institute of Nuclear Physics, 1/AF
  Bidhannagar, Calcutta 700064, INDIA} 
\affiliation{$^{d}$ Departamento de Física, 
Universidade Federal do Pará, 
Belém, Pará 66075-110, BRAZIL}

\bigskip

\begin{abstract}
We develop systematically to all orders the forward scattering
description for retarded amplitudes in field theories at zero
temperature. Subsequently,  through the application of the thermal
operator, we establish the forward scattering description at finite
temperature. We argue that, beyond providing a  graphical relation
between the zero temperature and the finite temperature amplitudes,
this method is calculationally quite useful. As an example, we derive
the important features of the  one loop retarded gluon self-energy in
the hard thermal loop approximation from the corresponding properties
of the zero temperature amplitude. 
\end{abstract}


\maketitle

\section{Introduction}

In a series of papers \cite{silvana1,silvana2,niegawa,silvana}, the idea of a thermal
operator representation \cite{Espinosa:2003af,Espinosa:2005gq} 
has been developed extensively in both the imaginary time formalism as
well as the real time formalism of closed time path. In simple terms,
the thermal operator 
representation relates a Feynman graph at finite temperature (with or
without a chemical potential) to the corresponding graph at zero
temperature. As we have argued earlier, the thermal operator
representation offers a powerful method for studying various questions
at finite temperature. As an example, we have shown in an earlier
paper \cite{silvana3} how the cutting rules at finite temperature
(with or without a 
chemical potential), in the closed time path formalism, can be derived
starting from those at zero temperature. This derivation also
clarifies the miraculous cancellations that arise in an explicit
demonstration of a cutting description for the imaginary part of a
thermal amplitude \cite{bedaque1,das:book97}.  

At finite temperature, retarded amplitudes play a significant role in
studying various physical phenomena. Plasma oscillations provide a
very simple example of this.  When a thermal plasma is perturbed
weakly, the subsequent response of the plasma to the perturbation is
studied using the linear response theory
\cite{fetter,kapusta:book89,lebellac:book96}. 
In particular, the damping
of the oscillation in the plasma is understood by analyzing the poles
of the retarded propagators of the particles moving through the
plasma. Of course, at finite temperature very few quantities can be
evaluated exactly, but the forms of thermal amplitudes simplify
considerably either in the low temperature or the high temperature
limits. In many phenomena of physical interest (such as quark-gluon
plasma phase transitions, early universe etc), it is the high
temperature behavior that is relevant. While there are many ways of
evaluating the high temperature behavior (also known as the hard
thermal loop approximation \cite{pisarski}) of thermal amplitudes, the forward
scattering description for the retarded amplitudes provide an efficient
calculational tool \cite{frenkel1}. This can be seen from the following simple
example. Let us consider a scalar field theory with a cubic
interaction in six dimensions (which is similar to non-Abelian gauge
theories in four dimensions). The thermal correction to the one
loop retarded self-energy can be directly calculated (see Fig. 1) in
this theory and, 
after doing the internal energy integral, leads to 

\begin{center}
\begin{figure}[ht!]
\includegraphics[scale=0.36]{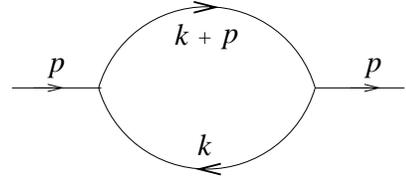}
\caption{One-loop retarded self-energy in $\phi^3$ theory.}
\end{figure}
\end{center}

\begin{eqnarray}
\lefteqn{\Sigma_{\rm R}^{(1) \beta} (p_{0},\vec{p})  =  \lambda^{2} \int
\frac{d^{5}k}{(2\pi)^{5}}\ \frac{n_{\rm
    B}(E_{k})}{4E_{k}E_{k+p}}}\nonumber\\ 
& &\quad \times\left[\frac{1}{p_{0}-E_{k}-E_{k+p}}-
\frac{1}{p_{0}+E_{k}+E_{k+p}}\right.\nonumber\\  
& &\qquad
\left.
+\frac{1}{p_{0}+E_{k}-E_{k+p}}-\frac{1}{p_{0}-E_{k}+E_{k+p}}\right], 
\label{direct} 
\end{eqnarray}
where $E_{k}= \sqrt{{\vec{k}}^{2} + m^{2}},\  E_{k+p} =
\sqrt{(\vec{k}+\vec{p})^{2} + m^{2}}$. Furthermore, $n_{\rm B}(E_{k})$
denotes the bosonic distribution function and $p_{0}$ is assumed to 
correspond to $p_{0}+i\epsilon$ which is necessary for the retarded
self-energy. At very 
high temperatures where $|\vec{k}|\gg p_{\mu}, m,$ the masses can be
neglected and we see from \eqref{direct} that the high temperature
limit needs to be calculated carefully since there are energy
differences in the denominator. 

On the other hand, the forward scattering description of the same
retarded self-energy involves diagrams where one of the internal
propagators in the loop is thermal and on-shell while the other
corresponds to a zero temperature retarded propagator. They are shown
in Fig. 2 and lead immediately to 

\begin{center}
\begin{figure}[ht!]
\includegraphics[scale=0.4]{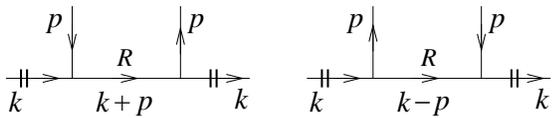}
\caption{Two forward scattering amplitudes for the one-loop retarded self-energy.}
\end{figure}
\end{center}

\begin{eqnarray}
\lefteqn{\Sigma_{\rm R}^{(1) \beta} (p_{0},\vec{p})  =  \lambda^{2}\int
\frac{d^{5}k}{(2\pi)^{5}}\  \frac{n_{\rm B} (E_{k})}{2E_{k}}}\nonumber\\ 
& &\qquad \times \left[\frac{1}{(k+p)^{2}-m^{2}} +
  \frac{1}{(k-p)^{2}-m^{2}}\right].\label{fsa} 
\end{eqnarray}
Here the ``$i\epsilon$" in $p_{0}$ (denoting a retarded propagator)
as well as $k_{0}=E_{k}$ are to be  understood. At high temperature
where masses can be neglected, this takes the simple form 
\begin{equation}
\Sigma_{\rm R}^{(1) \beta} (p_{0},\vec{p}) = -\lambda^{2}\int
\frac{d^{5}k}{(2\pi)^{5}}\ 
\frac{n_{\rm B} (|\vec{k}|)}{4|\vec{k}|} \frac{p^{2}}{(k\cdot
  p)^{2}}. \label{fsa1}
\end{equation}
The structure resulting directly from the forward scattering description 
is very simple and interesting (of course, the same structure
would also arise in a direct calculation, but limits have to be taken
carefully and terms have to be grouped properly before this simple
structure is obtained). First of all, we note that the coefficient of
the term $\frac{n_{\rm B}(|\vec{k}|)}{|\vec{k}|}$ in the integrand is
manifestly Lorentz covariant and is a homogeneous function of degree
zero in the external momentum and of degree $(-2)$ in the internal
momentum. The manifest Lorentz covariance is broken at high
temperature only when the angular integration is carried out. In fact, if
we carry out the integration over $|\vec{k}|$, the high temperature
limit is obtained to be 
\begin{equation}
\Sigma_{\rm R}^{(1) \beta} (p_{0},\vec{p}) =
-\frac{\lambda^{2}\pi^{2}T^{2}}{24} \int 
\frac{d\Omega}{(2\pi)^{5}} \left(\frac{p^{2}}{(p\cdot
  \hat{k})^{2}}\right), 
\end{equation}
where we have defined $\hat{k}^{\mu} = (1, \hat{\vec{k}})$. This
Lorentz covariant structure of the integrand in \eqref{fsa1} (which
results directly 
in the forward scattering description) is very helpful and has been used to
derive in a simple way,  in the hard thermal loop
approximation, the effective action for QCD as well as the
energy-momentum tensor for the 
quark-gluon plasma \cite{frenkel2}. This method is also convenient for the
analysis of the high temperature behavior of gauge field theories in a
curved space-time \cite{frenkel3}. It is also worth noting that the
study of the 
solution of the transport equation at high temperature leads to
structures naturally arising in the forward scattering method \cite{frenkel4}. 

In spite of its success, a simple and general derivation of the
forward scattering 
amplitudes to all orders at finite temperature is so far lacking. As
we have already 
argued, the thermal operator representation \cite{silvana1,silvana2,silvana3} 
provides a powerful method
for obtaining results at finite temperature starting from zero
temperature and in this paper we show how the forward scattering
amplitudes for retarded thermal $n$-point functions can be derived through the
use of the thermal operator representation. The thermal operator
representation is clearly  meaningful in studying this question if
there exists a forward scattering  description at zero
temperature. In this paper we derive the forward scattering
description for retarded amplitudes in 
zero temperature field theories. The thermal operator representation then
directly leads to the forward scattering description at finite
temperature and clarifies the origin of the nice structures observed
in the context of the forward scattering amplitudes at high temperature. 

The paper is organized as follows. In section {\bf II}, we develop the
forward scattering description for retarded amplitudes of a scalar
field  theory at zero 
temperature. In section {\bf III}, we show how the thermal operator
representation leads directly to the forward scattering description
for retarded thermal amplitudes. In this section, we also point out various
interesting features of retarded amplitudes both at zero as well as at
finite temperature. In section {\bf IV} we discuss the forward
scattering description for the Yang-Mills theory. In particular, we
emphasize that various nice 
properties in the hard thermal loop approximation such as
transversality, manifest Lorentz covariance  and gauge 
invariance of the integrand of the one loop retarded self-energy follow 
simply from the
properties of the zero temperature amplitude through the thermal
operator representation. We conclude with a brief summary in section {\bf V}. 

\section{Forward scattering description at zero temperature}

The idea of forward scattering description basically already exists
even at zero temperature although it is not as well developed
and is certainly not widely known. Therefore, in this section, we will
develop the idea of forward scattering amplitudes at zero temperature
systematically for retarded amplitudes so that the thermal operator
representation can lead directly to the forward scattering amplitudes
at finite temperature. The basic idea behind a forward scattering
description at zero temperature \cite{feynman} is the simple
observation that a (time ordered)  
Feynman propagator (for a massive scalar particle, for simplicity) can
be expressed as 
\begin{equation}
\frac{i}{k^{2}-m^{2}+i\epsilon} = \frac{i}{(k_{0}+i\epsilon)^{2} -
  E_{k}^{2}}  + 2\pi \theta (-k_{0}) \delta (k^{2}-m^{2}), 
\end{equation}
where we have identified $E_{k}^{2} = {\vec{k}}^{2} + m^{2}$. Namely,
the Feynman propagator is the sum of the retarded propagator and a
negative energy propagator. As a result, if we have a simple one loop
diagram with $n$ scalar propagators, the amplitude (neglecting vertex factors) 
can be written as 
\begin{widetext}
\begin{equation}
\Gamma_{n}^{(1)}  =  \int \frac{d^{4}k}{(2\pi)^{4}}\prod_{i=1}^{n}
 \frac{i}{k_{i}^{2} - m^{2} + i\epsilon} 
 =  \int \frac{d^{4}k}{(2\pi)^{4}}\prod_{i=1}^{n}\left(\frac{i}{(k_{i
 0}+i\epsilon)^{2}-E_{k_{i}}^{2}} + 2\pi \theta(-k_{i 0}) \delta
 (k_{i}^{2}-m^{2})\right),\label{feynman} 
\end{equation}
\end{widetext}
where we have denoted the momentum in the $i$th propagator as $k_{i}$
which is a sum of the loop momentum $k$ and some combination of the
external momenta whose explicit form is not relevant for our
discussion. The form of the integrand in \eqref{feynman} is
quite interesting. The first term in the product which will involve
only products of retarded propagators would vanish when integrated
over energy (which can be seen simply as a consequence of the fact
that  all the poles lie on the lower half of the complex plane and,
therefore,  the contour can be  closed in the upper half  plane to
yield zero). The other terms in the expansion of the right hand side
would involve terms with a number of retarded propagators and the rest
of the  propagators on-shell. If we assume that an on-shell propagator
can be thought of as a cut open line representing an on-shell particle
coming in and going out, this is very roughly a forward
scattering description, namely, a Feynman amplitude can be expressed
as a sum of diagrams that involve a number of on-shell particles
scattering in the forward direction (their momenta are unchanged in
the scattering) and retarded propagators. It is worth noting from the
form of \eqref{feynman}
that the series of forward scattering  diagrams may involve completely
disconnected diagrams (which is not the case for retarded amplitudes
at finite  temperature),
but we would like to point out that this is only a consequence of the
fact that we are looking at a time ordered Feynman amplitude. 

On the other hand, we are interested in retarded amplitudes and, as is
well known, these are quite hard to construct at zero temperature
within the context of the conventional Feynman
propagators. However, if we double the degrees of freedom (the theory
with the doubled degrees of freedom can be taken as the zero
temperature limit of the theory at finite temperature in the closed
time path formalism as described in \cite{silvana3,das:book97}), a
diagrammatic representation of retarded amplitudes can be 
constructed in a straight forward manner. With this in mind, let us
look at a scalar field theory with a $\phi^{3}$ interaction with
doubled degrees of freedom and we denote the two field degrees of
freedom as $\phi_{+}$ and $\phi_{-}$.  The propagator for the doubled
theory corresponds to a $2\times 2$ matrix 
\begin{equation}
\Delta = \left(\begin{array}{cc}
\Delta_{++} & \Delta_{+-}\\
\Delta_{-+} & \Delta_{--}
\end{array}\right),\label{prop}
\end{equation}
and in the momentum space, the components take the forms
\begin{eqnarray}
\Delta_{++} (p) & = & \lim_{\epsilon\rightarrow 0}\
\frac{i}{p^{2}-m^{2}+i\epsilon},\nonumber\\
\Delta_{+-} (p) & = & 2\pi \theta (-p_{0}) \delta
(p^{2}-m^{2}),\nonumber\\
\Delta_{-+} (p) & = & 2\pi \theta (p_{0}) \delta
(p^{2}-m^{2}),\nonumber\\
\Delta_{--} (p) & = & \lim_{\epsilon\rightarrow 0}\ -
\frac{i}{p^{2}-m^{2}-i\epsilon}.\label{zeroTmomprop}
\end{eqnarray}
(Unlike in our earlier
papers \cite{silvana1,silvana2,silvana3}, here we will follow the
simple convention of representing
quantities at zero temperature without any superscript $(T=0)$. We
will reserve the superscript $(T)$ only for quantities at nonzero
temperature in order to simplify the notation.) 

However, since the forward scattering amplitudes have a physical
description  in the mixed space (we will discuss this later), we will
analyze the problem in this context (the momentum space
analysis that is normally done  can be obtained from our results
through a Fourier transformation). In the mixed space, the components
of the propagator can be obtained from a Fourier transformation of
\eqref{zeroTmomprop} and have the forms (see also \cite{silvana3} for
various notations and conventions)
\begin{eqnarray}
\Delta_{++} (t,E) & = & 
\frac{1}{2E}\left[\theta (t) e^{-iEt} + \theta (-t)
e^{iEt}\right],\nonumber\\
\Delta_{+-} (t,E) & = & \frac{1}{2E}\ e^{iEt},\nonumber\\
\Delta_{-+} (t,E) & = & \frac{1}{2E}\ e^{-iEt},\nonumber\\
\Delta_{--} (t,E) & = & 
\frac{1}{2E}\left[\theta (t) e^{iEt} + \theta (-t)
e^{-iEt}\right], \label{zeroTprop}
\end{eqnarray}
where $E=\sqrt{\vec{p}^{\ 2} + m^{2}}$ describes the on-shell energy
of the particle and we are suppressing the ``$i\epsilon$" in the
exponents for simplicity. Thus, we see that $\Delta_{\pm\mp}$ describe
respectively the on-shell negative and positive energy
propagators. The vertices involving  the $\phi_{+}$ fields and the
$\phi_{-}$ fields differ by a relative negative sign. 

The time ordered components of the propagators in \eqref{zeroTprop}
satisfy the constraint 
\begin{equation}
\Delta_{++} + \Delta_{--} = \Delta_{+-} + \Delta_{-+}.\label{id1}
\end{equation} 
It is now simple to see that the retarded and the advanced propagators
of the theory can be identified with   
\begin{eqnarray}
\Delta_{\rm R} (t,E) & = & \Delta_{++} (t,E) - \Delta_{+-} (t,E)\nonumber\\
 & = & \theta (t)\ \frac{1}{2E}\left(e^{-iEt} - e^{iEt}\right),\nonumber\\
 \Delta_{\rm A} (t,E) & = & \Delta_{++} (t,E) - \Delta_{-+} (t,E)\nonumber\\
 & = & \theta(-t) \frac{1}{2E}\left(e^{iEt} - e^{-iEt}\right),\label{id2}
\end{eqnarray} 
much like at finite temperature \cite{das:book97}. From these definitions, we note that
\begin{equation}
\Delta_{\rm R} (-t, E) = \Delta_{\rm A} (t,E),\label{advancedretarded}
\end{equation} 
as we would expect. As a result of these relations, we can decompose
the matrix propagator in \eqref{prop} as 
\begin{equation}
\Delta = P + \Delta_{+-}\ Q,\label{decomp}
\end{equation}
where
\begin{eqnarray}
P (\Delta_{\rm R}, \Delta_{\rm A}) & = & \left(\begin{array}{ccc}
\Delta_{\rm R} & & 0\\
\Delta_{\rm R} - \Delta_{\rm A} & & -\Delta_{\rm A}
\end{array}\right),\nonumber\\
Q & = & \left(\begin{array}{cc}
1 & 1\\
1 & 1
\end{array}\right).\label{pq}
\end{eqnarray}

Let us next note that a retarded $n$-point amplitude
at any loop can be defined as follows. If we assume that the time
$t_{1}$ corresponding to the first index of the amplitude is the
largest among the time coordinates, then we have 
\begin{equation}
\Gamma_{n, {\rm R}} (t_{1},\cdots ,t_{n}) = \sum_{a_{i}=\pm}
\Gamma_{+a_{1}\cdots a_{n-1}} (t_{1},\cdots ,t_{n})\label{retarded0} 
\end{equation}
where we have suppressed the energy dependence of the amplitude for
simplicity. Here $a_{i}$ denote the ``thermal indices'' of the fields
which can take the values ``$\pm$''. For example, the retarded two point
function (self-energy)
at one loop would correspond to the sum of the two diagrams in Fig. 3.
Let us also note here for future use that, for any $n$-point
amplitude, 
\begin{equation}
\sum_{a_{i}=\pm} \Gamma_{a_{1}a_{2}a_{3}\cdots a_{n}} = 0,\label{largesttime}
\end{equation}
which follows from the largest time equation \cite{silvana3}. We are
now ready to derive the forward scattering description for retarded
amplitudes to all orders at zero temperature and we do so in two steps. 

\begin{widetext}
\begin{center}
\begin{figure}[h!]
\includegraphics[scale=0.36]{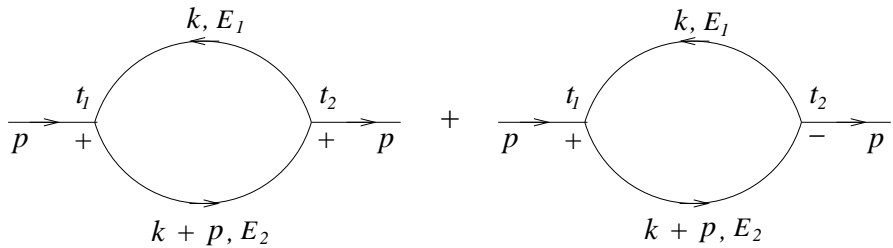}
\caption{Sum of two diagrams which gives the retarded self-energy at one-loop.}
\end{figure}
\end{center}
\end{widetext}

\subsection{Forward scattering amplitudes at one loop}

The forward scattering amplitudes for retarded $n$-point functions can
be derived algebraically at one loop (which is the reason for
separating the derivation into two cases). First we note from
\eqref{retarded0} that the
retarded amplitude consists of terms where each vertex other than the
largest time is summed over the thermal index ``$\pm$". Furthermore,
as we have already pointed out, the vertex for the $\phi_{-}$ field
has a relative negative sign compared to that for the $\phi_{+}$
field. Thus, summing over the thermal index of the graph at one loop can be
effected by multiplying the matrix propagator with a $2\times 2$
matrix $\sigma_{3}$ at the vertex where the thermal index is being
summed. For example, for the case of the retarded self energy at one
loop (see Fig. 3), we note that (once again we are neglecting factors associated
with the vertices as well as the dependence on external energies for
simplicity) 
\begin{eqnarray}
\lefteqn{\Sigma_{\rm R}^{(1)} (t_{1}-t_{2})}\nonumber\\
 & = & \int \frac{d^{3}k}{(2\pi)^{3}}\left[\Delta_{++} (t_{1}-t_{2},
 E_{1})\Delta_{++} (t_{2}-t_{1}, E_{2})\right.\nonumber\\ 
 & & \quad \left. - \Delta_{+-} (t_{1}-t_{2},E_{1})\Delta_{-+}
 (t_{2}-t_{1},E_{2})\right]\nonumber\\ 
 & = & \int \frac{d^{3}k}{(2\pi)^{3}}\left[\Delta (t_{1}-t_{2}, E_{1})
 \sigma_{3} \Delta (t_{2}-t_{1}, E_{2})\right]_{++}\!\!. 
 \end{eqnarray}
Note also that the retarded amplitude is obtained by taking the ``$++$''
 component in the matrix product (simply because we start from a ``$+$''
 vertex  and end at the same vertex).
 As a result of this simplification, the retarded $n$-point amplitude
 at one loop, shown in Fig. 4, can be written as 
 
 \begin{center}
\begin{figure}[h!]
\includegraphics[scale=0.36]{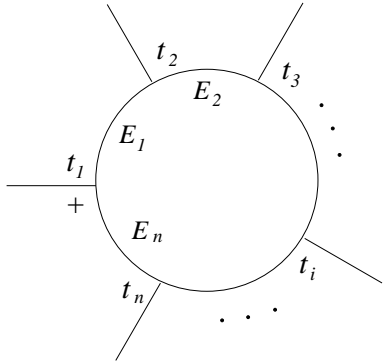}
\caption{One-loop diagram for the retarded $n$-point function. Sum
  over the thermal indices from $t_2$ to $t_n$ is to be understood.}
\end{figure}
\end{center}
 
 \begin{eqnarray}
 \lefteqn{\Gamma_{n,{\rm R}}^{(1)} (t_{1}, \cdots , t_{n})  =  \int
  \frac{d^{3}k}{(2\pi)^{3}}} \nonumber\\ 
  &\times&\left[\left(\prod_{i=1}^{n-1} \Delta (t_{i}-
  t_{i+1}, E_{i}) \sigma_{3}\right)\Delta (t_{n}-t_{1},
  E_{n})\right]_{++}\!\!\!\!\!\!\!\!.\label{npoint} 
 \end{eqnarray}
 
We can use the decomposition \eqref{decomp} of the propagator in terms of
the $P,Q$ matrices which satisfy many interesting relations. We list
below some of the relations that are useful for our discussion. 
\begin{widetext}
\begin{eqnarray}
& & Q\sigma_{3}Q = 0,\quad
 Q\sigma_{3}P(\Delta_{\rm R},\Delta_{\rm A}) =  \Delta_{\rm A}
  Q,\quad
 P (\Delta_{\rm R},\Delta_{\rm A})\sigma_{3}Q = \Delta_{\rm R} Q,\nonumber\\
& & P (\Delta_{1,{\rm R}},\Delta_{1,{\rm A}})\sigma_{3}P
  (\Delta_{2,{\rm R}},\Delta_{2,{\rm A}}) = P (\Delta_{1,{\rm
      R}}\Delta_{2,{\rm R}}, \Delta_{1,{\rm A}}\Delta_{2,{\rm A}}).  
\end{eqnarray}
\end{widetext}
Using these relations, the $n$-point amplitude in \eqref{npoint} can
be simplified considerably. First, we note that the expression on the
right hand side can at most be linear in $Q$ and, therefore, can only
have at most a single propagator of the type $\Delta_{+-}$ which as we
have seen can describe on-shell particles (see, for example,
\eqref{zeroTmomprop}). Furthermore, if we assume that an on-shell
propagator can be thought of as a cut open line (representing an
on-shell particle), it is clear that the retarded $n$-point amplitude
will involve only connected diagrams (not disconnected as can be the
case in a time ordered Feynman amplitude which we have seen earlier).
In fact, an explicit evaluation of \eqref{npoint} leads to 
\begin{widetext}
\begin{eqnarray}
\Gamma_{n,{\rm R}}^{(1)} & = & \int
\frac{d^{3}k}{(2\pi)^{3}}\left[\prod_{i=1}^{n} \Delta_{\rm R}
  (t_{i}-t_{i+1},E_{i}) + \sum_{m=0}^{n-1} \left(\prod_{i=1}^{n-m-1}
  \Delta_{\rm R} (t_{i}-t_{i+1},E_{i})\right)\Delta_{+-}
  (t_{n-m}-t_{n-m+1},E_{n-m})\right.\nonumber\\ 
&  & \quad \times \left.\left(\prod_{j=1}^{m} \Delta_{\rm A}
  (t_{n-m+j}-t_{n-m+j+1},E_{n-m+j})\right)\right], 
\end{eqnarray}
\end{widetext}
where we are identifying $t_{n+1}=t_{1}, E_{n+1}=E_{1}$. We are also using the
convention that when $m=0$ (or $m=n-1$), the term in the parenthesis has the value
\begin{equation}
\left(\prod_{i=1}^{0} \Delta (t_{i})\right)  = 1.
\end{equation}
It is obvious
that the first term in the bracket that involves only a product of
retarded propagators would vanish when integrated. Furthermore, using
\eqref{advancedretarded} we can convert all the advanced propagators
into retarded ones and write 
\begin{widetext}
\begin{eqnarray}
\Gamma_{n,{\rm R}}^{(1)} & = & \int \frac{d^{3}k}{(2\pi)^{3}}\
\sum_{m=0}^{n-1} \left(\prod_{i=1}^{n-m-1} \Delta_{\rm R}
(t_{i}-t_{i+1},E_{i})\right)\Delta_{+-}
(t_{n-m}-t_{n-m+1},E_{n-m})\nonumber\\ 
& & \quad \times \left(\prod_{j=1}^{m} \Delta_{\rm R}
(-t_{n-m+j}+t_{n-m+j+1},E_{n-m+j})\right).\label{forwardoneloop} 
\end{eqnarray}
\end{widetext}
This gives a forward scattering description for the retarded
$n$-point amplitude at one loop at zero temperature. Each term in the
series is a number
of retarded propagators with one on-shell propagator ($\Delta_{+-}$)
leading to  the forward scattering of a single on-shell particle in
all possible manner in a connected causal manner. Unlike the Feynman
amplitude in \eqref{feynman}, the forward scattering description for
the retarded $n$-point function does not involve disconnected diagrams
which is also reflected in the basic definition of the retarded
amplitudes in terms of nested commutators that we will discuss in
section {\bf III}. From
\eqref{forwardoneloop}, we can easily derive a recursion relation for
the integrands of the one loop retarded amplitudes of the form 
\begin{eqnarray}
\gamma_{n+1,{\rm R}}^{(1)} & = & \prod_{i=1}^{n}\Delta_{\rm
 R}(t_{i}-t_{i+1},E_{i}) \Delta_{+-}
 (t_{n+1}-t_{1},E_{n+1})\nonumber\\ 
 & & \quad+ \gamma_{n,{\rm R}}^{(1)} \Delta_{\rm R}
 (-t_{n+1}+t_{1},E_{n+1}).\label{onelooprelation} 
 \end{eqnarray}
 
\subsection{Forward scattering amplitude at higher loops}

The simple algebraic derivation of the one loop forward scattering
amplitudes for the zero temperature retarded $n$-point function does
not carry over to higher loops in general. This is simply because of
the fact that at higher loops, more than two propagators (internal
lines) may be connected to a given vertex. In such a case, the
convenient matrix structure for retarded amplitudes that arises in one
loop (because only two propagators can be connected to a vertex) is
not present. Nonetheless, the forward scattering description for some
simple higher loop graphs can be easily derived algebraically as
follows. Let us
consider the scalar $\phi^{n+2}$ theory. In this case, the retarded
self-energy at  $n$-loops (see Fig. 5) can be written as 

\begin{eqnarray}
\lefteqn{\Sigma_{\rm R}^{(n)}}\nonumber\\
& = & \int
  \frac{d^{3}k_{n}}{(2\pi)^{3}}\left[\Sigma_{++}^{(n-1)}\Delta_{++}(E_{n+1}) 
+ \Sigma_{+-}^{(n-1)} \Delta_{+-} (E_{n+1})\right]\nonumber\\ 
 & = & \int \frac{d^{3}k_{n}}{(2\pi)^{3}}\left[\Sigma_{\rm R}^{(n-1)}
    \Delta_{+-}(E_{n+1}) - \Sigma_{+-}^{(n-1)} \Delta_{\rm R}
    (E_{n+1})\right]\nonumber\\ 
 & = & \int \frac{d^{3}k_{n}}{(2\pi)^{3}}\ \Sigma_{\rm R}^{(n-1)}
  \Delta_{+-}(E_{n+1}) \nonumber\\ 
 & & \quad + \int \prod_{i=1}^{n}
  \frac{d^{3}k_{i}}{(2\pi)^{3}}\Delta_{+-}(E_{i}) \Delta_{\rm
    R}(E_{n+1}).\label{higherlooprelation} 
\end{eqnarray}
Here $k_{i}, i=1,2,\cdots , n$ denote the $n$ independent momenta of
the loops and in the intermediate steps, we have used various relations
such as  \eqref{id2} and have neglected terms involving products of
retarded quantities in the integrand (which will vanish upon
integration). The recursion relation in \eqref{higherlooprelation} is
interesting for two reasons. First, it shows the generic feature in
higher loops that any retarded amplitude at $n$ loops can be given a
forward scattering description in terms of retarded amplitudes at
lower order. Second, the recursion relation \eqref{higherlooprelation}
can be thought of as a recipe for opening up loops \cite{feynman} for
this  particular diagram.

\begin{widetext}
\begin{center}
\begin{figure}[h!]
\includegraphics[scale=0.36]{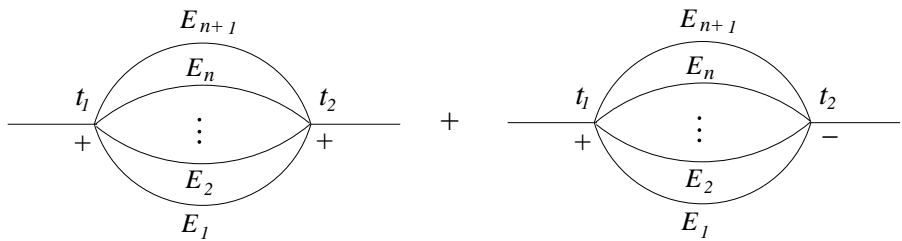}
\caption{Sum of diagrams which gives the  $n$-loop retarded self-energy
  in $\phi^{n+2}$ theory.}
\end{figure}
\end{center}
\end{widetext}

Although the forward scattering description for some simple higher
loop diagrams can be derived algebraically, for a general higher loop
amplitude, this is best established diagrammatically. For this purpose, let
us introduce the graphical representation for the two parts of the
matrix propagator in \eqref{decomp} as 
\begin{eqnarray}
P_{ab} & = &  \begin{array}{c}\includegraphics[scale=0.3]{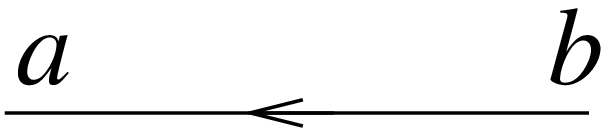}\end{array} 
   ,\nonumber\\
\Delta_{+-} Q_{ab} & = & \begin{array}{c}\includegraphics[scale=0.3]{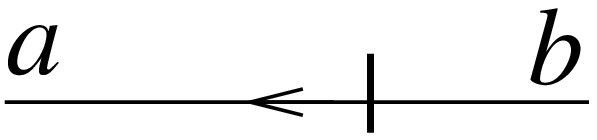}\end{array}   .\label{graph}
\end{eqnarray}
There are two important things to note here. First, the ``cut"
propagator corresponds to the on-shell propagator and all the elements
of the matrix $Q$ have the value unity so that the form of the ``cut"
propagator is the same independent of the indices
$a,b=\pm$. Second,
since the propagator $P$ is directional, we choose the convention of
taking the direction of time flow to be towards the ``+" vertex in a
retarded amplitude (which corresponds to the largest time). This
simplifies the derivation and is physically meaningful to give a
causal evolution for the amplitudes. The direction of the time flow at
other vertices, where the thermal indices are being summed over, is
unimportant. It is clear that a graphical decomposition of the
propagator in the manner as described in \eqref{graph} would allow us
to write any diagram as a sum of diagrams each consisting of certain
numbers of ``$P$" propagators and the remaining ones ``cut"
propagators. Each ``cut" propagator corresponds to an on-shell
propagator and, therefore, can be thought of as a cut open line
representing forward scattering of an on-shell particle. This 
can, therefore, also be thought of as a graphical description of the
opening up of loops. 

In this process of ``opening up of loops", we may run into
disconnected diagrams. 
As we will discuss in more detail in the next section, the retarded
amplitudes correspond to vacuum expectation values of products of
nested commutators and as such cannot have disconnected diagrams
(which would correspond to products of vacuum expectation
values). This can, of course, be checked graph by graph at any
order as was done at finite temperature in \cite{adilson}. However,
this can also be seen graphically as follows.
Suppose, in this process of opening up of a diagram, it separates into
two disconnected parts as shown in Fig. 6. 

\begin{widetext}
\begin{center}
\begin{figure}[h!]
\includegraphics[scale=0.62]{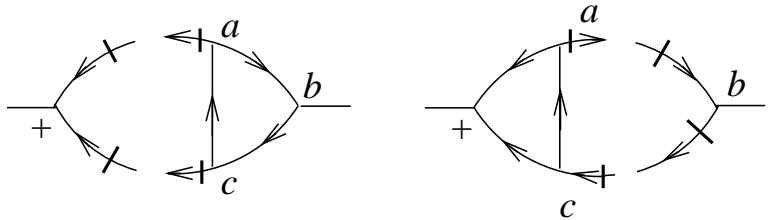}
\caption{Two ways of cutting the two-loop self-energy diagram with two
  cut propagators which render the graph disconnected. The thermal
  indices $a$, $b$ and $c$ are being summed}
\end{figure}
\end{center}
\end{widetext}

\noindent In this case, there are two distinct possibilities. First,  one
of the disconnected parts contains the ``$+$" vertex corresponding to
the largest time and the other a connected part involving only
vertices whose thermal indices are being summed over. In this case,
the second part would vanish because of the identity
\eqref{largesttime}. The other possibility is that one of the
disconnected parts is a connected diagram involving the ``$+$" vertex
and the other simply consists of a vertex whose thermal index is being summed
over. Once again, when we sum over the thermal index of this
disconnected vertex (with a fixed distribution of the other thermal
indices), the diagram would sum to zero (since the ``$+$'' and the
``$-$'' 
vertices have a relative negative sign). The crucial ingredient that
allows this argument to go through is the special property that a
``cut" propagator is the same for any value of the thermal indices. 
As a consequence of this nice result, it follows now that for an arbitrary
retarded amplitude at $n$ loops, there can at the most be $n$ number
of ``cut" propagators in a diagram because more cuts than that would
render the graph disconnected. Furthermore, only those propagators in
a diagram can be ``cut" propagators  (even when their number is less
than or equal to $n$) if they do not render the diagram disconnected.  
The propagators that are not cut correspond to the ``$P$" propagators
which can be seen explicitly from \eqref{pq} to be lower triangular with 
\begin{equation}
P_{++} = \Delta_{\rm R},\quad P_{+-} = 0,\quad P_{\rm R} =
P_{++}-P_{+-} = \Delta_{\rm R}.\label{lowertriangular} 
\end{equation} 
As a result, in a retarded amplitude, the uncut propagators simply
correspond to $\Delta_{\rm R}$ propagators (which can be thought of as
retarded ``$P$" propagators).  

From these interesting properties follows a simple recipe for
constructing the forward scattering amplitudes for a retarded graph at
$n$ loops. Start with a given graph and write it as sum of all
possible diagrams involving ``uncut" and ``cut" propagators (open
lines) such that none of the diagrams is disconnected and that there are at
the most $n$ number of ``cut" propagators. The diagram with $n$ number
of ``cut'' propagators would correspond to a tree level forward scattering
diagram with intermediate retarded propagators. Any diagram with the
number of ``cut" propagators less than $n$, would involve vertex
diagrams of lower order (loop) as well as intermediate propagators
that are retarded. The vertex diagrams of lower order would correspond
to retarded diagrams with respect to the ``$P$" propagators and,
therefore, would involve only $\Delta_{\rm R}$ propagators. This is the forward
scattering description for a retarded diagram at any loop at zero
temperature. 

The above recipe is already  obvious in the examples that we have
discussed before. Let us illustrate these as well as some 
nontrivial examples at higher loops in a graphical manner. First, let
us look at the 
one loop retarded self-energy in the $\phi^{3}$ theory which can be
written as 
\begin{widetext}
\begin{eqnarray}
\Sigma_{\rm R}^{(1)} & = & 
\begin{array}{c}\includegraphics[scale=0.2]{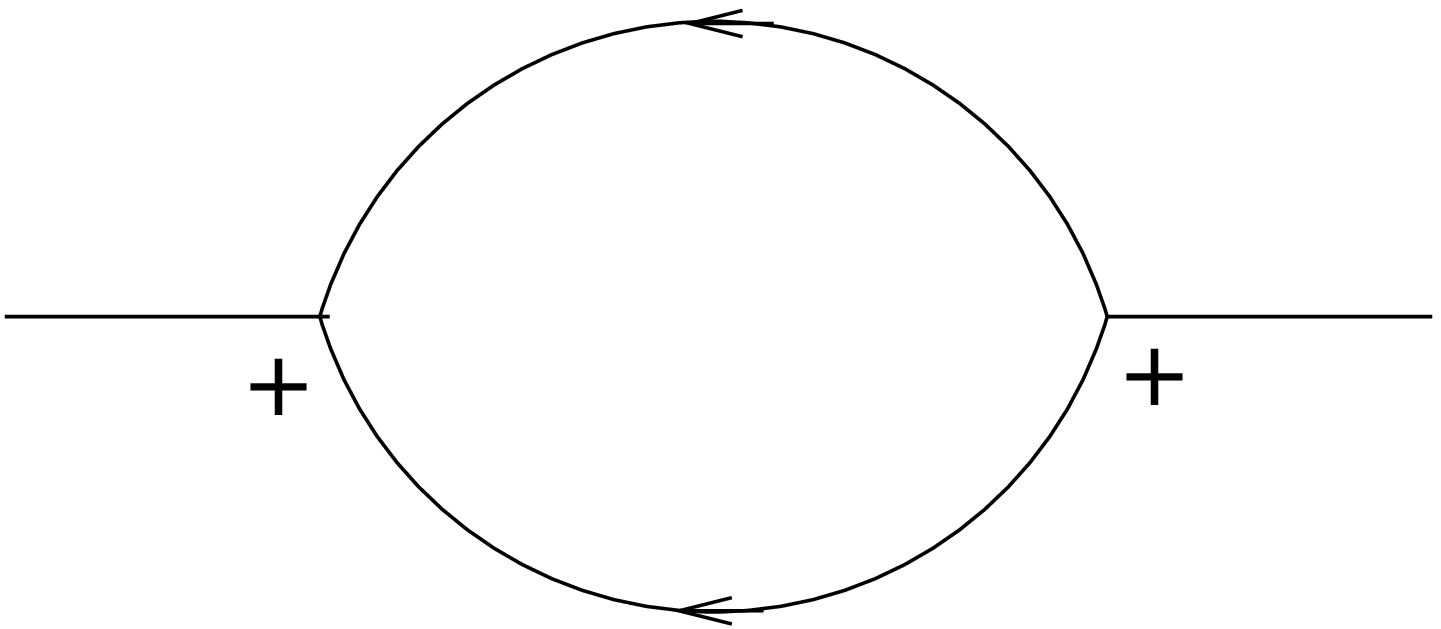}\end{array}+
\begin{array}{c}\includegraphics[scale=0.2]{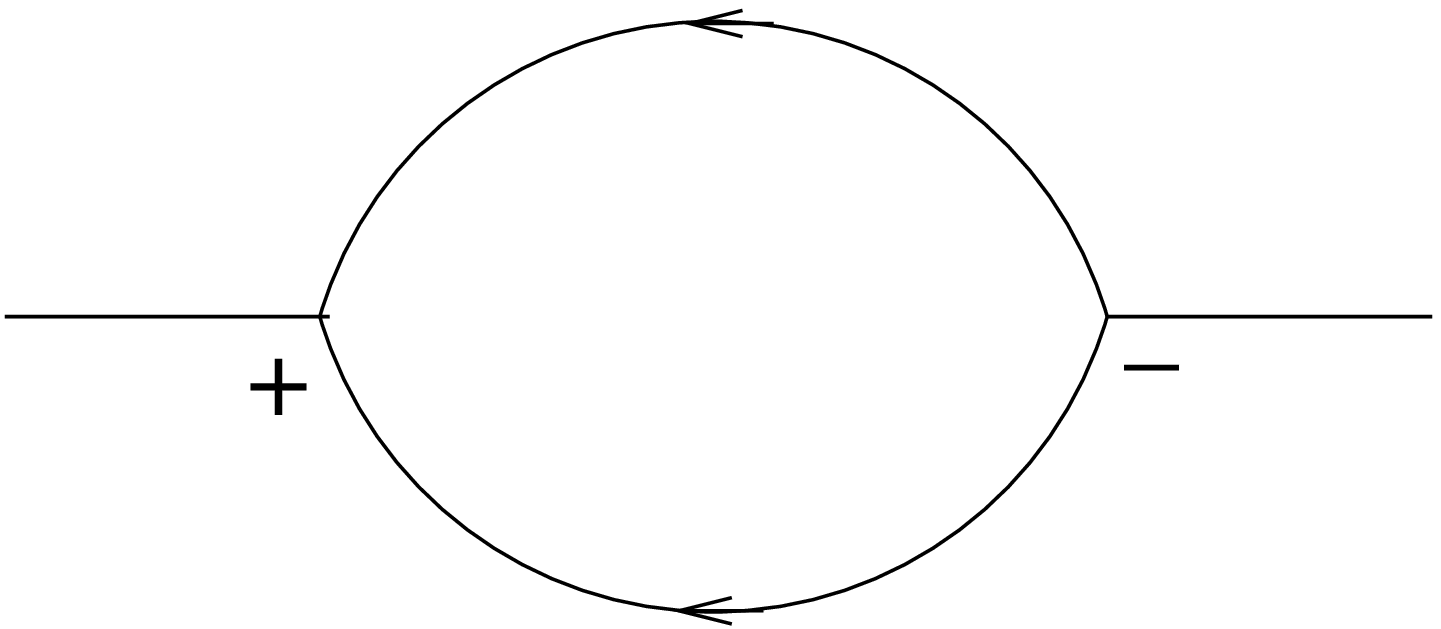}\end{array} 
+  
\begin{array}{c}\includegraphics[scale=0.2]{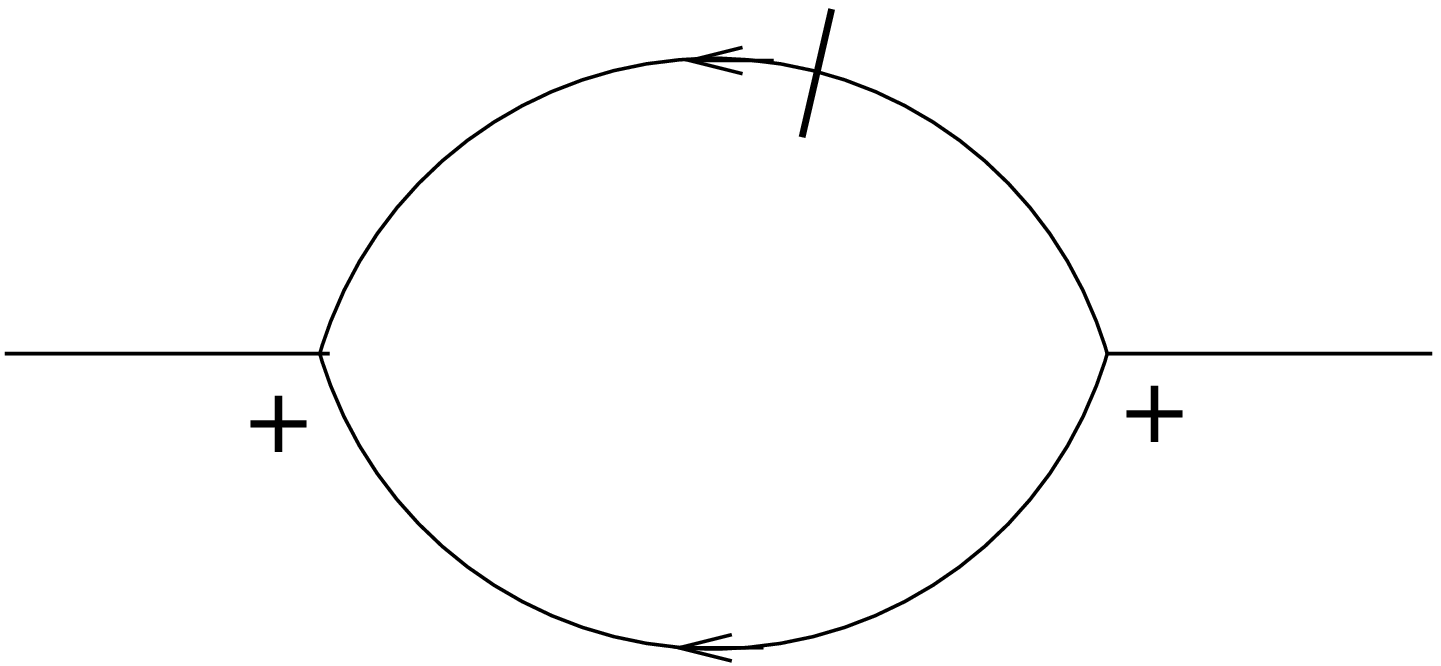}\end{array}\nonumber\\
 &  & +
\begin{array}{c}\includegraphics[scale=0.2]{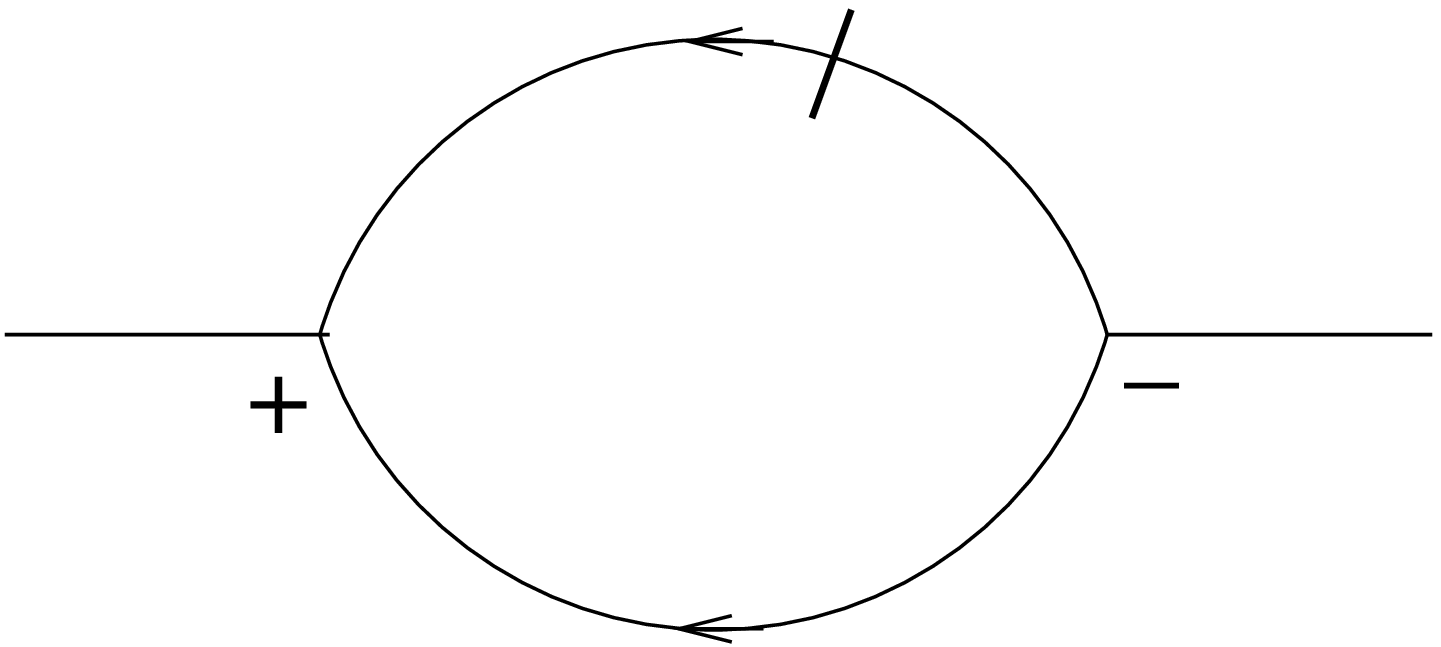}\end{array} 
 +  
\begin{array}{c}\includegraphics[scale=0.2]{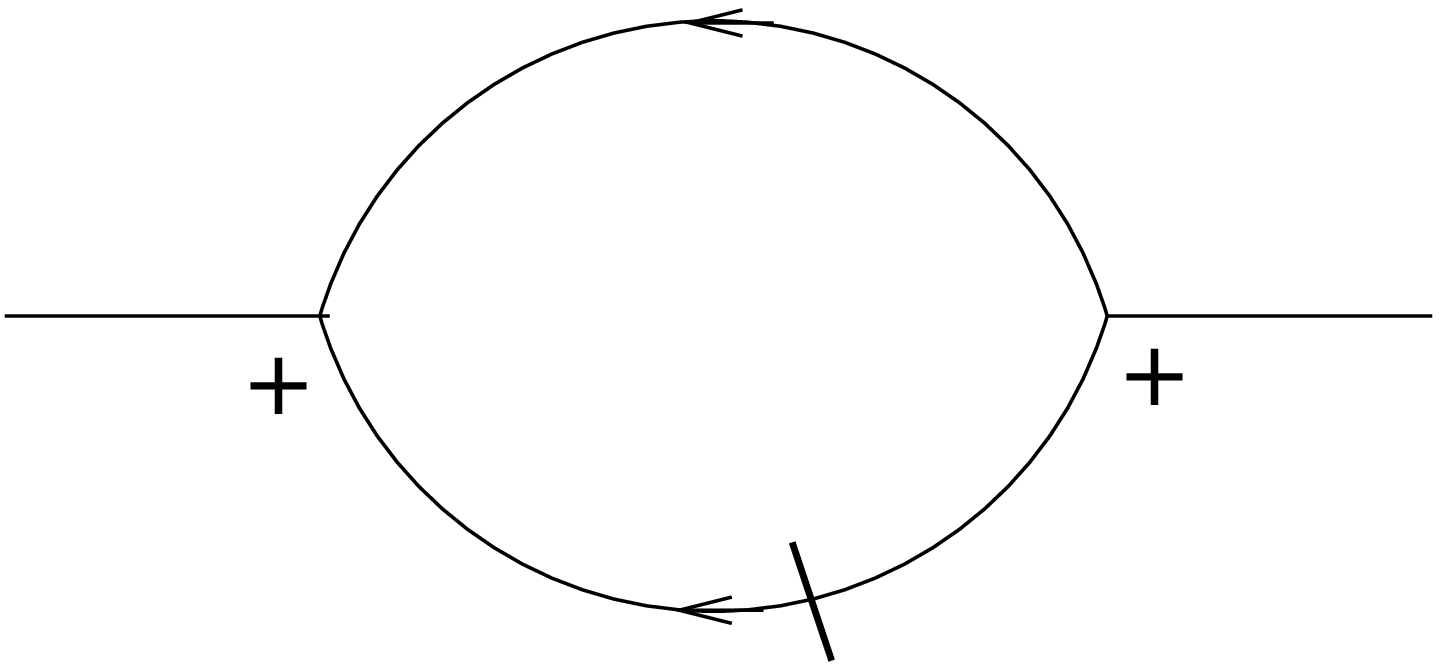}\end{array}+
\begin{array}{c}\includegraphics[scale=0.2]{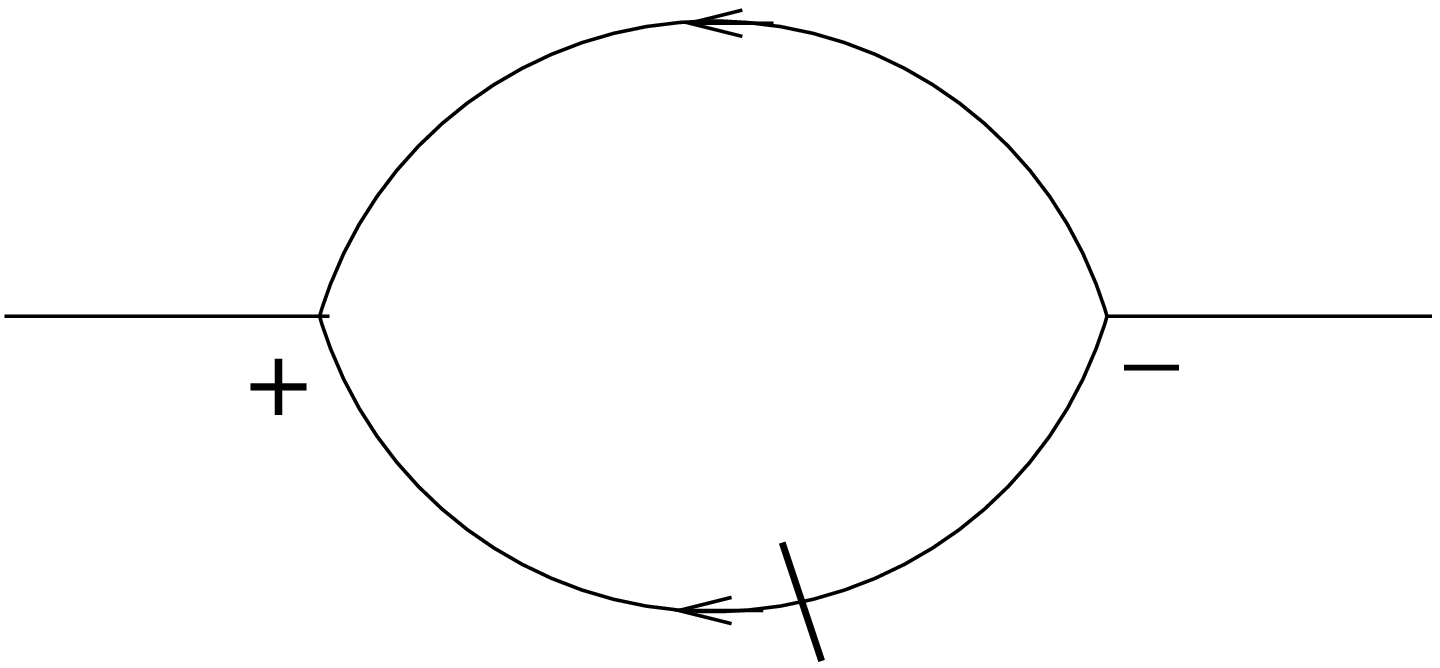}\end{array} 
\nonumber\\
& = &
\Sigma_R^{(1)(P)}+
\begin{array}{c}\includegraphics[scale=0.24]{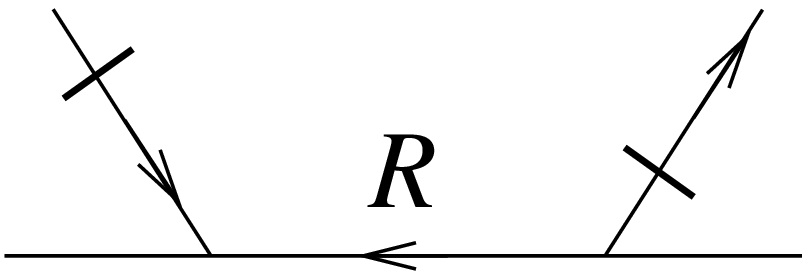}\end{array}+
\begin{array}{c}\includegraphics[scale=0.24]{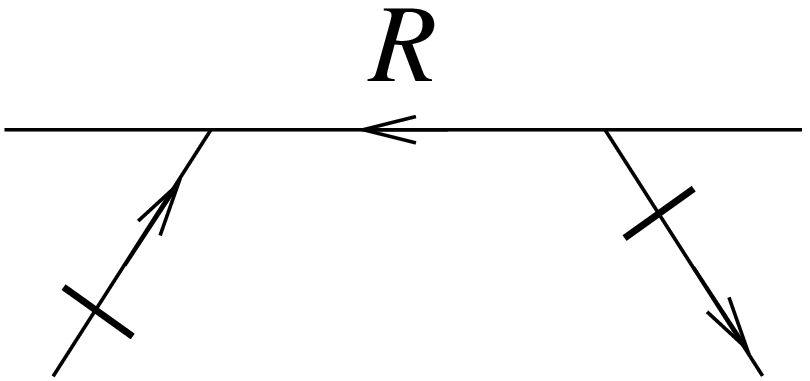}\end{array} .
\label{2pt}
\end{eqnarray}
\end{widetext}
Similarly, the retarded  three point function at one loop in the
$\phi^{3}$ theory takes the form
\begin{widetext} 
\begin{eqnarray}
\Gamma_{3,{\rm R}}^{(1)} & = & 
\begin{array}{c}\includegraphics[scale=0.2]{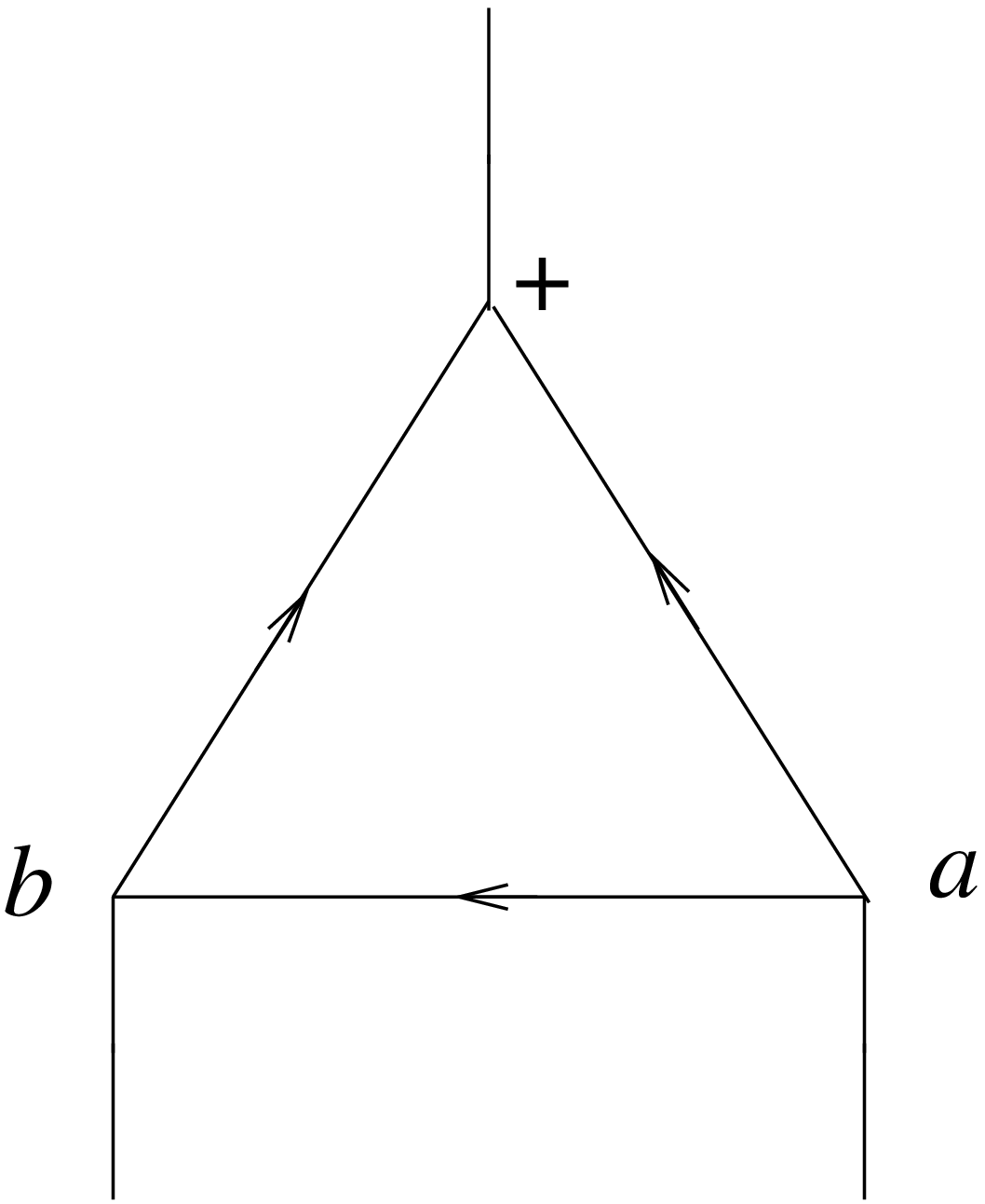}\end{array}+
\begin{array}{c}\includegraphics[scale=0.2]{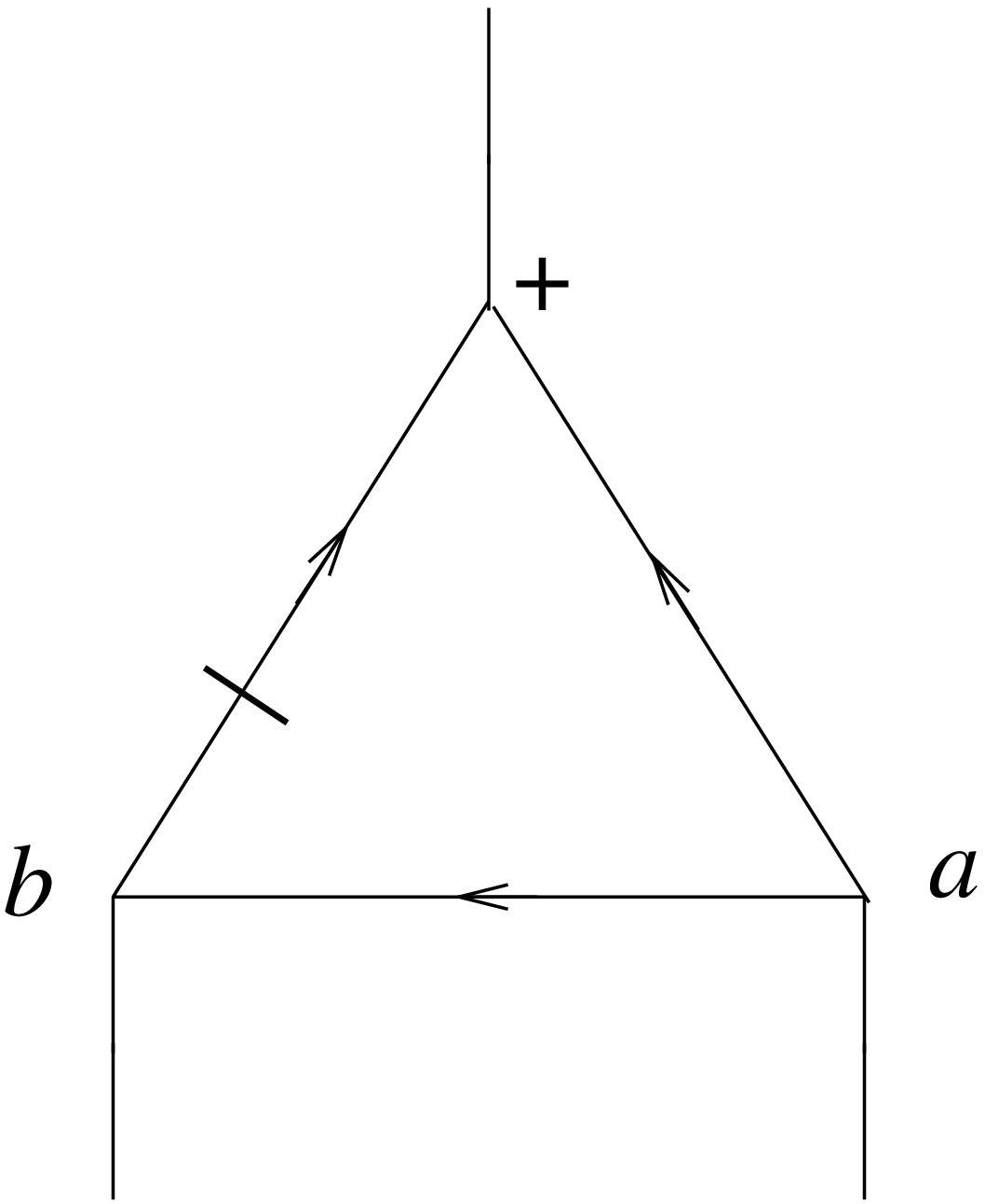}\end{array} 
 +  
\begin{array}{c}\includegraphics[scale=0.2]{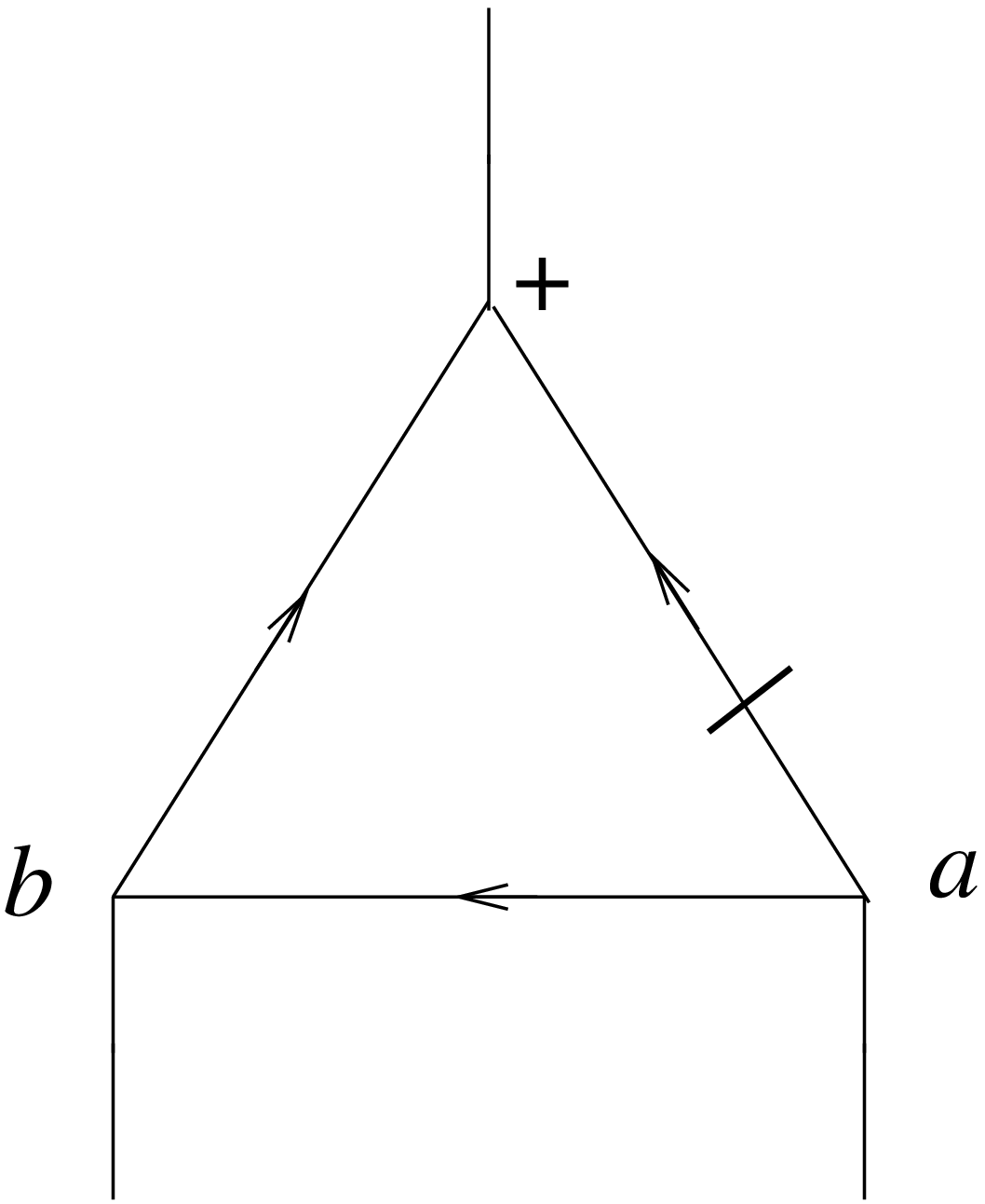}\end{array}+
\begin{array}{c}\includegraphics[scale=0.2]{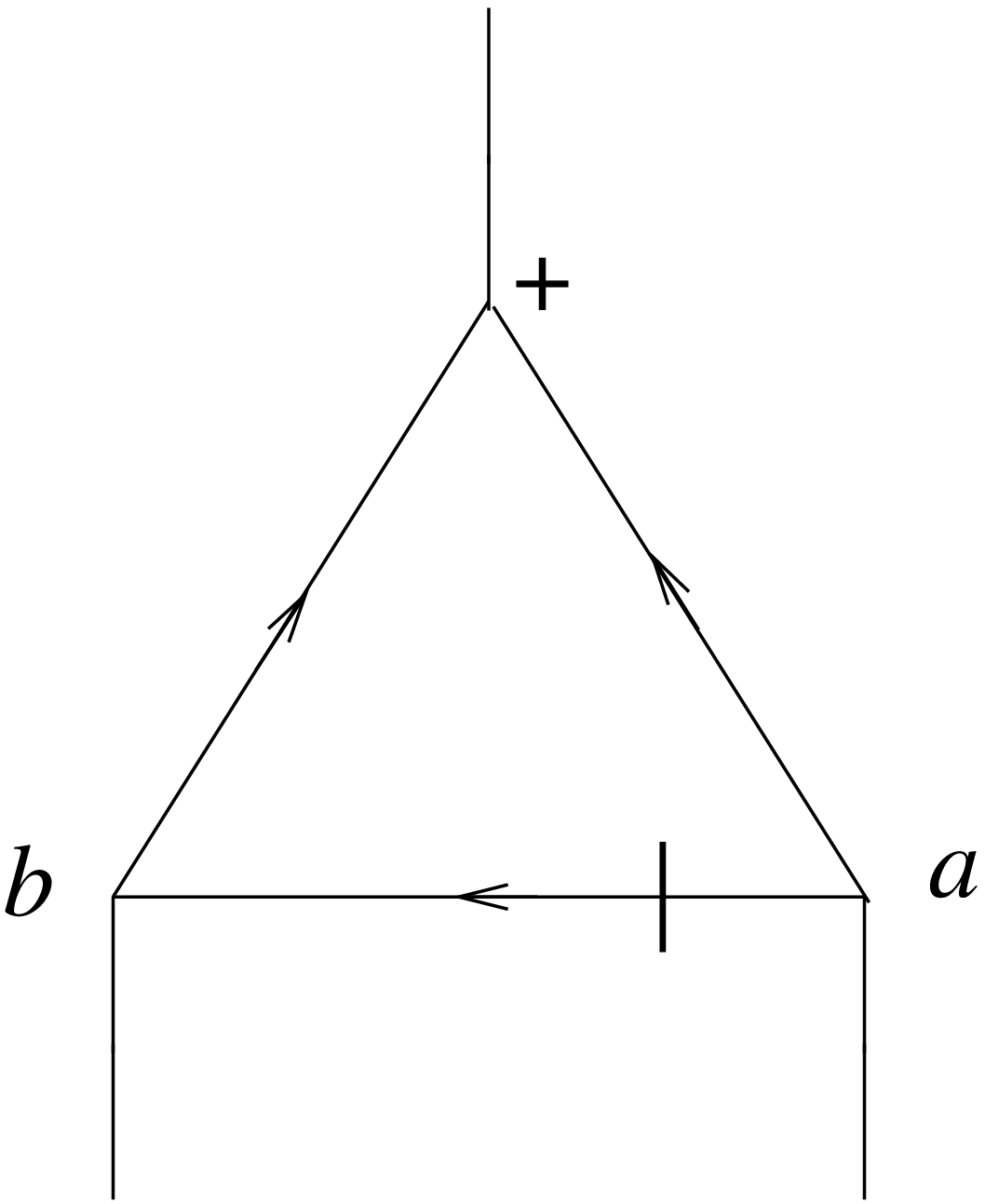}\end{array} 
\nonumber\\
& = & 
\Gamma_{3,R}^{(1)(P)}+
\begin{array}{c}\includegraphics[scale=0.24]{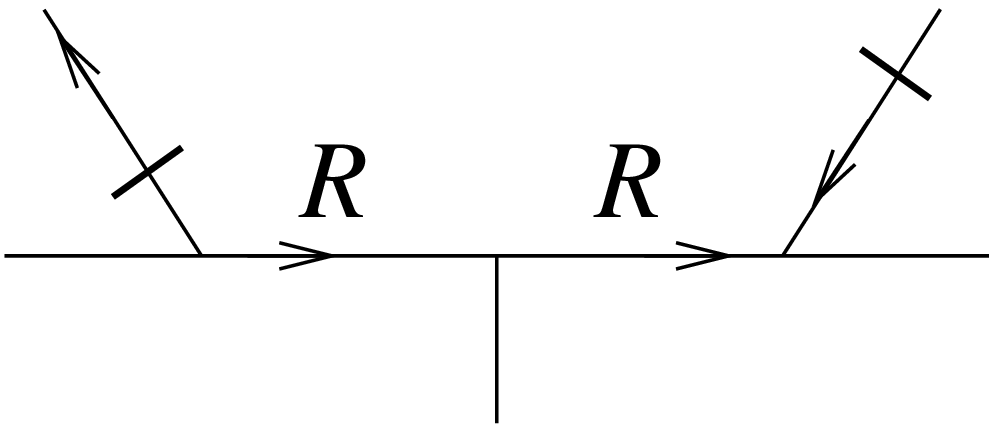}\end{array}
+  
\begin{array}{c}\includegraphics[scale=0.24]{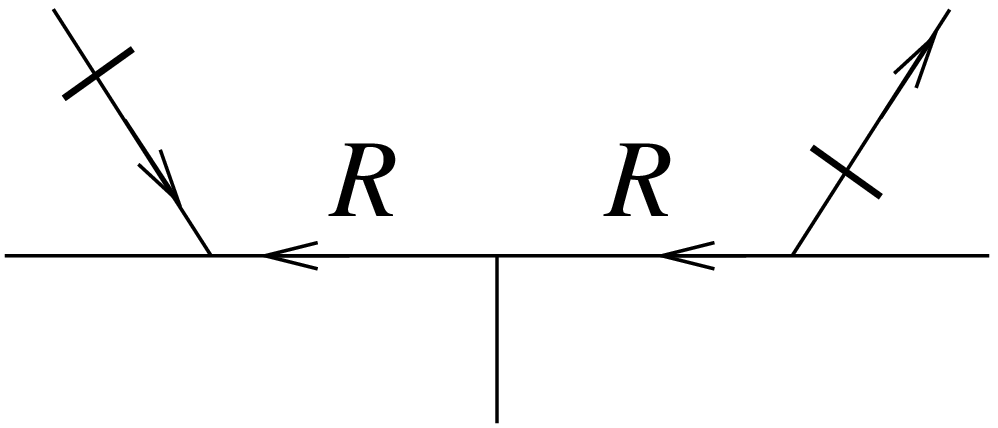}\end{array} 
\begin{array}{c}\includegraphics[scale=0.24]{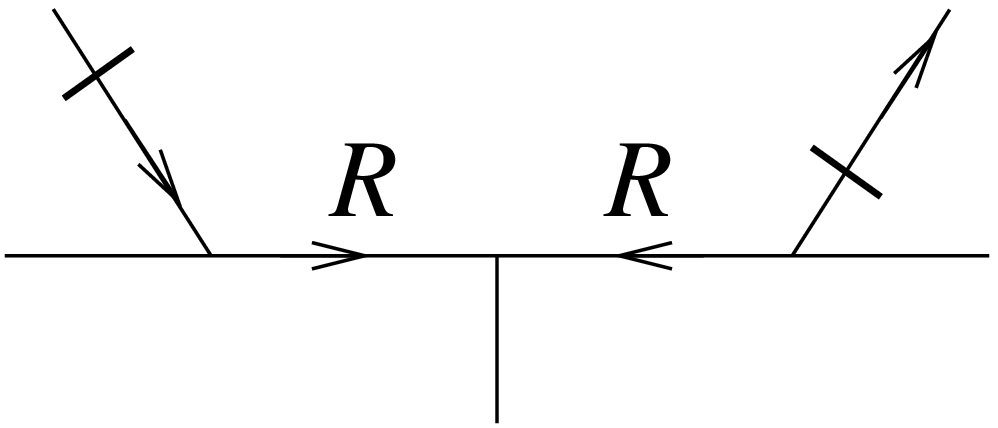}\end{array} .
\label{3pt}
\end{eqnarray}
\end{widetext}
Let us next look at the retarded self-energy at two loops in the
$\phi^{4}$ theory, which takes the form
\begin{widetext} 
\begin{eqnarray}
\Sigma_{\rm R}^{(2)} & = & 
\begin{array}{c}\includegraphics[scale=0.16]{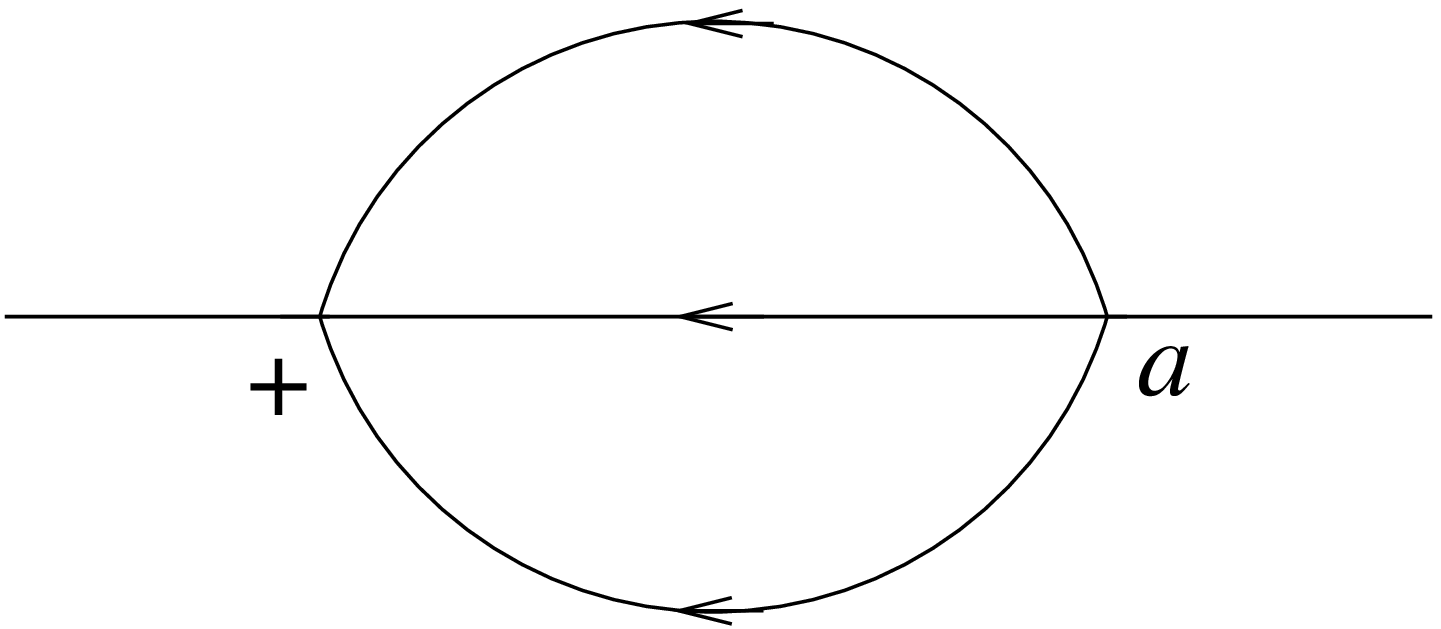}\end{array}+
\begin{array}{c}\includegraphics[scale=0.16]{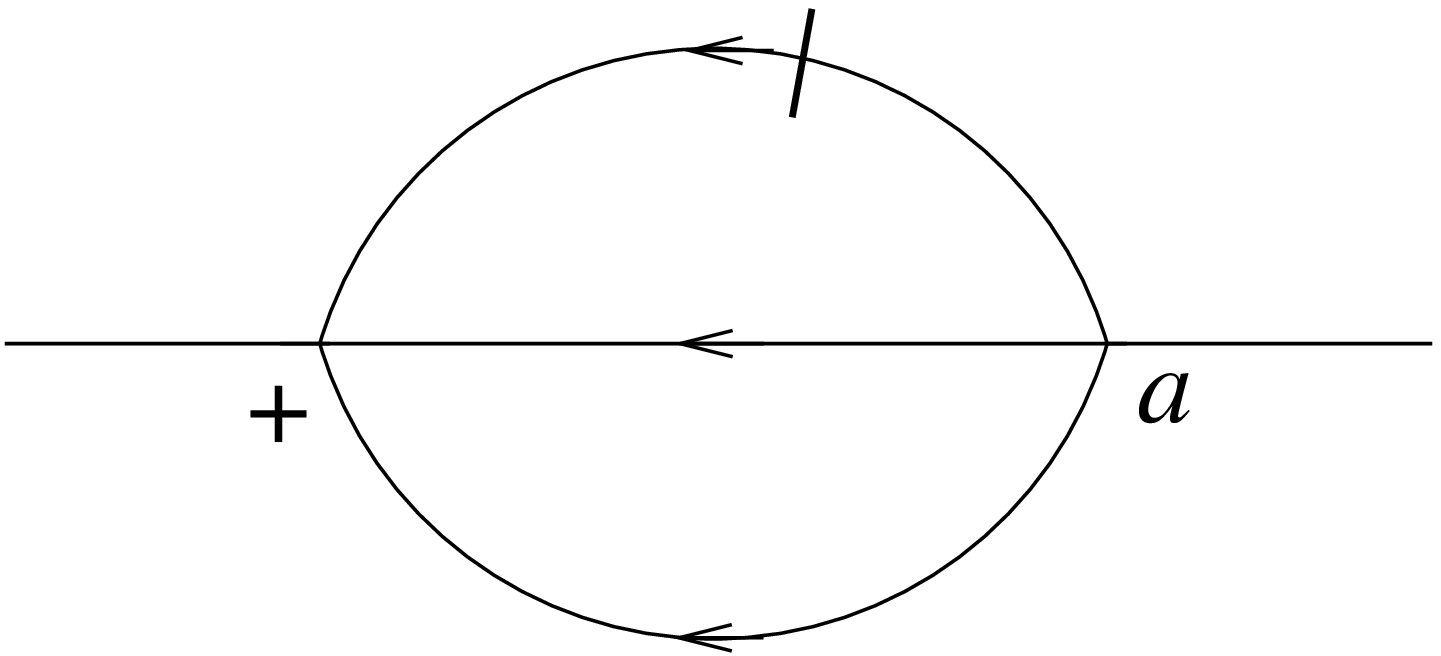}\end{array}
 + 
\begin{array}{c}\includegraphics[scale=0.16]{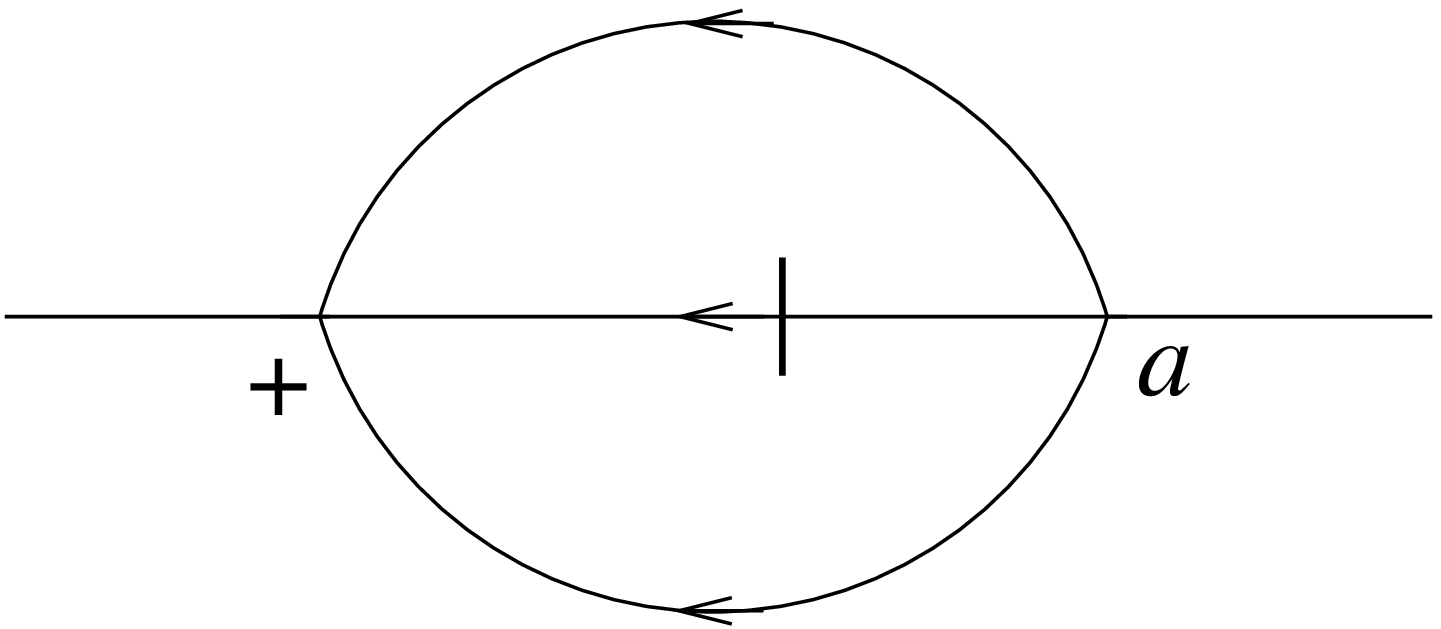}\end{array}+
\begin{array}{c}\includegraphics[scale=0.16]{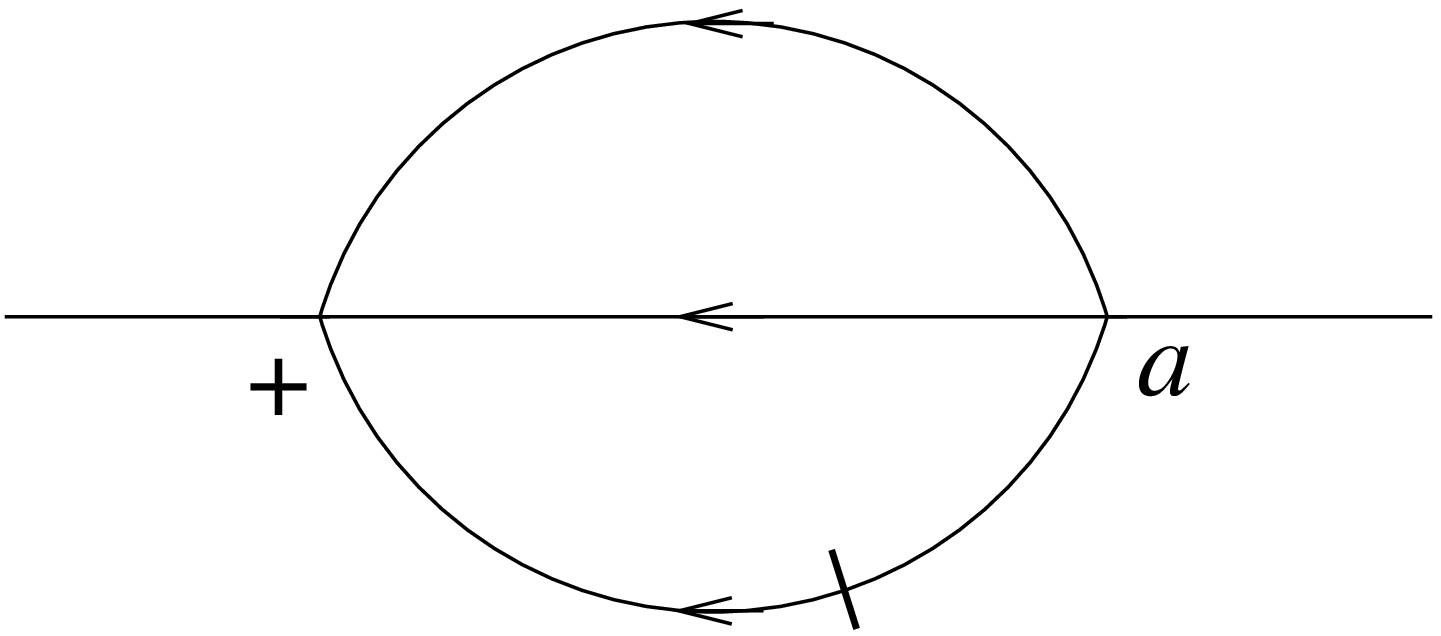}\end{array} \nonumber\\
&  & +
\begin{array}{c}\includegraphics[scale=0.16]{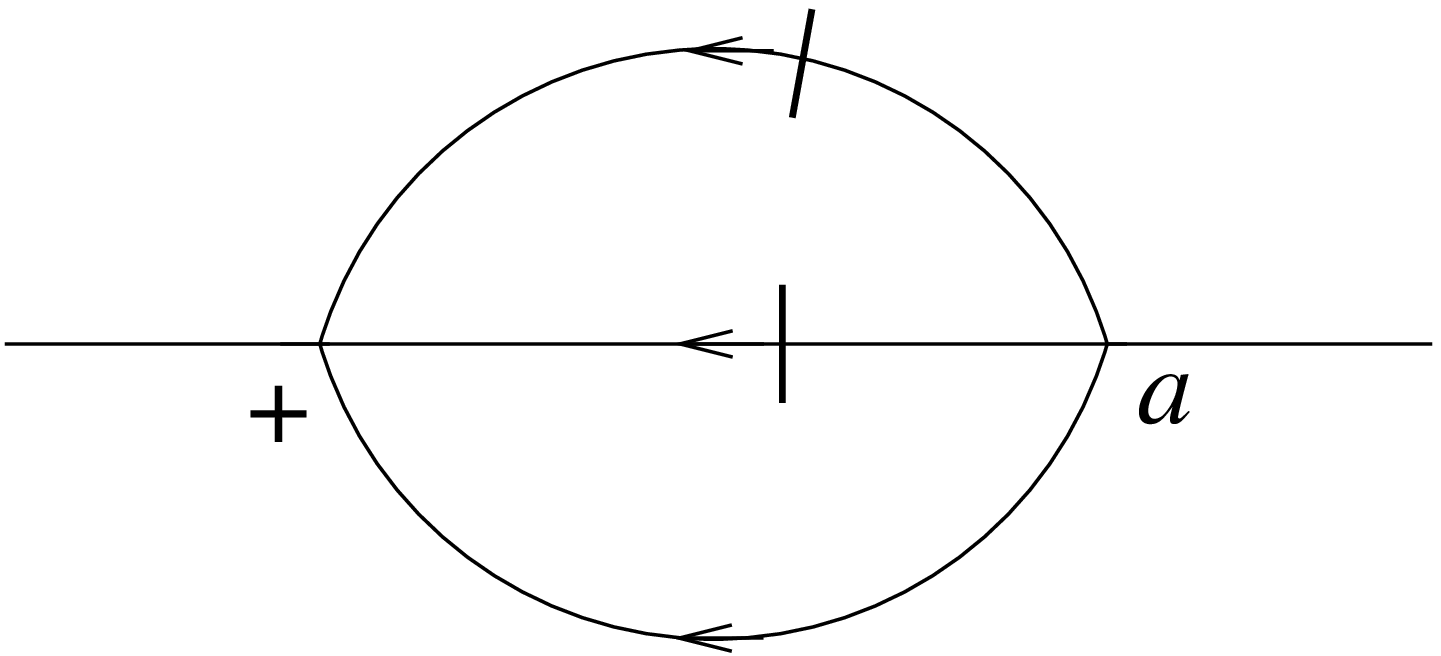}\end{array}
 + 
\begin{array}{c}\includegraphics[scale=0.16]{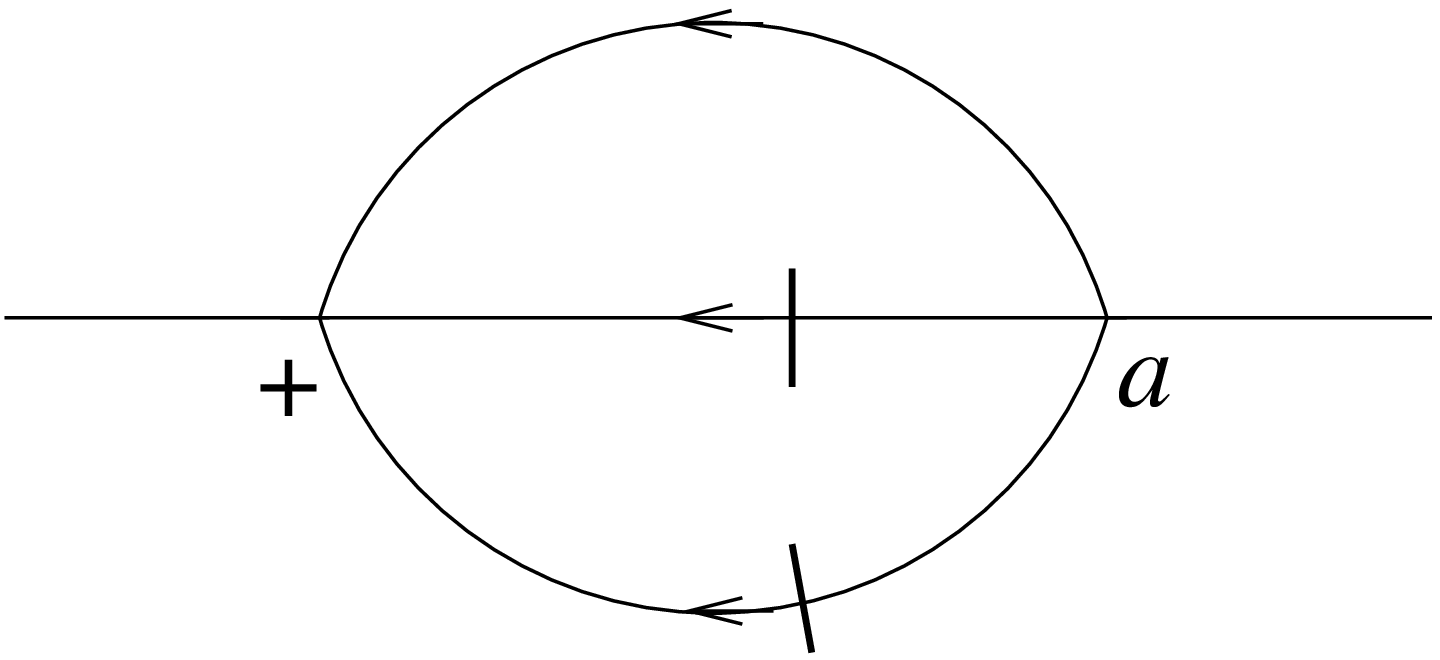}\end{array}+
\begin{array}{c}\includegraphics[scale=0.16]{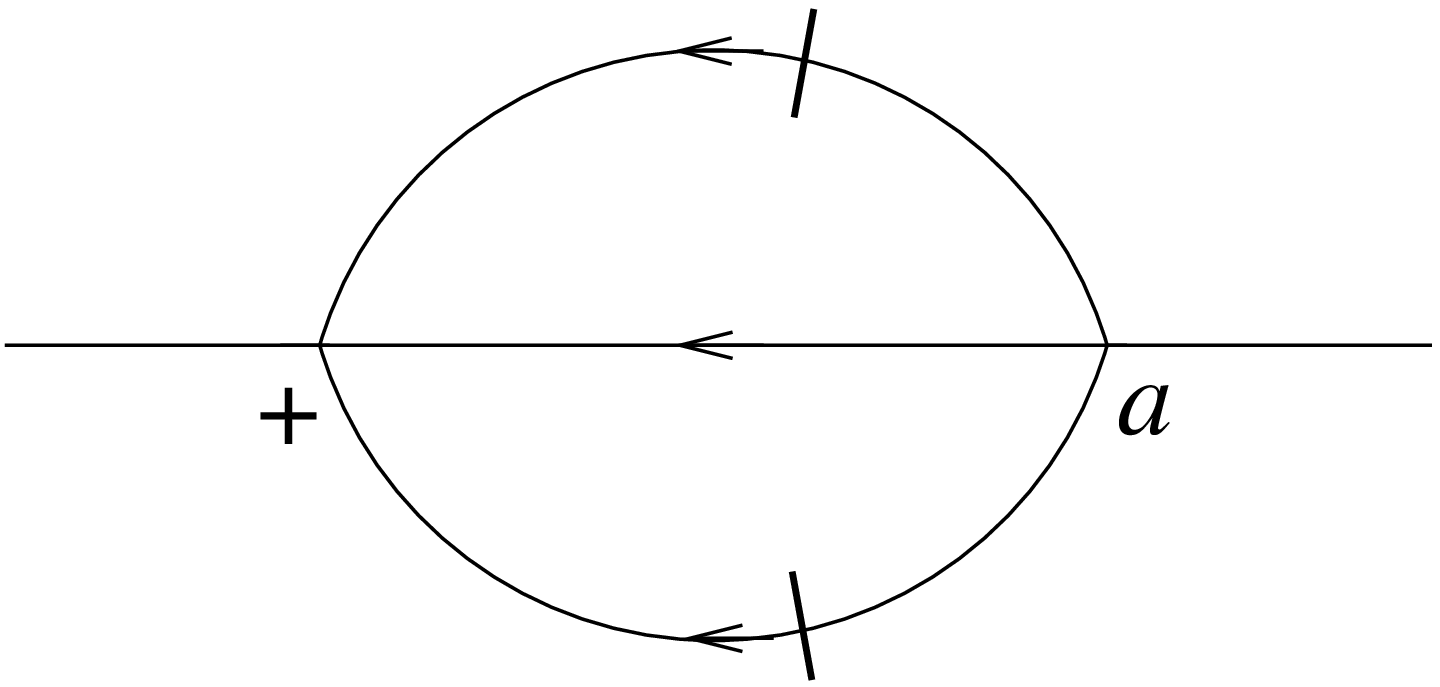}\end{array}
\nonumber\\ 
& = & 
\Sigma_R^{(2)(P)} + \;\;3\;\;
\begin{array}{c}\includegraphics[scale=0.16]{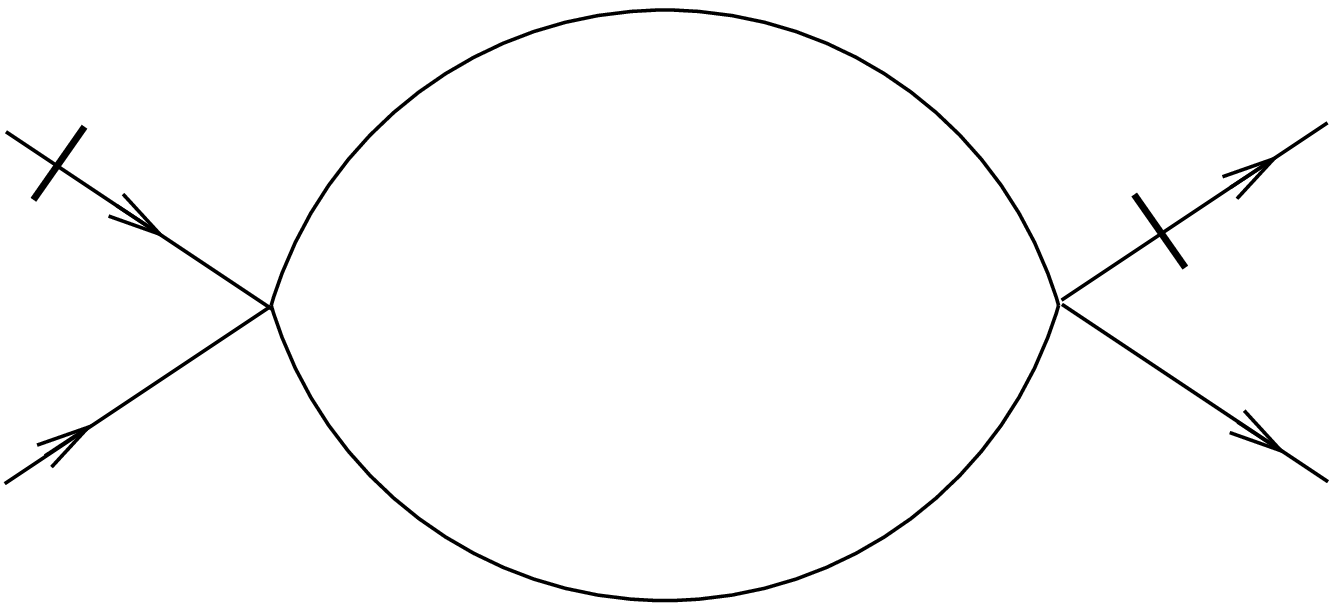}\end{array} 
\!\!\!\!\!\!\!\!\!\!\!\!\!\!\!\!\!\!\!\!\!\!\!\!\!\!\!\!\!
\Sigma_R^{(1)(P)} 
\;\;\;\;\;\;\;
\;\;+\;\; 3\;\;
\begin{array}{c}\includegraphics[scale=0.16]{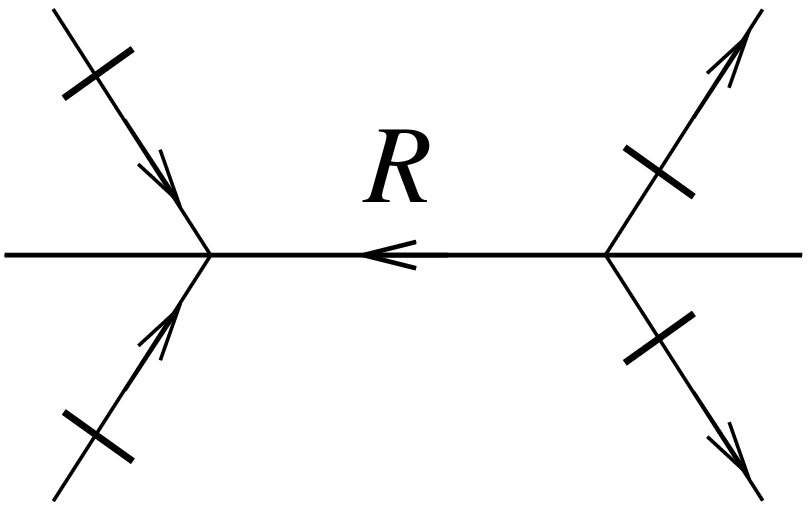}\end{array}.
\label{2pt1}
\end{eqnarray}
\end{widetext}
Here and in what follows a multiplicative factor in a graph denotes
symbolically the number of distinct graphs of the same topology that
can be drawn. 
Let us next look at a nontrivial diagram for the retarded self-energy at two loops 
in the $\phi^{3}$ theory which takes the form
\begin{widetext} 
\begin{equation}
\Sigma_{\rm R}^{(2)}  =  
\begin{array}{c}\includegraphics[scale=0.19]{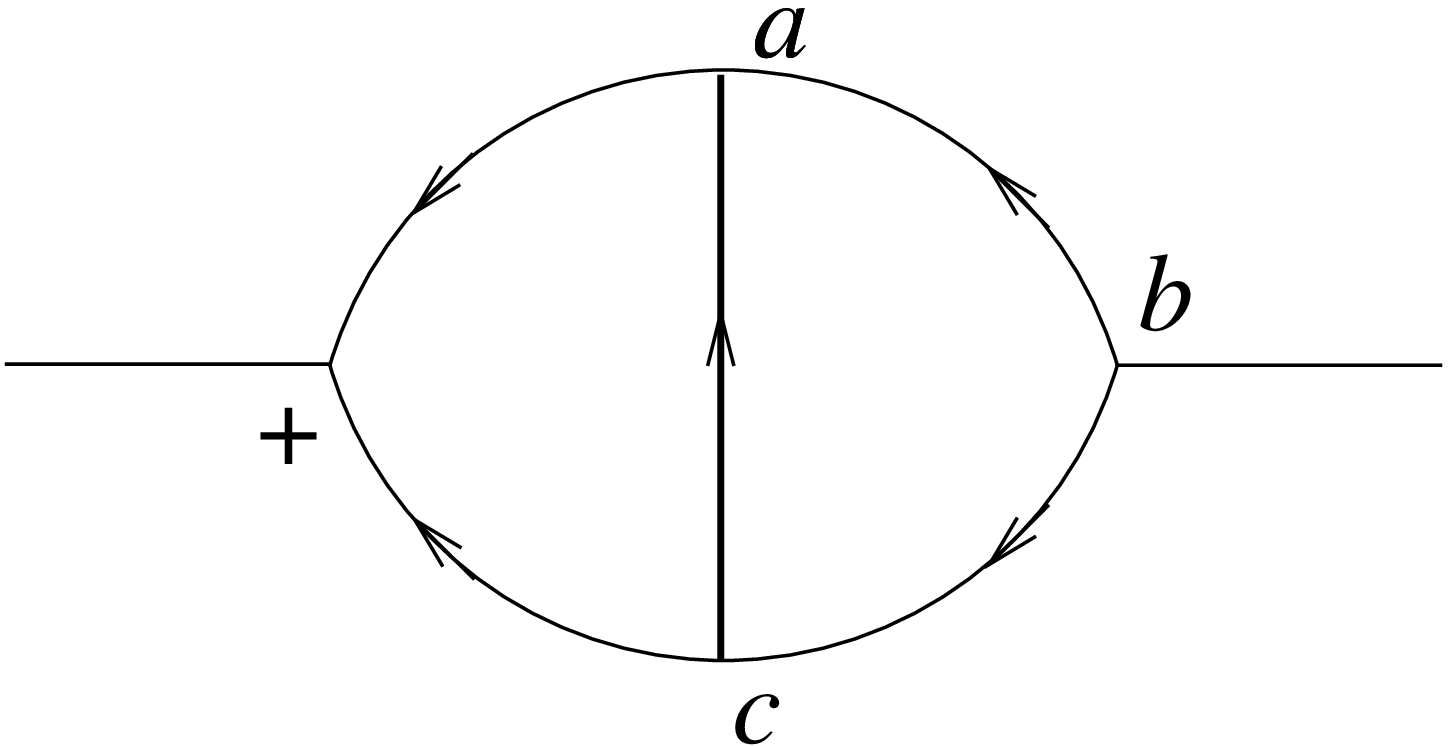}\end{array}
 =  
\Sigma_{\rm R}^{(2)(P)}+
\begin{array}{c}\includegraphics[scale=0.20]{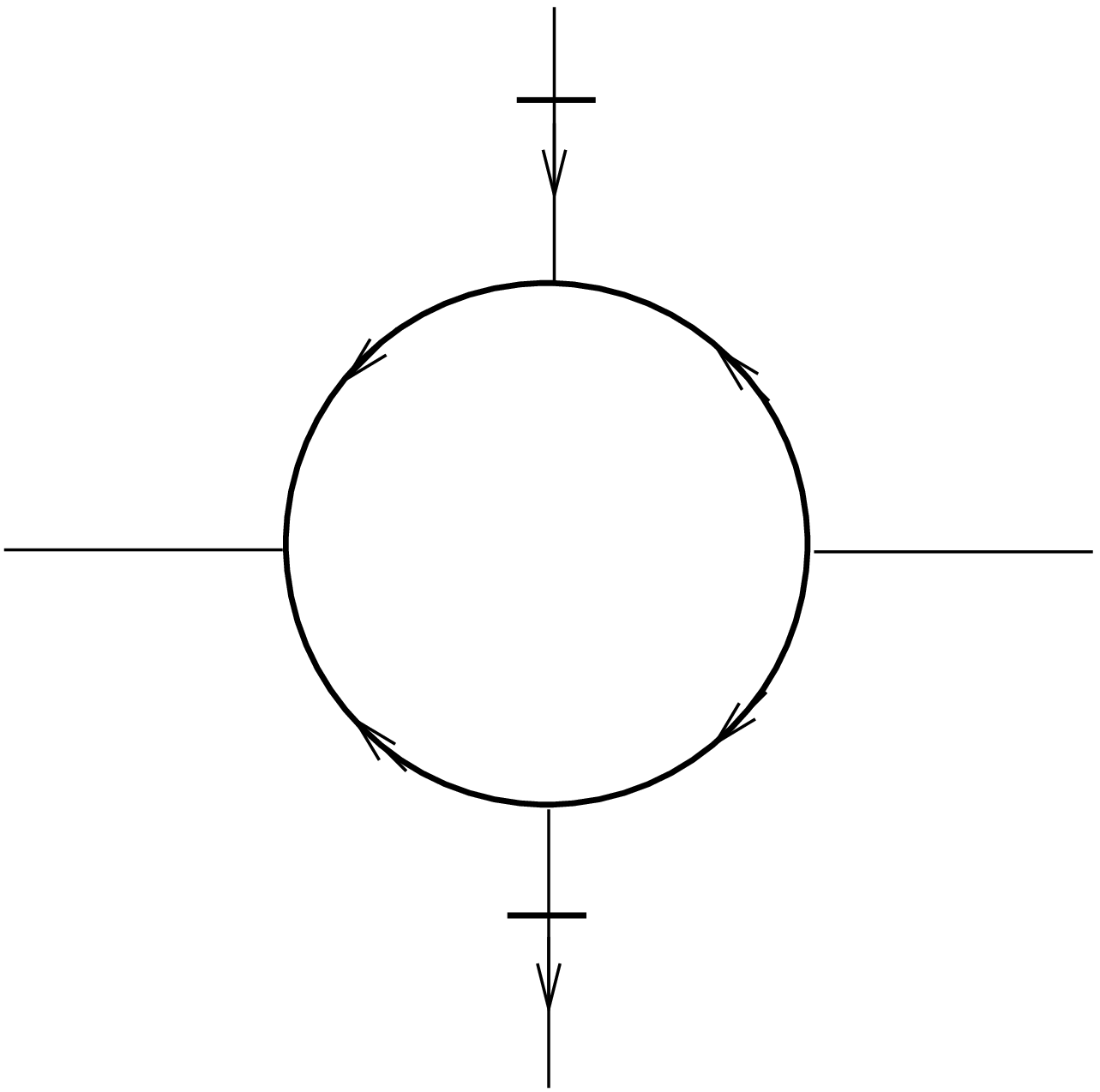}\end{array}
\!\!\!\!\!\!\!\!\!\!\!\!\!\!\!\!\!\!\!\!\!\!\!\!\!\!\!\!\!\!\!\!\!
\Gamma_{(4),R}^{(1)(P)} 
\;\;\;\;\;\;\;
  \;\;+\;\;  4\;\;
\begin{array}{c}\includegraphics[scale=0.23]{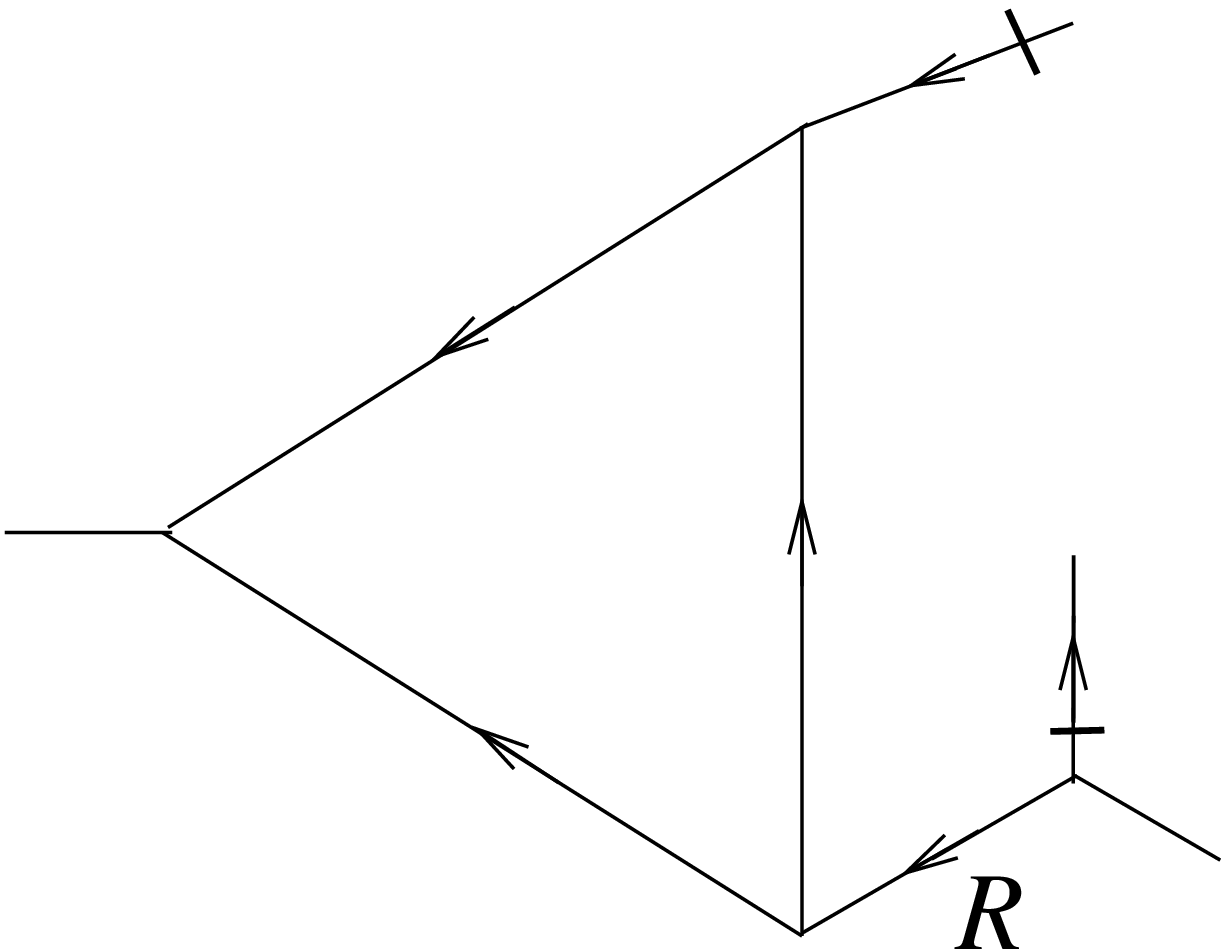}\end{array}
\!\!\!\!\!\!\!\!\!\!\!\!\!\!\!\!\!\!\!\!\!\!\!\!\!\!\!\!\!\!\!\!\!\!\!\!\!
\Gamma_{(3),R}^{(1)(P)} 
\;\;\;\;\;\;\;
\;\;+\;\;8\;\;
\begin{array}{c}\includegraphics[scale=0.20]{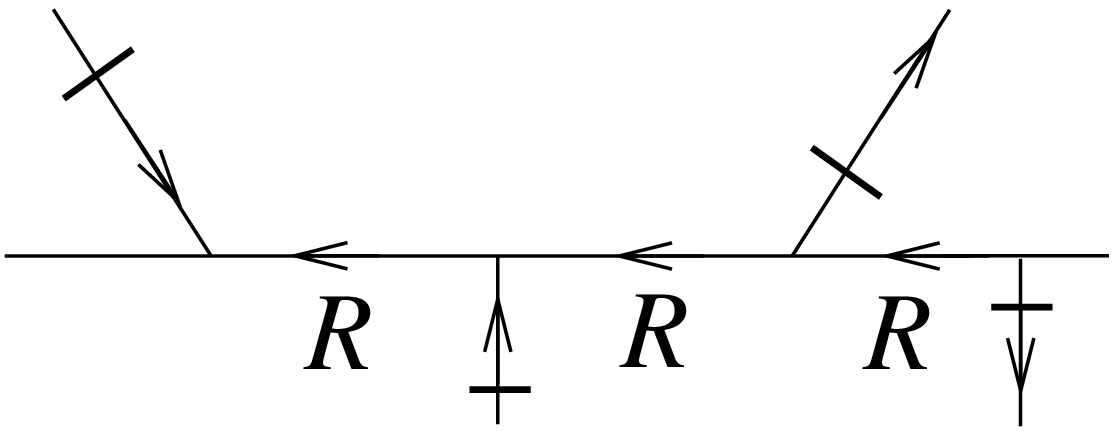}\end{array}.
\label{2pt2}
\end{equation}
\end{widetext}
All these examples illustrtate how the recipe works for an arbitrary retarded 
amplitude at $n$-loops and demonstrate the forward scattering
description for a retarded amplitude at zero temperature. 

\section{Forward scattering description at finite temperature}

Given the forward scattering description for retarded amplitudes at
zero temperature, it is now straight forward to derive the forward
scattering description at finite temperature through the use of the
thermal operator. Let us recall that in the closed time path
formalism, the thermal propagator for a scalar field can be related to
the zero temperature one through the thermal operator as \cite{silvana1}
\begin{equation}
\Delta^{(T)} (t, E) = {\cal O}^{(T)} (E) \Delta (t,E),\label{tor}
\end{equation}
where
\begin{equation}
{\cal O}^{(T)} (E) = 1 + n_{\rm B} (E) (1 - S(E)).\label{tor1}
\end{equation}
Here $n_{\rm B} (E)$ represents the bosonic distribution function and
$S(E)$ is a reflection operator that takes $E\rightarrow -E$. The
thermal operator is the same for each component of the propagator (it
is a scalar multiplicative operator) and leads to  
\begin{eqnarray}
\Delta_{+-}^{(T)} (t,E)  & = & {\cal O}^{(T)} (E) \Delta_{+-}(t,E)\nonumber\\
& = &
\Delta_{+-}(t,E) + \Delta_{+-}^{\beta}(t,E),\label{tor2} 
\end{eqnarray}
where we have identified the temperature dependent part of the
propagator to be (this notation is in an attempt to be consistent with
the notation in \cite{adilson}, although we have denoted the propagator in those
papers by $G$) 
\begin{equation}
\Delta_{+-}^{\beta} (t,E) = \frac{n_{\rm B}(E)}{2E}\left(e^{-iEt} +
e^{iEt}\right).\label{thermal} 
\end{equation}
The other interesting thing to note is that
\begin{eqnarray}
\Delta_{\rm R}^{(T)} (t, E) & = & {\cal O}^{T}(E) \Delta_{\rm R} (t,
E) = \Delta_{\rm R}(t,E),\nonumber\\ 
\Delta_{\rm A}^{(T)} (t,E)  & = & {\cal O}^{(T)}(E) \Delta_{\rm A}
(t,E) = \Delta_{\rm A} (t,E).\label{physical} 
\end{eqnarray}
Namely, physical propagators such as the retarded and the advanced
propagators (at the tree level) are independent of
temperature. Graphically, these results can be written as 
\begin{eqnarray}
{\cal O}^{(T)} (E) 
\begin{array}{c}\includegraphics[scale=0.2]{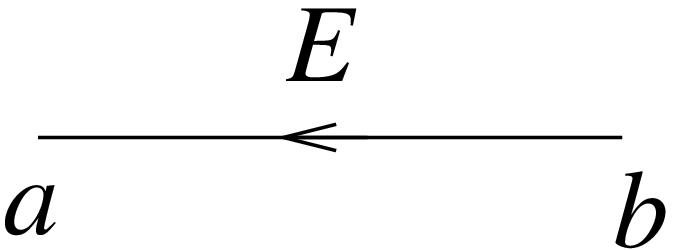}\end{array}
& = & 
\begin{array}{c}\includegraphics[scale=0.2]{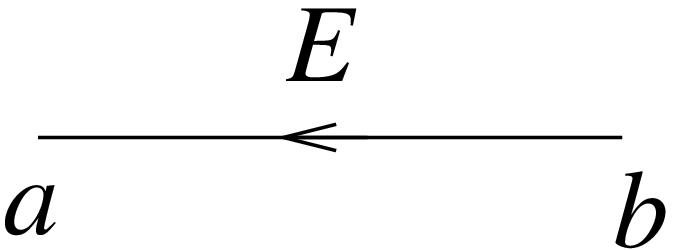}\end{array},
\nonumber\\
{\cal O}^{(T)} (E) 
\begin{array}{c}\includegraphics[scale=0.2]{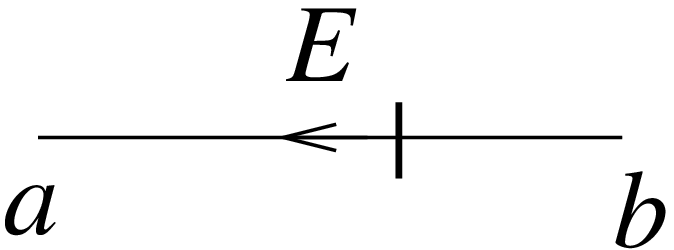}\end{array}
& = & 
\begin{array}{c}\includegraphics[scale=0.2]{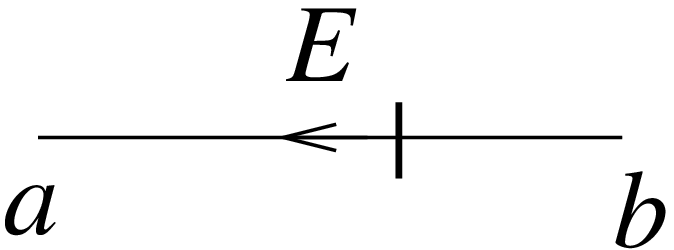}\end{array}+
\begin{array}{c}\includegraphics[scale=0.2]{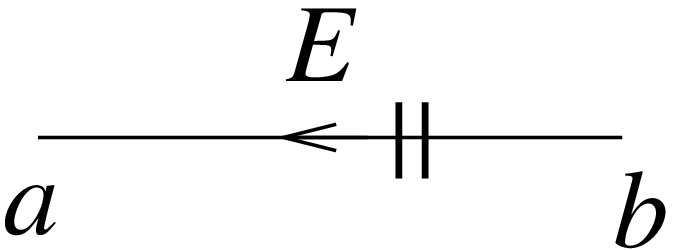}\end{array}
\label{tor3}
\end{eqnarray}
where the ``double cut" propagator can be identified with
$\Delta_{+-}^{\beta} Q$ (completely parallel with the notation of
\cite{adilson}). 

We know that any Feynman graph at finite temperature is related to the
corresponding zero temperature graph through a thermal operator 
\cite{silvana1,silvana2} that
can be built out of the basic thermal operator in \eqref{tor1}. For
example, the integrand of a graph (after the internal time integrations are 
done in the mixed space or the energy integrations are done in the 
energy-momentum space) with $N$ scalar propagators carrying
energy $E_{i}, i=1,2,\cdots , N$ at finite temperature is related to
the integrand of the corresponding graph at zero temperature by the
thermal operator 
\begin{equation}
{\cal O}^{(T)} = \prod_{i=1}^{N} {\cal O}^{(T)} (E_{i}).\label{TOR}
\end{equation}
This is a consequence of the simple fact that at finite temperature
only the propagators of the theory are modified (because of the
periodicity properties) while the interaction vertices remain
unaltered. As a result, the finite temperature forward scattering
description for a retarded diagram can be obtained directly from the
zero temperature one by simply applying the thermal operator
appropriate to the particular diagram. Of course, this also clarifies
the origin of the finite temperature forward scattering description,
namely, it exists because there is a corresponding description at zero
temperature.  

It is clear from \eqref{tor3} that since the thermal operator does not
change the ``$P$" propagators, all the uncut lines in a diagram in the
forward scattering description will continue to be the zero
temperature retarded propagator, $\Delta_{\rm R}$. The thermal
operator will only change the ``cut" propagators (the open lines) to a
``cut" plus a ``double cut" propagator.  (Like the ``cut" propagators, the 
``double cut" propagators are also on-shell with a factor of the distribution
function $n_{\rm B}$ and correspondingly can be thought of as representing
thermal on-shell incoming and outgoing particles.) Thus, one can organize the
graphs in the number of ``double cut" propagators. There will be
diagrams with no ``double cut" propagator, a single ``double cut"
propagator and so on and the maximum number of ``double cut"
propagators will be $n$ for a retarded amplitude at $n$ loops. The
diagrams without any ``double cut" propagator will, of course,
correspond to the zero temperature retarded amplitude. The diagrams
with the maximum number of ``double cut" propagators ($n$ in the case
under study) will represent tree level forward scattering amplitudes
for thermal on-shell particles. The diagrams with the number of
``double cut" propagators less than $n$ will all arrange into forward
scattering amplitudes for thermal on-shell particles with (zero
temperature) retarded vertices of lower order ($n-1$ and
lower). Namely, the effect of applying the thermal operator to a
retarded amplitude at zero temperature is to change all the ``$P$"
retarded vertices in the forward scattering description to genuine
retarded vertices at zero temperature and replace all the on-shell
forward scattering particles by thermal on-shell forward scattering
particles. If we ignore the zero temperature retarded amplitude, the rest
of the diagrams yield the temperature dependent forward scattering
amplitudes. 

The fact that the graphs will arrange as described above can be seen 
symbolically as follows. Let us identify
$\Delta_{\rm R} = P, \Delta_{+-}=y,
\Delta_{+-}^{\beta}=y^{\beta}$. Then, a retarded graph with $N$
propagators at zero temperature can be symbolically represented as $(P+y)^{N}$. 
If the
graph is at $n$ loops, then we can expand it in terms of the number of
on-shell propagators ($y$) and write symbolically as 
\begin{equation}
\Gamma_{(N),{\rm R}}^{(n)} =  (P+y)^{N} = P^{N} + a_{1}P^{N-1}y + \cdots
+ a_{n}P^{N-n}y^{n}. 
\end{equation}
Here the multiplicities $a_{i}, i=1,2,\cdots , n$ are assumed to
denote the number of ways the forward scattering on-shell particles
can occur in a graph without disconnecting the diagram (which is why
these cannot correspond to the pure binomial
coefficients). Furthermore, the terms involving powers of $P$
represent intermediate retarded propagators as well ``$P$" retarded
vertices. As we have seen, under the action of the thermal operator 
\begin{equation}
{\cal O}^{(T)} (P+y) = (P+y+y^{\beta}).
\end{equation}
As a result, using the thermal operator representation, we obtain
\begin{eqnarray}
\Gamma_{(N),{\rm R}}^{(n) (T)} & = & {\cal O}^{(T)} \Gamma_{(N),{\rm
 R}}^{(n)} = (P+y+y^{\beta})^{N}\nonumber\\ 
 & = & (P+y)^{N} + a_{1}(P+y)^{N-1} y^{\beta} + \cdots \nonumber\\
 & & +\cdots + a_{n}(P+y)^{N-n}(y^{\beta})^{n}\nonumber\\
 & = & \Gamma_{(N),{\rm R}}^{(n)} + y^{\beta}\left(\Gamma_{(N-1),{\rm
 R}}^{(n-1)} + P \Gamma_{(N-2),{\rm
 R}}^{(n-1)}+\cdots\right)\nonumber\\ 
 & & + (y^{\beta})^{2}\left(\Gamma_{(N-2),{\rm R}}^{(n-2)} +
 \cdots\right) + \cdots . \label{symbolic}
 \end{eqnarray}
 It is useful to remember that the sum of the power of $y$ and $y^{\beta}$ in any
 term on the right hand side of \eqref{symbolic} can at the most be $n$ at $n$ loops.

Let us illustrate these relations with some examples. Applying the thermal
operator to the forward scattering description of the retarded
self-energy in the $\phi^{3}$ theory at one loop (see \eqref{2pt}), we
obtain 
\begin{widetext}
\begin{eqnarray}
\Sigma_{\rm R}^{(1)(T)} & = & {\cal O}^{(T)} \Sigma_{\rm R}^{(1)}
  =  
\Sigma_{\rm R}^{(1)(P)}+
\begin{array}{c}\includegraphics[scale=0.2]{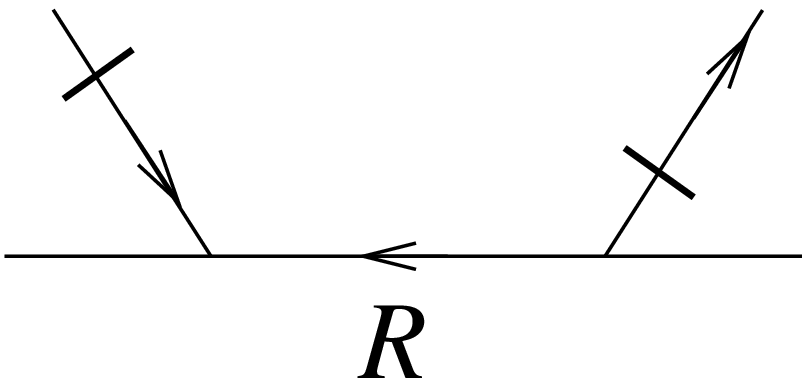}\end{array}+
\begin{array}{c}\includegraphics[scale=0.2]{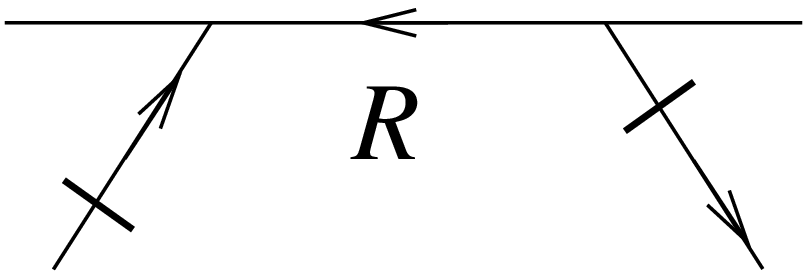}\end{array}
 +  
\begin{array}{c}\includegraphics[scale=0.2]{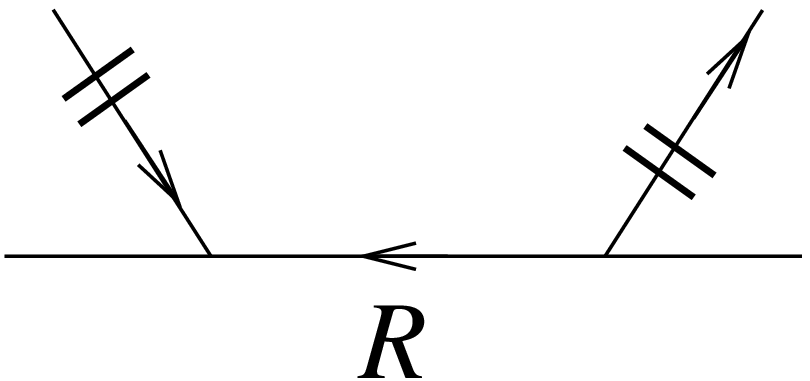}\end{array}+
\begin{array}{c}\includegraphics[scale=0.2]{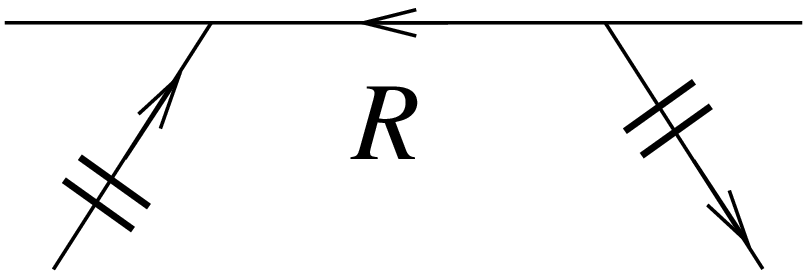}\end{array}
\nonumber\\
& = & \Sigma_{\rm R}^{(1)} + \begin{array}{c}\includegraphics[scale=0.2]{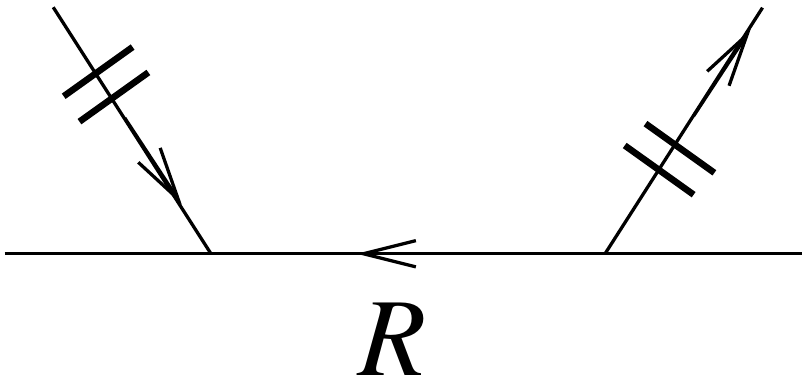}\end{array}+
\begin{array}{c}\includegraphics[scale=0.2]{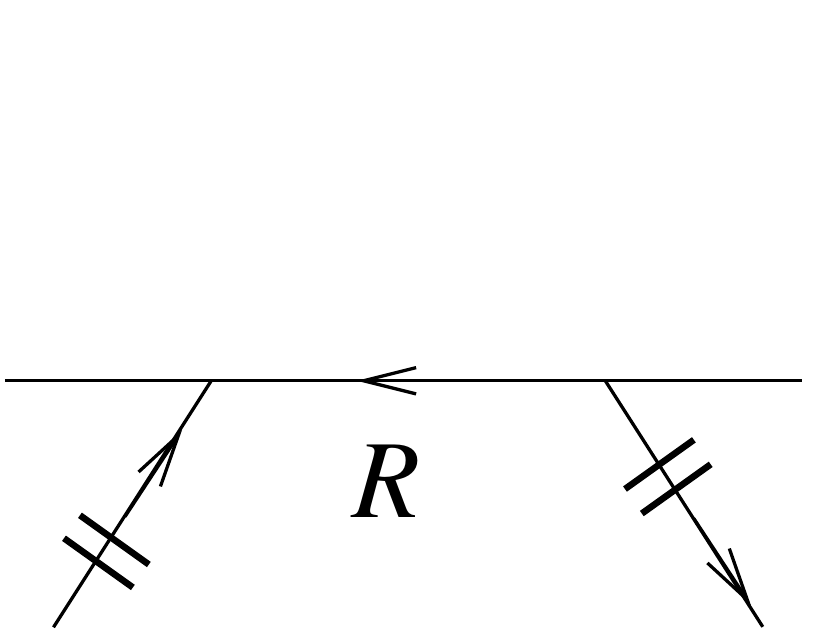}\end{array}
 = \Sigma_{\rm R}^{(1)} + \Sigma_{\rm R}^{(1) \beta}.\label{2ptbeta}
\end{eqnarray}
\end{widetext}
Similarly, the one loop retarded three point function (see
\eqref{3pt}) at finite temperature takes the form
\begin{widetext} 
\begin{eqnarray}
\Gamma_{3,{\rm R}}^{(1)(T)} & = & {\cal O}^{(T)} \Gamma_{3,{\rm R}}^{(1)}
=\Gamma_{3,{\rm R}}^{(1)(P)}
 + 
\begin{array}{c}\includegraphics[scale=0.2]{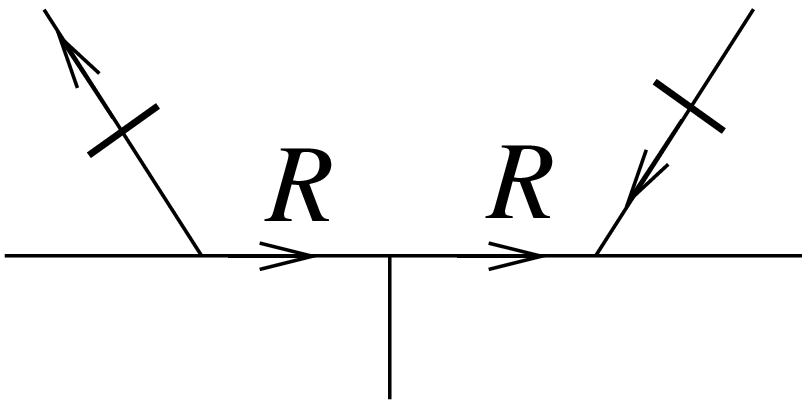}\end{array}+
\begin{array}{c}\includegraphics[scale=0.2]{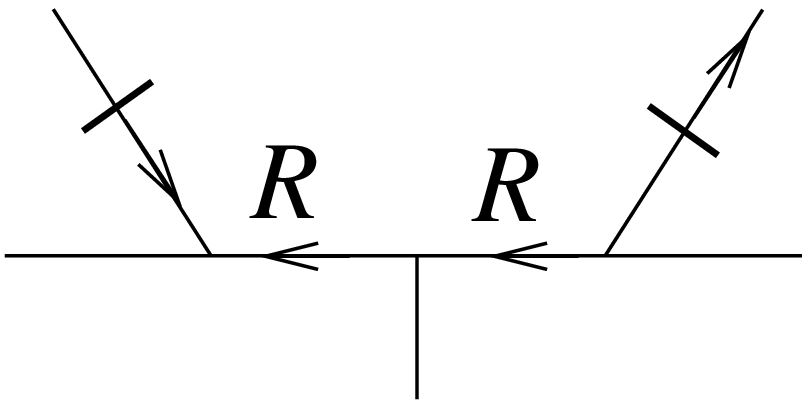}\end{array}
 + 
\begin{array}{c}\includegraphics[scale=0.2]{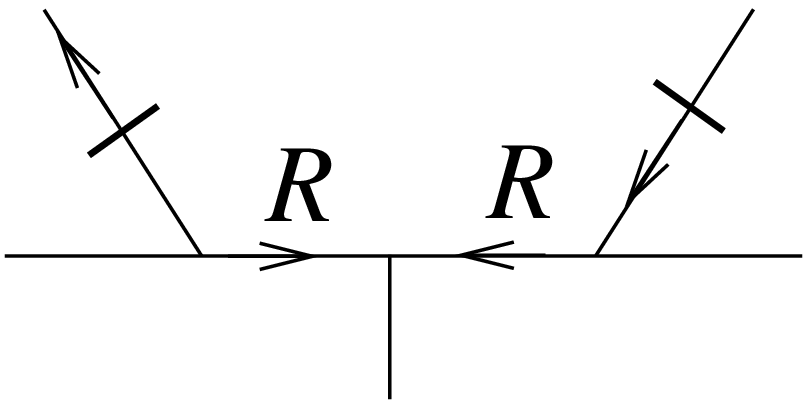}\end{array}\nonumber\\
& & +
\begin{array}{c}\includegraphics[scale=0.2]{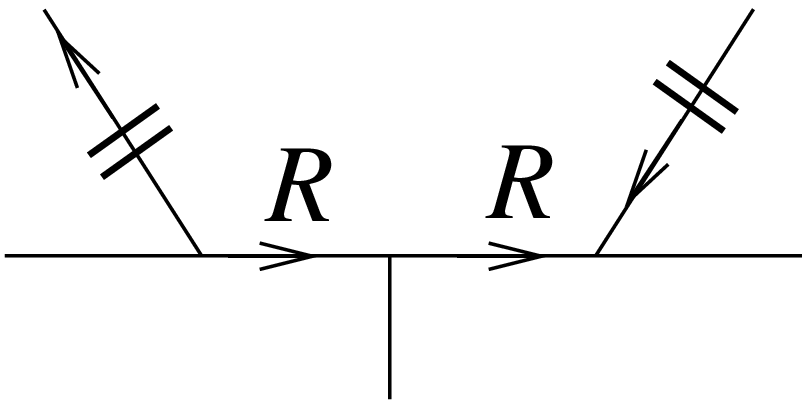}\end{array}
 + 
\begin{array}{c}\includegraphics[scale=0.2]{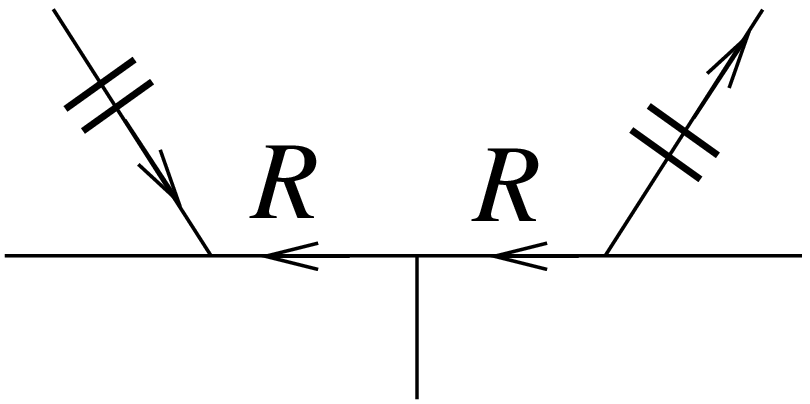}\end{array}+
\begin{array}{c}\includegraphics[scale=0.2]{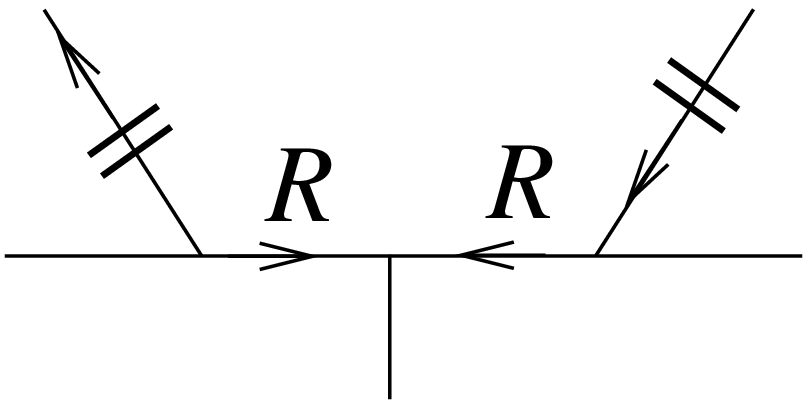}\end{array}
\nonumber\\
& = & \Gamma_{3,{\rm R}}^{(1)} + \begin{array}{c}\includegraphics[scale=0.2]{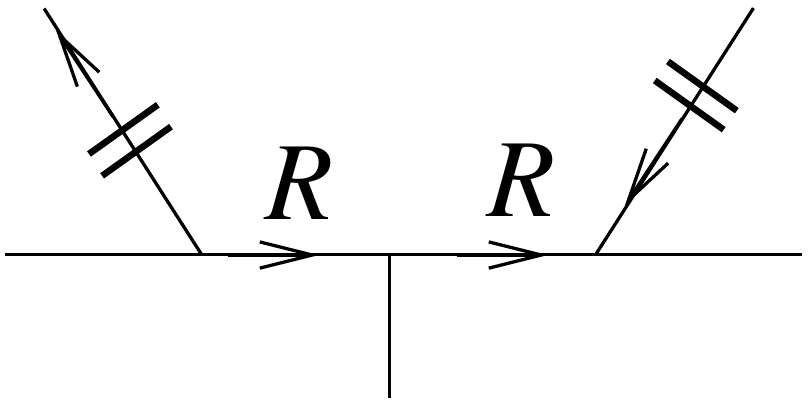}\end{array}
 + 
\begin{array}{c}\includegraphics[scale=0.2]{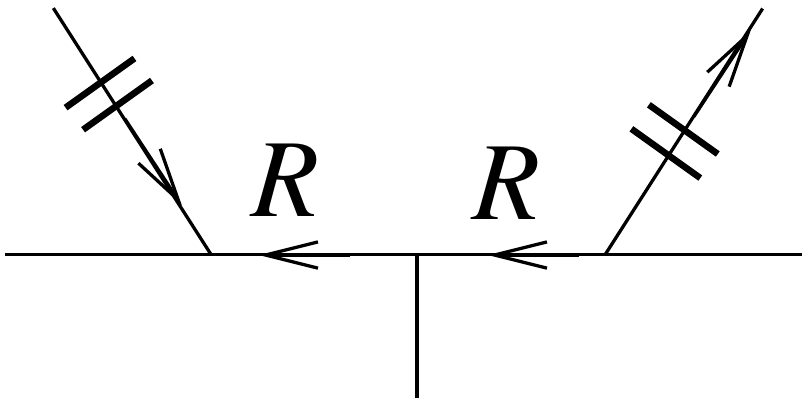}\end{array}+
\begin{array}{c}\includegraphics[scale=0.2]{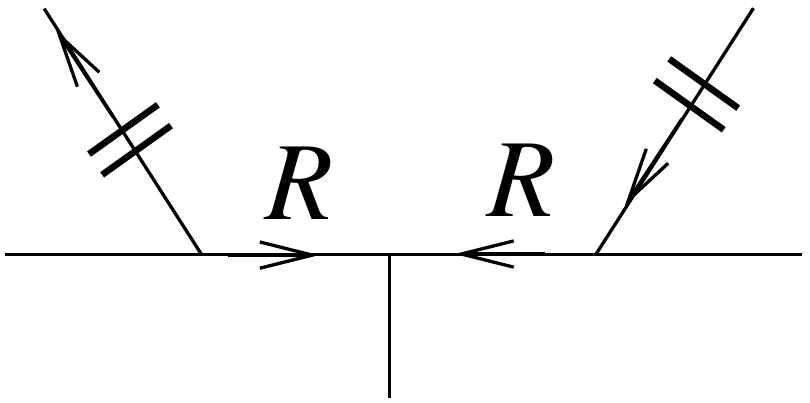}\end{array}
 = \Gamma_{3,{\rm R}}^{(1)} + \Gamma_{3,{\rm R}}^{(1) \beta}.\label{3ptbeta}
\end{eqnarray}
\end{widetext}
A slightly more complicated example would correspond to the two loop
retarded self-energy in the $\phi^{4}$ theory (see \eqref{2pt1}) at
finite temperature which takes the form 
\begin{widetext}
\begin{eqnarray}
\Sigma_{\rm R}^{(2)(T)} & = & {\cal O}^{(T)} \Sigma_{\rm R}^{(2)} 
 =  \Sigma_{\rm R}^{(2)(P)} + \;\;3\;\;
\begin{array}{c}\includegraphics[scale=0.2]{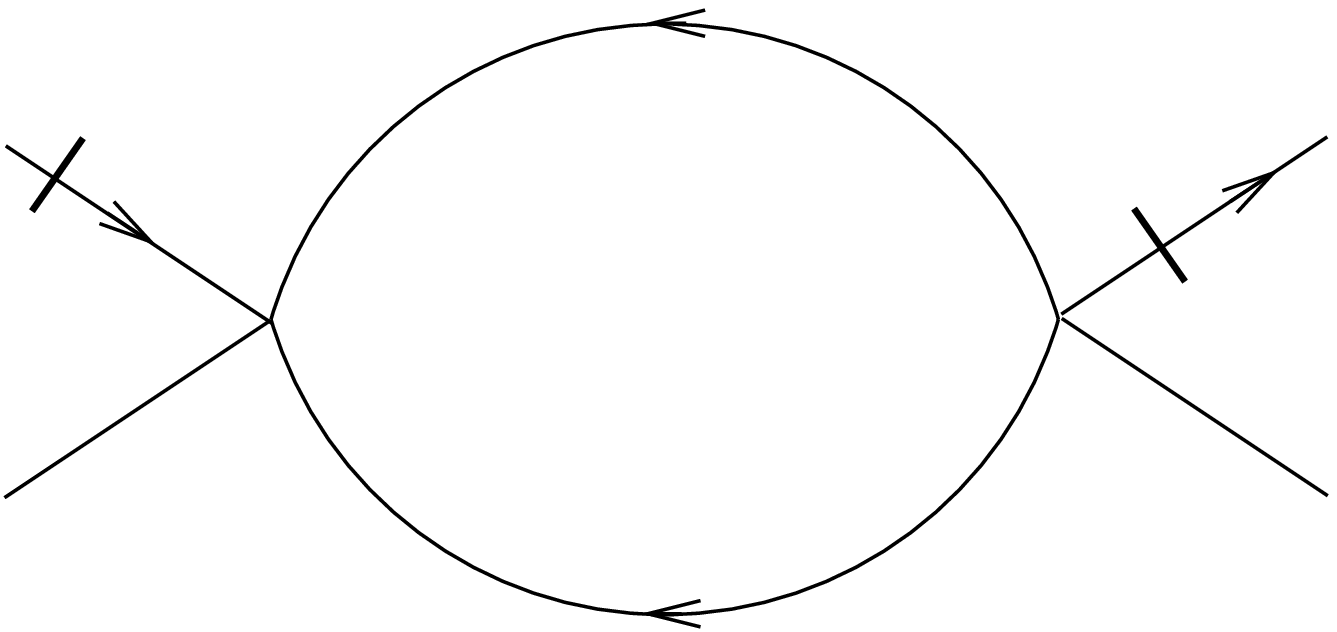}\end{array}
\!\!\!\!\!\!\!\!\!\!\!\!\!\!\!\!\!\!\!\!\!\!\!\!\!\!\!\!\!\!\!\!\!\!
\Sigma_{R}^{(1)(P)} 
\;\;\;\;\;\;\;
 \quad+  \;\;3\;\;
\begin{array}{c}\includegraphics[scale=0.2]{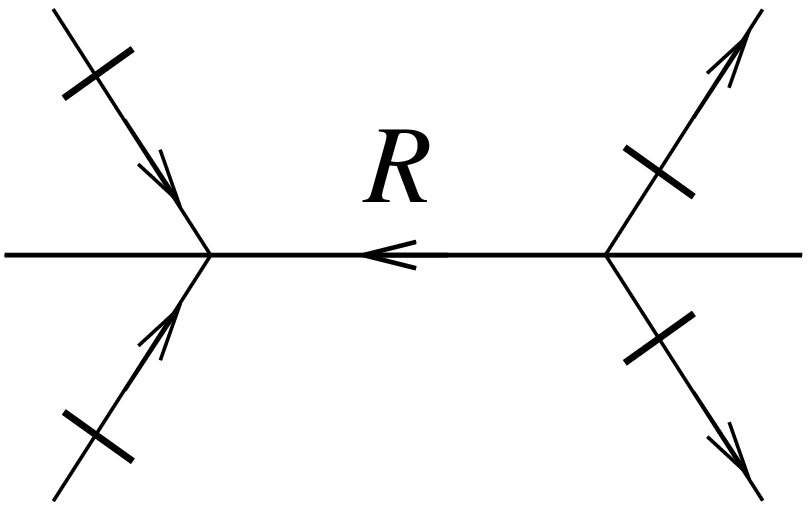}\end{array}\nonumber\\
&  & +\;\;3\;\;
\begin{array}{c}\includegraphics[scale=0.2]{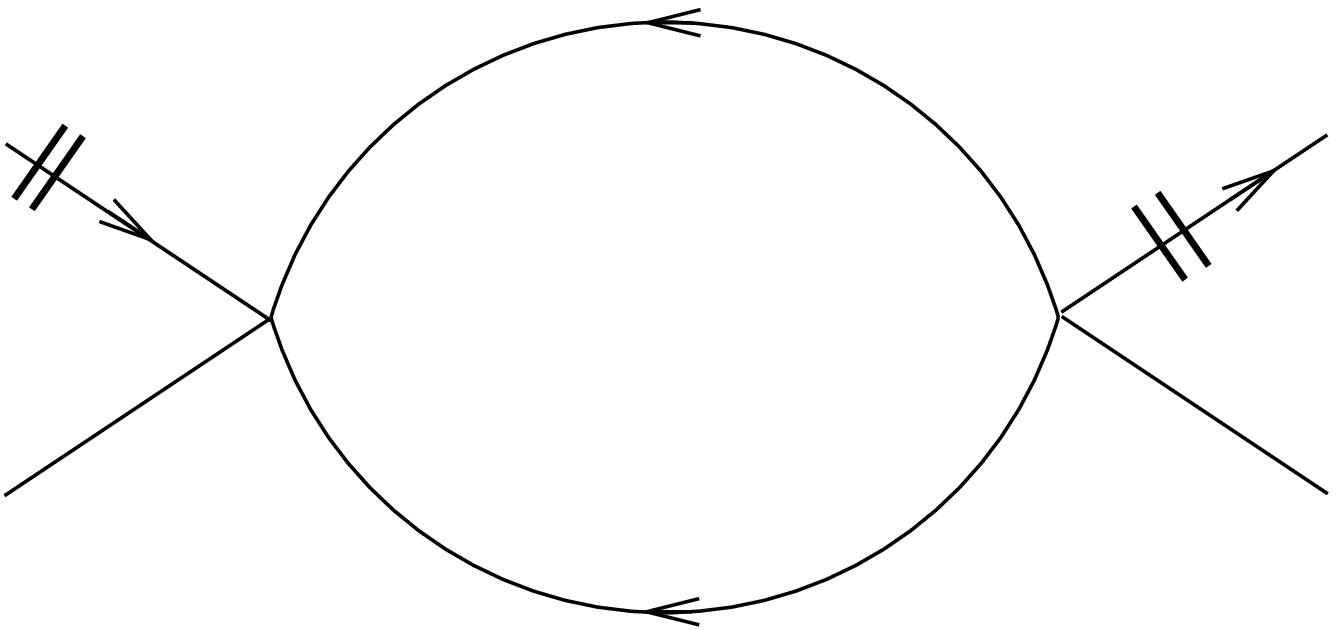}\end{array}
\!\!\!\!\!\!\!\!\!\!\!\!\!\!\!\!\!\!\!\!\!\!\!\!\!\!\!\!\!\!\!\!\!\!
\Sigma_{R}^{(1)(P)} 
\;\;\;\;\;\;\;
 \quad +\;\;  6\;\;
\begin{array}{c}\includegraphics[scale=0.2]{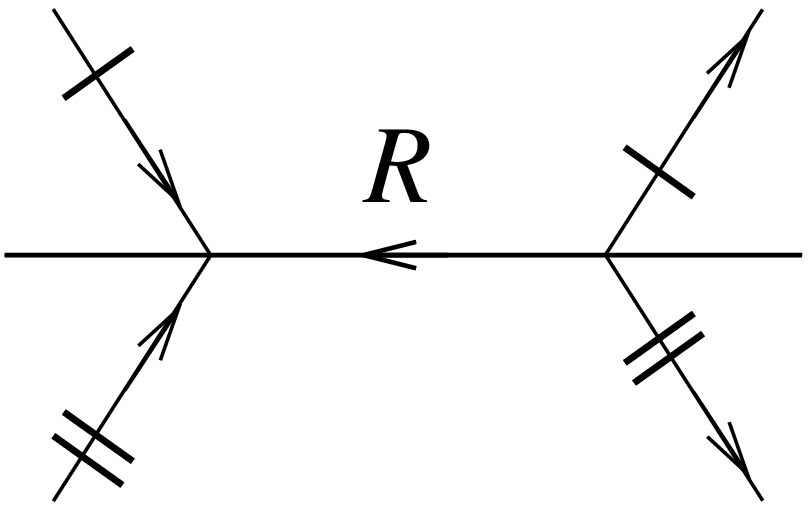}\end{array}+\;\;3\;\;
\begin{array}{c}\includegraphics[scale=0.2]{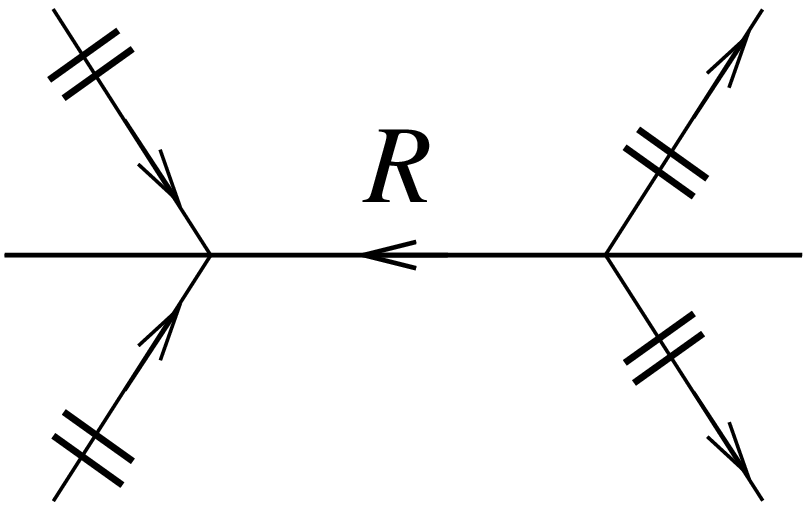}\end{array}
\nonumber\\
& = & 
\Sigma_{\rm R}^{(2)} +\;\; 3\;\;
\begin{array}{c}\includegraphics[scale=0.2]{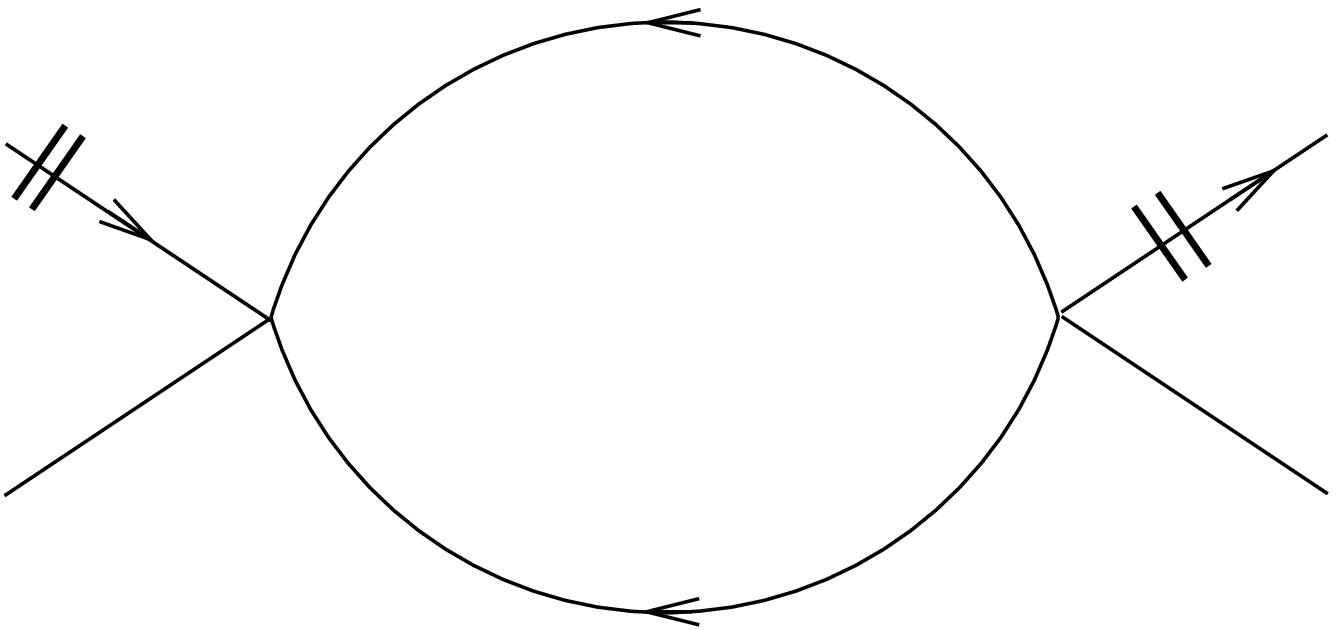}\end{array}
\!\!\!\!\!\!\!\!\!\!\!\!\!\!\!\!\!\!\!\!\!\!\!\!\!\!\!\!\!
\Sigma_{R}^{(1)} 
 \qquad\quad + \;\; 3\;\;
\begin{array}{c}\includegraphics[scale=0.2]{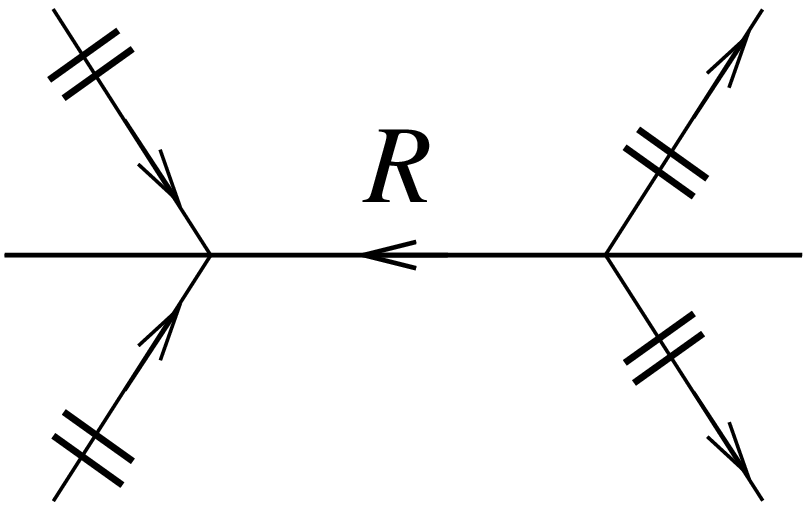}\end{array}
=\Sigma_{\rm R}^{(2)} + \Sigma_{\rm R}^{(2) \beta}.\label{2pt1beta}
\end{eqnarray}
\end{widetext}
This shows explicitly how the ``$P$" retarded vertices of lower order rearrange themselves
into full retarded vertices.
Finally, let us consider the nontrivial example of the two loop
retarded self-energy diagram (see \eqref{2pt2}) at finite temperature
\begin{widetext} 
\begin{eqnarray}
\Sigma_{\rm R}^{(2)(T)} & = & {\cal O}^{(T)} \Sigma_{\rm R}^{(2)}
 =  
\Sigma_{\rm R}^{(2)(P)}+
\begin{array}{c}\includegraphics[scale=0.2]{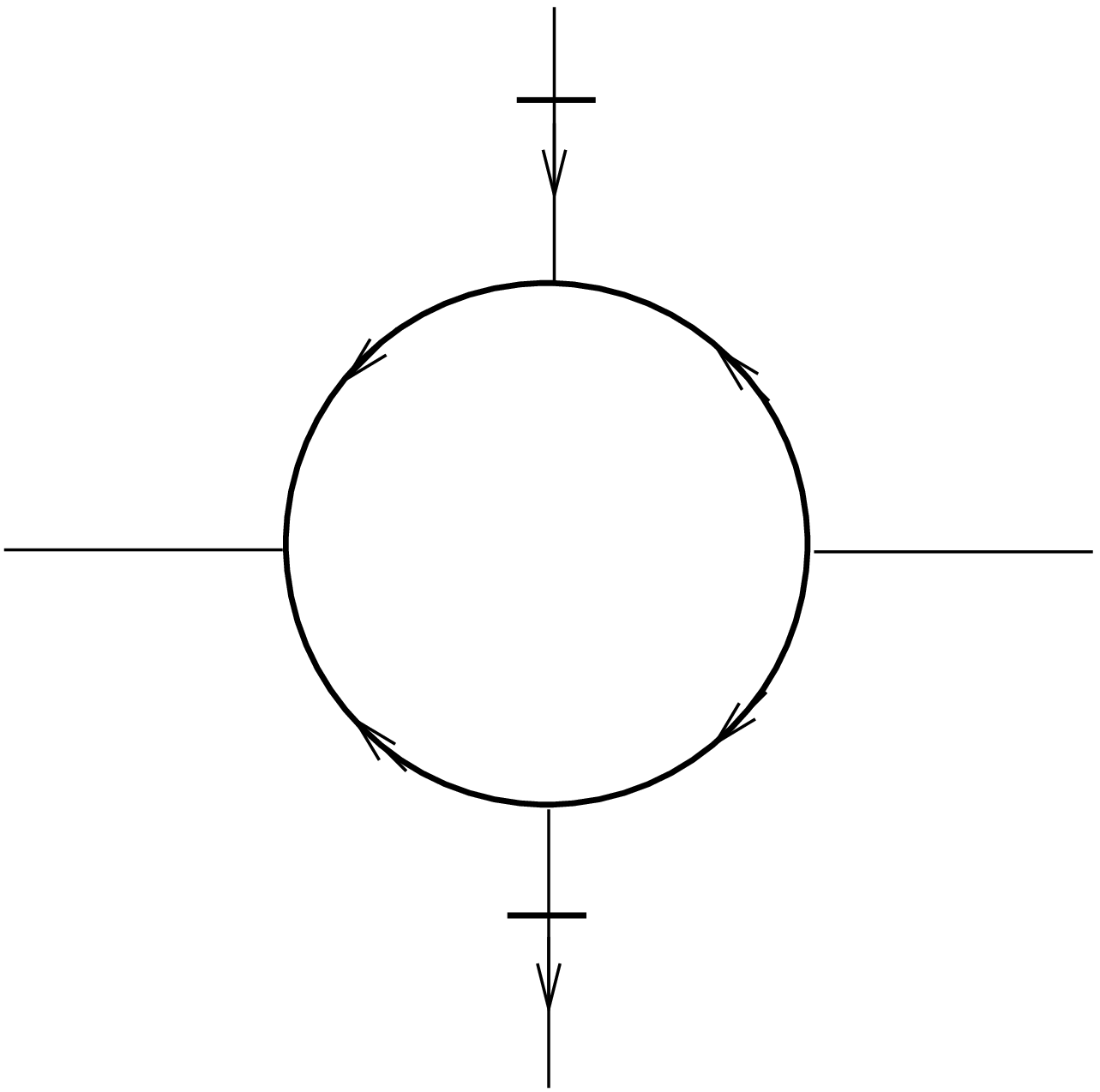}\end{array}
\!\!\!\!\!\!\!\!\!\!\!\!\!\!\!\!\!\!\!\!\!\!\!\!\!\!\!\!\!\!\!\!
\Gamma_{(4),R}^{(1)(P)} 
 \qquad\;\; +\;\;  4\;\;
\begin{array}{c}\includegraphics[scale=0.23]{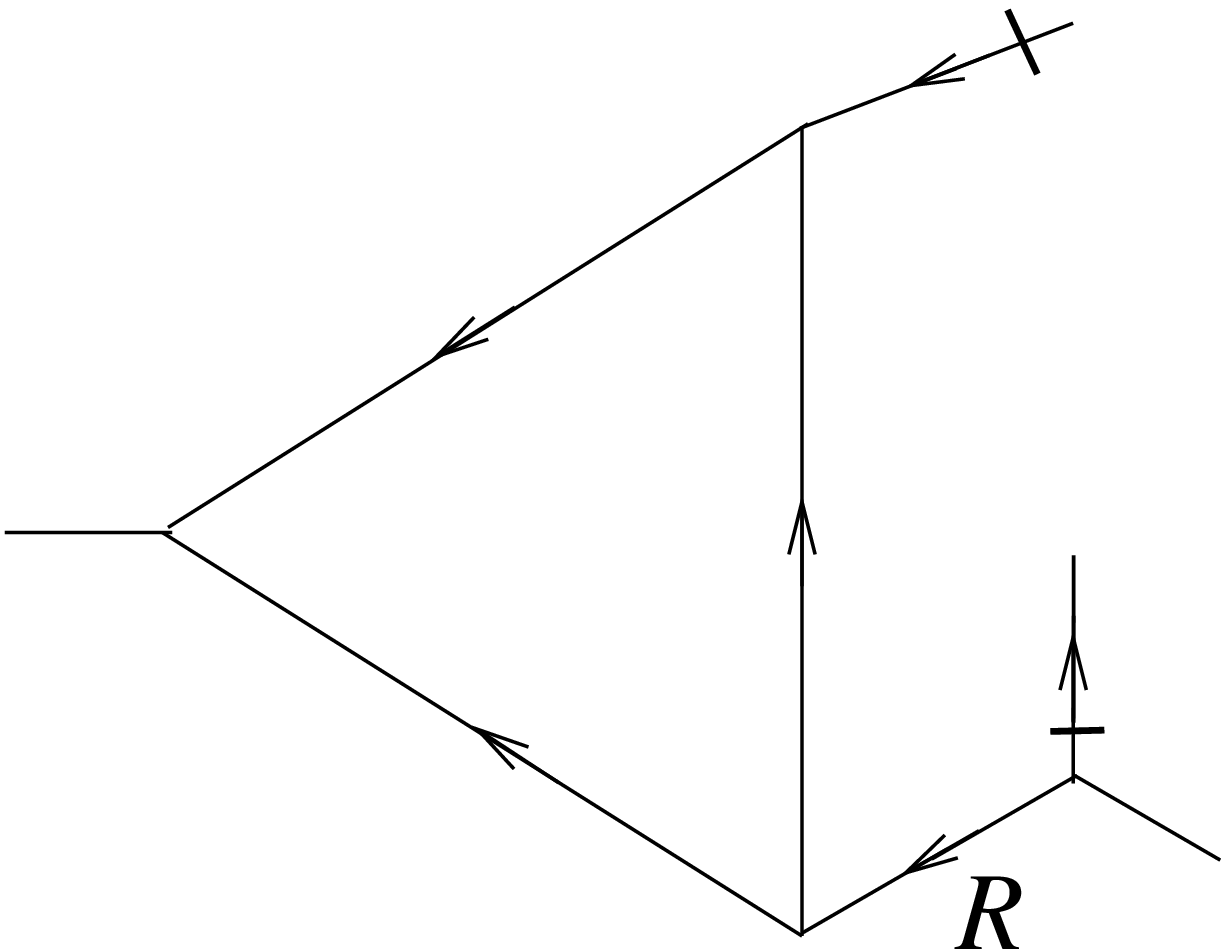}\end{array}
\!\!\!\!\!\!\!\!\!\!\!\!\!\!\!\!\!\!\!\!\!\!\!\!\!\!\!\!\!\!\!\!
\!\!\!\!\!
\Gamma_{(3),R}^{(1)(P)} 
\;\;\;\;\;\;\;\;\;\;\;\; \quad + \;\;8\;\; 
\begin{array}{c}\includegraphics[scale=0.2]{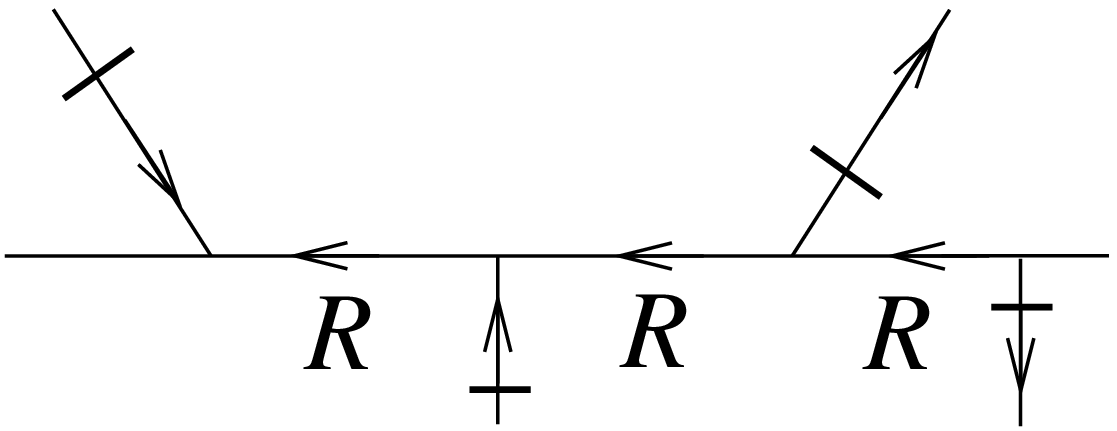}\end{array}
\nonumber\\
&  & + 
\begin{array}{c}\includegraphics[scale=0.23]{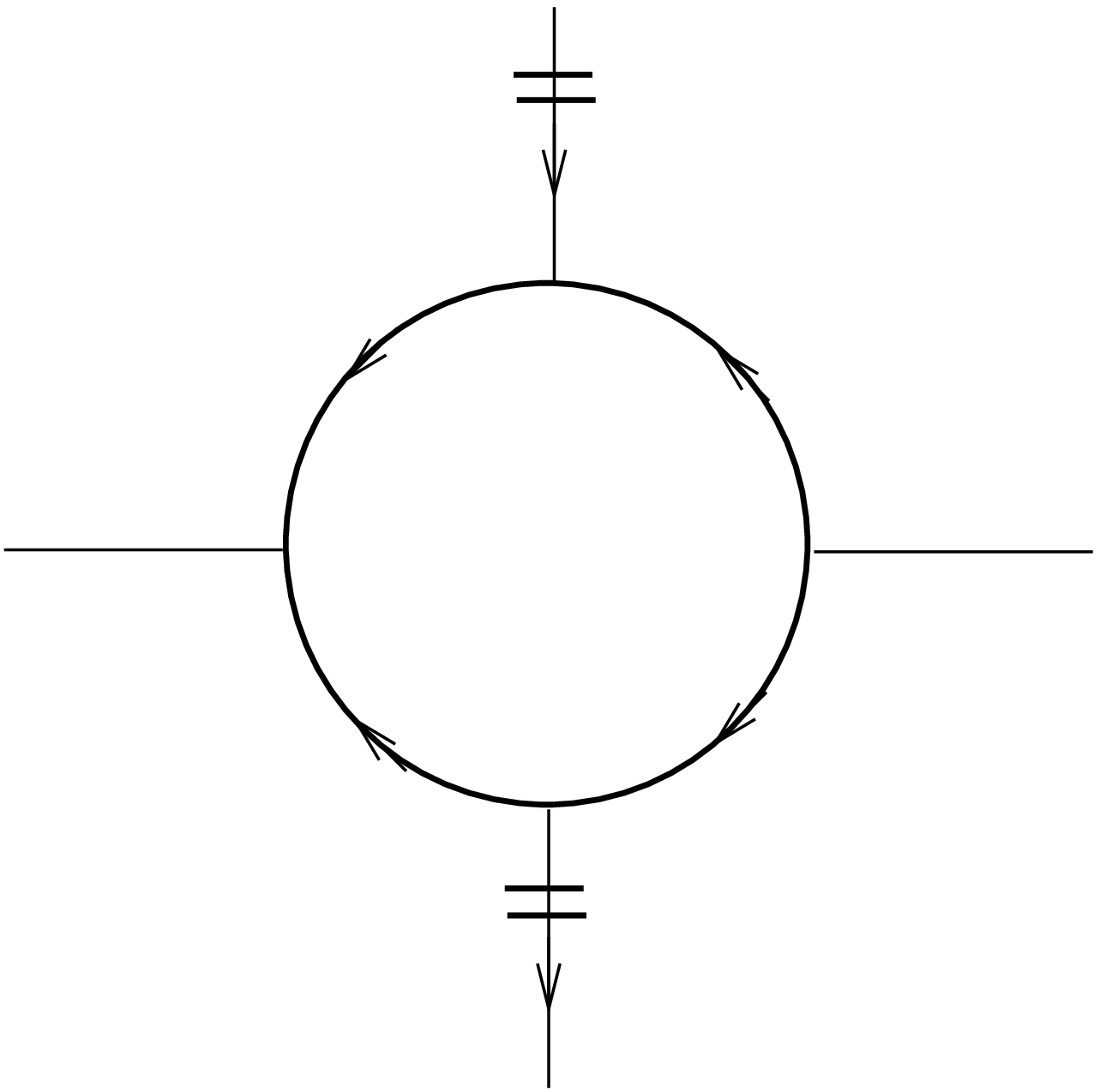}\end{array}
\!\!\!\!\!\!\!\!\!\!\!\!\!\!\!\!\!\!\!\!\!\!\!\!\!\!\!\!\!\!\!\!
\!\!\!\!\!
\Gamma_{(4),R}^{(1)(P)} 
\;\;\;\;\;\;\;\;\;\;\;\; +\;\; 4\;\;
\begin{array}{c}\includegraphics[scale=0.23]{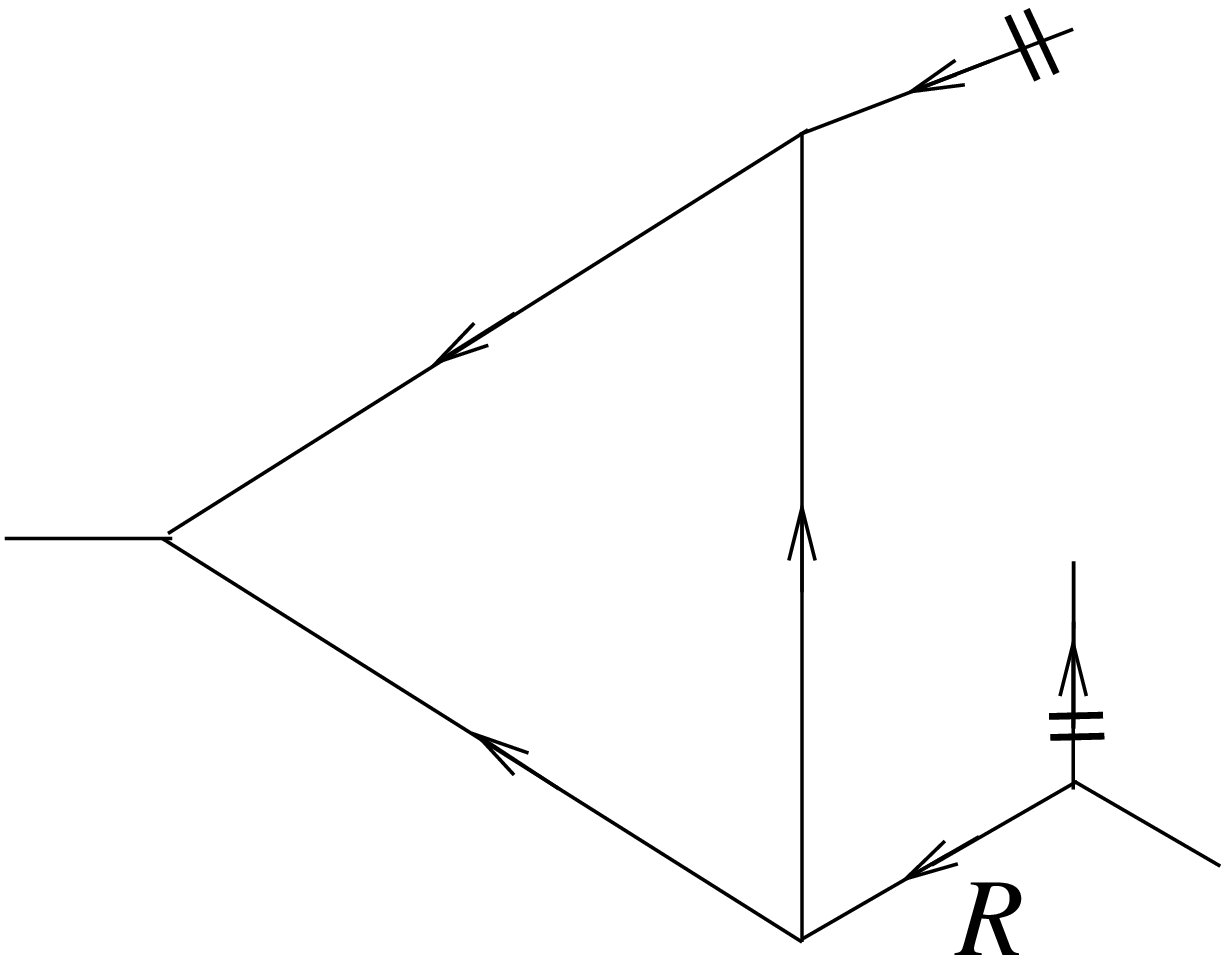}\end{array}
\!\!\!\!\!\!\!\!\!\!\!\!\!\!\!\!\!\!\!\!\!\!\!\!\!\!\!\!\!\!\!\!
\!\!\!\!\!
\Gamma_{(3),R}^{(1)(P)} 
 \qquad+  16\;\;
\begin{array}{c}\includegraphics[scale=0.23]{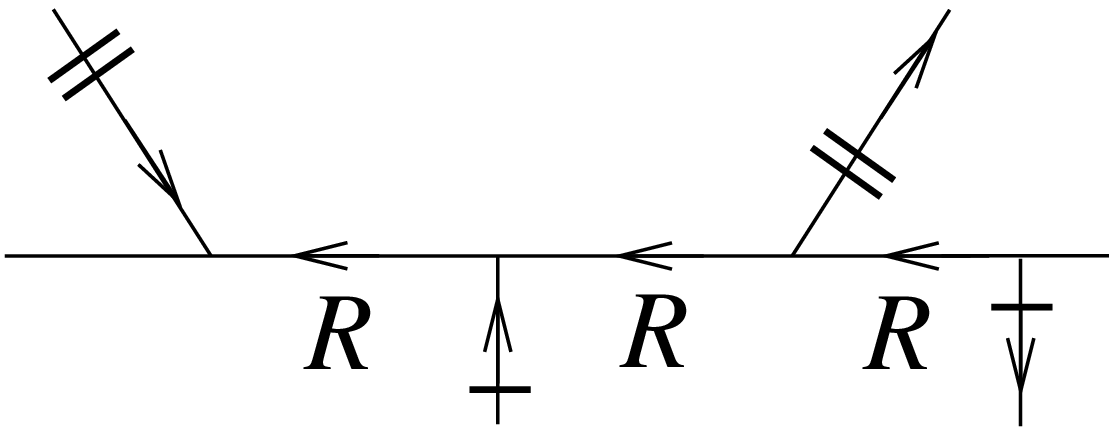}\end{array}+8
\begin{array}{c}\includegraphics[scale=0.23]{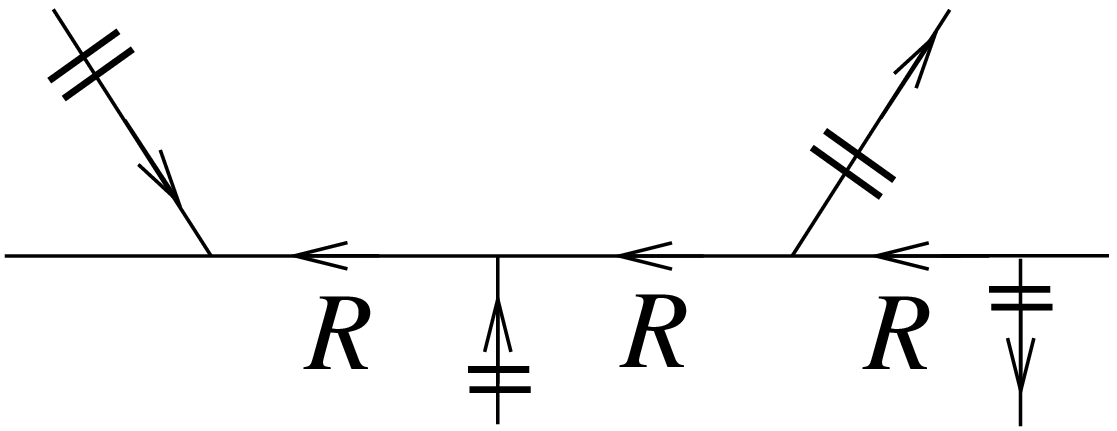}\end{array}
\nonumber\\
& = &
\Sigma_{\rm R}^{(2)} + 
\begin{array}{c}\includegraphics[scale=0.23]{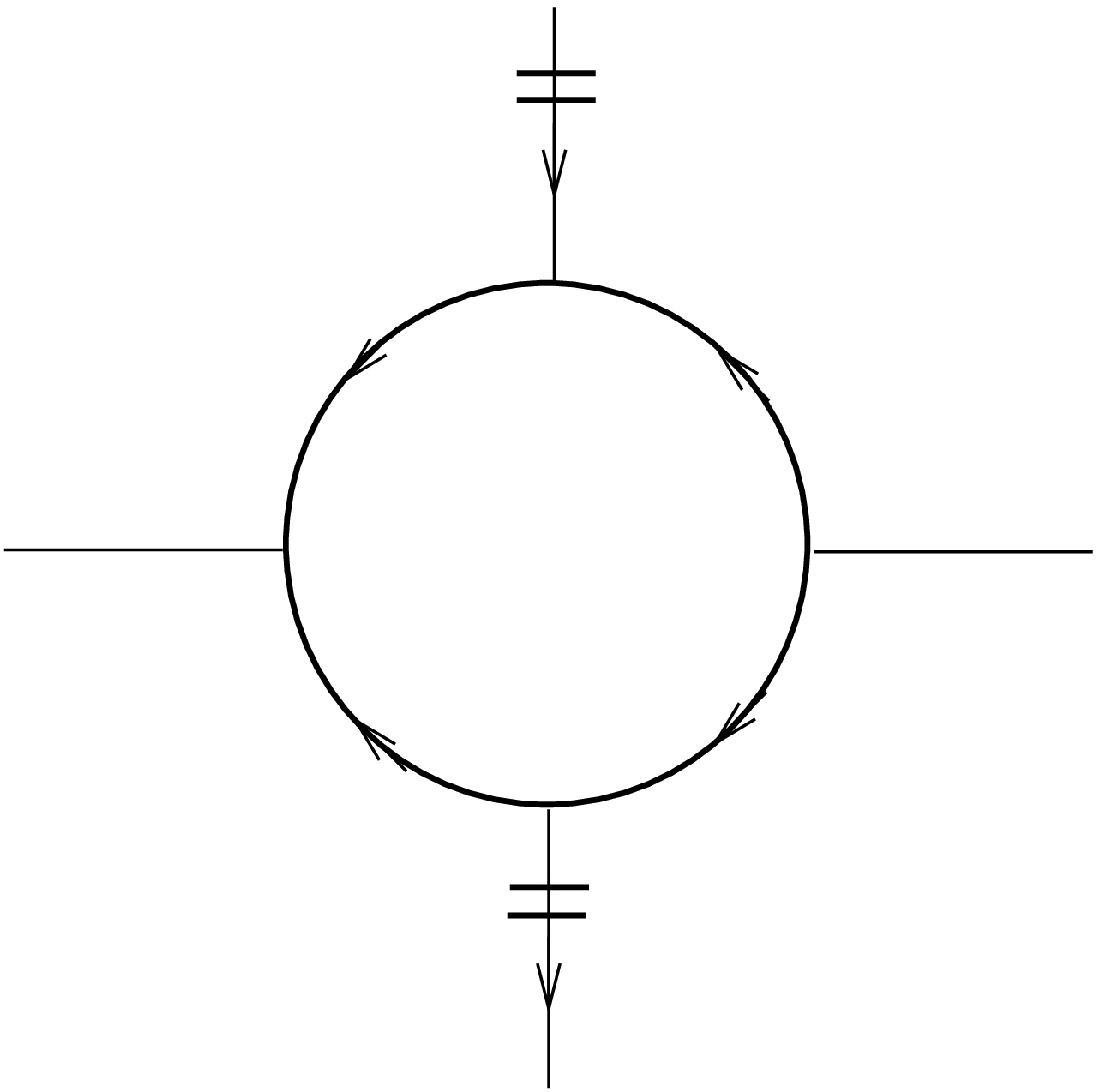}\end{array}
\!\!\!\!\!\!\!\!\!\!\!\!\!\!\!\!\!\!\!\!\!\!\!\!\!\!\!\!\!\!\!\!
\!\!\!\!\!
\Gamma_{(4),R}^{(1)} 
 \qquad\quad\;\;\;+  4\;\;
\begin{array}{c}\includegraphics[scale=0.23]{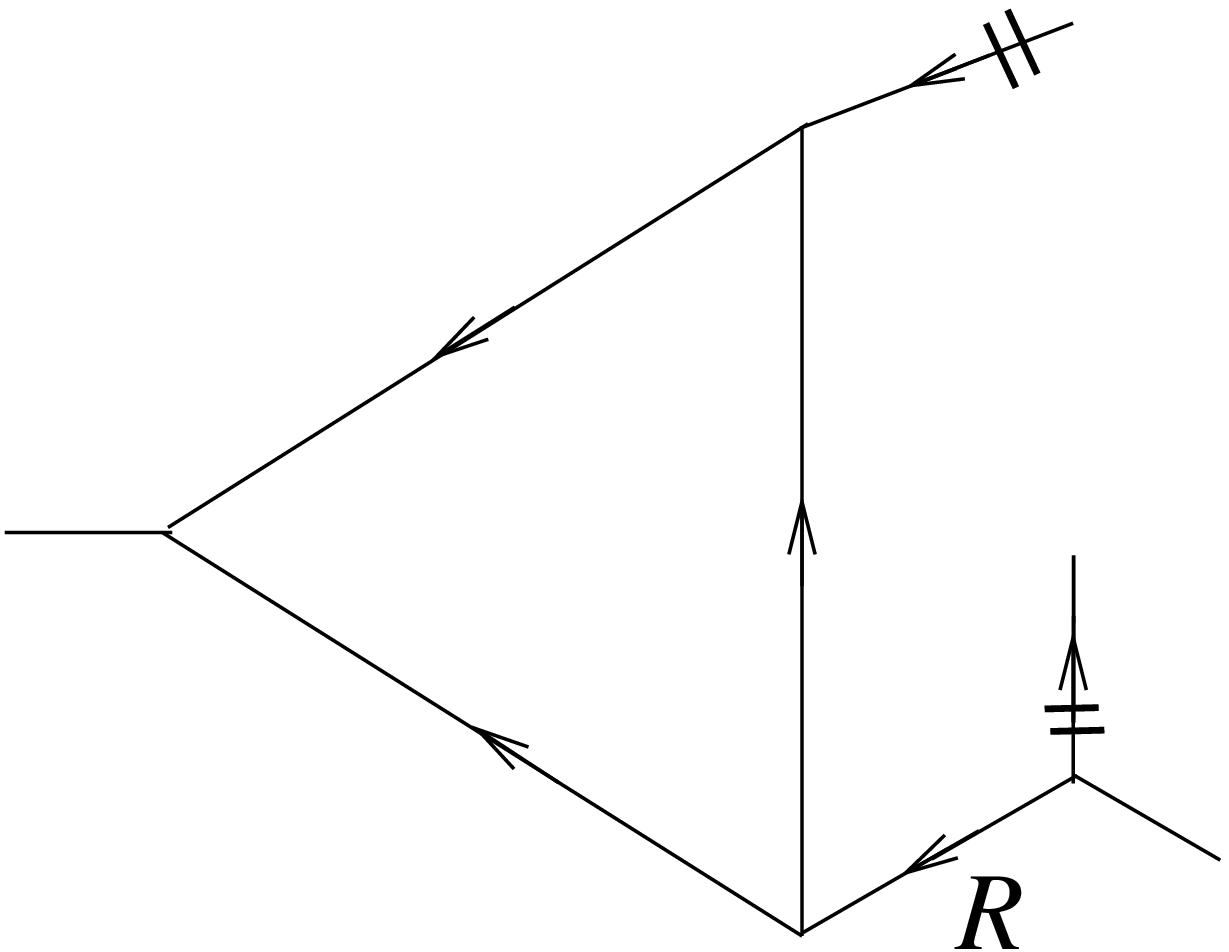}\end{array}
\!\!\!\!\!\!\!\!\!\!\!\!\!\!\!\!\!\!\!\!\!\!\!\!\!\!\!\!\!\!\!\!
\!\!\!\!\!
\Gamma_{(3),R}^{(1)} 
\;\;\;\;\;\;\;\;\;\;\;\; + 8
\begin{array}{c}\includegraphics[scale=0.23]{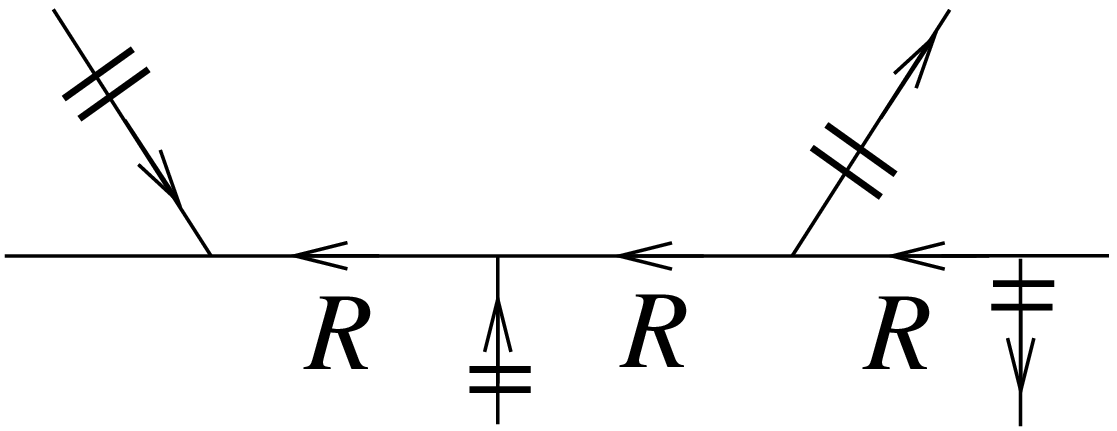}\end{array}
 =  \Sigma_{\rm R}^{(2)} + \Sigma_{\rm R}^{(2) \beta}.\label{2pt2beta}
\end{eqnarray}
\end{widetext}
This nontrivial example once again demonstrates how the ``$P$" retarded
vertices of lower order rearrange themselves into full retarded vertices.
We note that the thermal operator ${\cal O}^{(T)}$ for the different
amplitudes in \eqref{2ptbeta},\eqref{3ptbeta},\eqref{2pt1beta} and
\eqref{2pt2beta} are different and their appropriate forms can be obtained from
\eqref{TOR}. Furthermore, the temperature dependent forward scattering
amplitudes are denoted with a superscript $\beta$ to coincide with
the definition in \cite{adilson} and in the applicable examples can be checked
to agree with the results there.

We would now like to make some observations on the structure of
retarded amplitudes in general which are not directly related to the
main goal of this paper, but are quite important in understanding
their structures. First, we note that the retarded $N$-point  amplitude
at any loop is defined algebraically in the coordinate space in terms
of the original fields of the theory as the vacuum expectation value of 
the nested commutators \cite{schweber} 
\begin{eqnarray}
\lefteqn{\Gamma_{N,{\rm R}} (t_{1}, t_{2},\cdots , t_{N})}\nonumber\\
 & = &
(-i)^{N-1}\theta(t_{1}-t_{2})\theta(t_{2}-t_{3})\cdots
\theta(t_{N-1}-t_{N})\nonumber\\
& & \times \langle 0|\left[\left[\cdots
  [\phi(x_{1}),\phi(x_{2})],\phi(x_{3})\right]\cdots
  ,\phi(x_{N})\right]|0\rangle\nonumber\\
& & + {\rm permutations}.\label{retarded}
\end{eqnarray}
Here we have assumed that the time coordinate $t_{1}$ is the largest
among all the coordinates and the ``permutations'' refer to
symmetrizing in all the other coordinates (other than the largest time) and 
fields. As a result, the
retarded amplitude is symmetric in all the coordinates other than the
largest time coordinate. For a real scalar field, $\phi (x)$ is a
Hermitian operator and, therefore, the factor $(-i)^{N-1}$ shows that the
retarded amplitudes are real in coordinate space. Under Hermitian
conjugation, the change in the sign of each factor of $i$ is compensated by the
change in sign coming from each commutator. We have already argued in
section {\bf II} graphically that a retarded amplitude cannot have
disconnected parts which would correspond to products of vacuum
expectation values. This can also be seen from the above definition as
follows. Let us denote the nested commutator involving the first $N-1$
fields as 
\begin{equation}
A = \left[\left[\cdots
    [\phi(x_{1}),\phi(x_{2})],\phi(x_{3})\right]\cdots
    ,\phi(x_{N-1})\right].
\end{equation}
Then, we can write the vacuum expectation value of the nested
commutator in \eqref{retarded} as
\begin{equation}
\langle 0|\left[A, \phi(x_{N})\right]|0\rangle.
\end{equation}
Inserting a complete set of intermediate states (say, discrete energy
eigenstates or the particle number states), this can be written as
\begin{eqnarray}
\lefteqn{\langle 0|\left[A,\phi(x_{N})\right]|0\rangle}\nonumber\\
 & = &
\sum_{n=0}^{\infty} \left(\langle 0|A|n\rangle \langle
n|\phi(x_{N})|0\rangle - \langle 0|\phi(x_{N})|n\rangle
\langle n|A|0\rangle\right)\nonumber\\
& = & \sum_{n=1}^{\infty} \left(\langle 0|A|n\rangle \langle
n|\phi(x_{N})|0\rangle - \langle 0|\phi(x_{N})|n\rangle
\langle n|A|0\rangle\right).\nonumber\\
& & 
\end{eqnarray}
Namely, the intermediate vacuum states cancel out in the vacuum
expectation value of the commutator and as a result, the retarded
amplitudes contain only connected graphs which we have explicitly seen
earlier.

It was noted earlier \cite{silvana3} that the retarded self-energy for
a real scalar field in the mixed space is a real quantity. As we have
already argued, the retarded amplitudes, by definition, are 
real in the coordinate space. However, in going to the mixed space, one Fourier 
transforms the
spatial coordinates into spatial momenta and Fourier transformation
does not maintain the reality of a function in general. Let us comment
here briefly on  when the retarded amplitudes will be real in the
mixed space for a  scalar theory. The definition of the retarded
amplitudes in the mixed space take the form (see \eqref{retarded}) 
\begin{eqnarray}
\lefteqn{\tilde{\Gamma}_{N,{\rm R}} (t_{1}, t_{2},\cdots , t_{N})}\nonumber\\
 & = &
(-i)^{N-1}\theta(t_{1}-t_{2})\theta(t_{2}-t_{3})\cdots
\theta(t_{N-1}-t_{N})\nonumber\\
& & \times \langle 0|\left[\cdots
  [\phi(t_{1},\vec{p}_{1}),\phi(t_{2},\vec{p}_{2})],\cdots
  ,\phi(t_{N},\vec{p}_{N})\right]|0\rangle\nonumber\\
& & + {\rm permutations}.\label{retarded1}
\end{eqnarray}
Under Hermitian conjugation, the change in sign in each factor of
``$i$" is still compensated for by the change in sign coming from each
commutator. However, since under Hermitian conjugation 
\begin{equation}
\phi(t,\vec{p})\rightarrow \phi(t,-\vec{p}),
\end{equation}
the retarded amplitude is not real in general. In fact, let us note
explicitly that under Hermitian conjugation, 
\begin{eqnarray}
\lefteqn{\tilde{\Gamma}_{N,{\rm R}}^{*} (t_{1}, t_{2},\cdots ,
 t_{N})}\nonumber\\ 
 & = &
(-i)^{N-1}\theta(t_{1}-t_{2})\theta(t_{2}-t_{3})\cdots
\theta(t_{N-1}-t_{N})\nonumber\\
& & \times \langle 0|\left[\cdots
  [\phi(t_{1},-\vec{p}_{1}),\phi(t_{2},-\vec{p}_{2})],\cdots
  ,\phi(t_{N},-\vec{p}_{N})\right]|0\rangle\nonumber\\
& & + {\rm permutations}.\label{retarded2}
\end{eqnarray}
We note that if the scalar field transforms under parity as
\begin{equation}
\phi(t,\vec{p})\rightarrow \phi(t,-\vec{p}) = (-1)^{\alpha} \phi(t,\vec{p}),
\end{equation}
where $\eta=(-1)^{\alpha}$ denotes the intrinsic parity of the scalar field, then
we can write \eqref{retarded2} as 
\begin{eqnarray}
\lefteqn{\tilde{\Gamma}_{N,{\rm R}}^{*} (t_{1}, t_{2},\cdots ,
 t_{N})}\nonumber\\ 
 & = & (-1)^{N\alpha}
(-i)^{N-1}\theta(t_{1}-t_{2})\theta(t_{2}-t_{3})\cdots
\theta(t_{N-1}-t_{N})\nonumber\\
& & \times \langle 0|\left[\cdots
  [\phi(t_{1},\vec{p}_{1}),\phi(t_{2},\vec{p}_{2})],\cdots
  ,\phi(t_{N},\vec{p}_{N})\right]|0\rangle\nonumber\\
& & + {\rm permutations}\nonumber\\
& = & (-1)^{N\alpha} \tilde{\Gamma}_{N,{\rm R}} (t_{1},t_{2},\cdots ,t_{N}).
\end{eqnarray}
For a scalar field of even parity, $\alpha=0$ and we see that the retarded
amplitudes will continue to be real even in the mixed space. However,
for a pseudoscalar field, $\alpha=1$ and we note that only parity
conserving retarded amplitudes will be real while the parity violating
retarded amplitudes will be purely imaginary in the mixed space. We
would like to emphasize that the reality of an amplitude in the
coordinate space/mixed space is not contradictory to the existence of
dispersion relations in the energy-momentum space since the imaginary
parts of the amplitudes in energy-momentum space arise from the
imaginary parts of the step functions ($\theta(t)$) in the integral
representation. 

\section{Forward scattering description for Yang-Mills theory}

The results of the earlier sections show that the forward scattering
description at finite temperature can be obtained from the forward
scattering description at zero temperature by the use of the thermal
operator. Although we have done this explicitly for scalar field
theories, this can be generalized easily to other theories. We would
like to emphasize that this correspondence between the finite
temperature and zero temperature forward scattering descriptions
should not be thought of as useful only in establishing  a graphical
identification. It is also quite useful as a calculational tool as well as
in clarifying various aspects of field theories at high
temperature. To give an example of this, we will next derive the retarded 
gluon self-energy in the Yang-Mills theory (belonging to $SU(N)$) at 
one loop in 
the hard
thermal loop approximation from the zero temperature result, which will
also clarify the structure of this thermal amplitude. 

A lot is known about the structure of Yang-Mills theories at high
temperature \cite{kapusta:book89,lebellac:book96}. It is known, for
example, that in the hard thermal loop approximation the  one loop retarded 
 self-energy in the forward scattering description is
independent of the gauge fixing parameter and has a manifestly  gauge
invariant (transverse) and Lorentz covariant structure (before carrying
out the angular integrations). However, the reason for such a structure
at finite temperature is not well understood. We will see below that
such a structure of the integrand for the retarded self-energy already
exists at zero temperature in the appropriate regime and since the
thermal operator is gauge invariant, the thermal amplitude obtained
through the application of the thermal operator representation
preserves these properties.  

From the discussions in section {\bf II}, we can immediately write
down the forward scattering description for the one loop retarded
self-energy for the Yang-Mills field (including the ghost
contributions) easily. Our discussion in section {\bf II} has been
completely within the context of the mixed space simply because we wanted to bring
out the physical nature of the retarded amplitudes as evolving forward
in time in a connected manner. However, the thermal operator
representation holds equally well in the energy-momentum space (which
can be seen simply by Fourier transforming the external time
coordinates) where the thermal operator acts on the integrand after
all the energy integrations have been carried out. Since all the
calculations of thermal amplitudes in the hard thermal loop
approximation have been carried out in the energy-momentum space, in
this section we will also work in the energy-momentum space and use
the thermal operator representation in this space.  

The forward scattering description for the one loop retarded
self-energy for the Yang-Mills field (belonging to $SU(N)$) at zero
temperature can be written as 
\begin{widetext}
\begin{eqnarray}
\Pi_{\mu\nu,{\rm R}}^{ab (1)} (p) & = & 
\begin{array}{c}\includegraphics[scale=0.85]{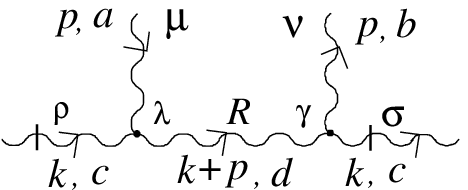}\end{array}+
\begin{array}{c}\includegraphics[scale=0.85]{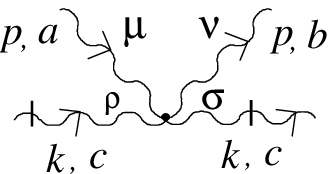}\end{array}
 + 
\begin{array}{c}\includegraphics[scale=0.85]{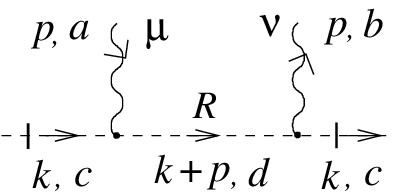}\end{array}+
\left(k\longrightarrow -k\right)
\nonumber\\
& = & \frac{g^{2}}{2}\int \frac{d^{4}k}{(2\pi)^{3}}
\delta^{+}(k^{2})\Big[f^{adc}f^{bcd}V_{\mu\lambda\rho}(p,-k-p,k)
\frac{d^{\lambda\gamma}(k+p)}{(k+p)^{2}}d^{\rho\sigma}(k)V_{\nu\sigma\gamma}
  (-p,-k,k+p)\nonumber\\ 
& & \qquad\qquad 
+ W_{\mu\nu\rho\sigma}^{abcc} d^{\rho\sigma} (k) -
2f^{adc}f^{bcd} \frac{k_{\mu}(k+p)_{\nu}}{(k+p)^{2}} + (k\rightarrow
-k)\Big],\label{YM} 
\end{eqnarray}
\end{widetext}
where wavy and dotted lines in the graphs denote respectively the gluon and the 
ghost propagators, which in an arbitrary covariant gauge have the forms 
\begin{eqnarray}
D_{\mu\nu}^{ab} (k) & = &-\frac{i\delta^{ab}}{k^{2}} d_{\mu\nu} (k)\nonumber\\
& = &  -\frac{i\delta^{ab}}{k^{2}}\left(\eta_{\mu\nu} - (1-\xi)
\frac{k_{\mu}k_{\nu}}{k^{2}}\right),\nonumber\\ 
D^{ab} (k) & = & \frac{i\delta^{ab}}{k^{2}}.\label{YMprop}
\end{eqnarray}
Here $\xi$ denotes the gauge fixing parameter and the appropriate
``$i\epsilon$" factors in the denominators of the propagators are
understood. We also note here that a product such as $\delta^{+}
(k^{2}) \frac{1}{k^{2}}$ in \eqref{YM} has to be understood in a
regularized sense as has been described in \cite{bedaque}. Similarly,
the gauge and the ghost vertices have the forms (we are suppressing
the energy-momentum conserving delta functions and are defining
$V_{\mu\nu\lambda}^{abc} = -g f^{abc} V_{\mu\nu\lambda}$ and
$V_{\mu}^{abc} = g f^{abc} V_{\mu}$) 
\begin{eqnarray}
V_{\mu\nu\lambda} (p,k,q) & = & \eta_{\mu\nu} (p-k)_{\lambda} +
\eta_{\nu\lambda}(k-q)_{\mu} + \eta_{\lambda\mu}
(q-p)_{\nu},\nonumber\\ 
V_{\mu} (p,k,q) & = &  k_{\mu},\label{vertices}
\end{eqnarray}
The quartic vertex, $(-ig^{2}W_{\mu\nu\rho\sigma}^{abcd})$,  can be read out from
\cite{gross}. Furthermore, we have defined 
\begin{equation}
\delta^{+} (k^{2}) = \theta(k_{0}) \delta (k^{2}).
\end{equation}
Since the ghost propagator as well as the gauge interaction vertex are
independent of $\xi$, for the purpose of understanding the $\xi$
dependence in \eqref{YM} we can restrict ourselves only to the
diagrams involving intermediate gauge field propagators. 

The tree level gauge field vertices \eqref{vertices} satisfy the identity 
\begin{equation}
k^{\nu} V_{\mu\nu\lambda} (p,k,q) = \left(\eta_{\mu\lambda} p^{2} -
p_{\mu}p_{\lambda} - \eta_{\mu\lambda}q^{2} +
q_{\mu}q_{\lambda}\right),\label{identity} 
\end{equation}
where we have used the fact that $k=-q-p$ because of energy-momentum
conservation. In the region where $k_{\mu}\gg p_{\mu}$ (where we are
assuming that $k,q$ are the internal momenta and $p$ is the external
one), the first two terms in \eqref{identity} can be neglected for our
purpose and what remains is a structure that is transverse to the
other internal momentum. As a result, we see that the terms quadratic
in the parameter $\xi$ vanish so that the self-energy in \eqref{YM}
can at most depend linearly on the gauge fixing parameter $\xi$. An
explicit evaluation of the diagrams involving the gauge field
propagators shows that the linear terms in $\xi$ cancel out among the
two classes of diagrams at the integrand level. Consequently,  in the
limit $k_{\mu}\gg p_{\mu}$, the leading order contribution to the
self-energy is independent of the gauge fixing parameter $\xi$.  

At zero temperature, the amplitudes are manifestly Lorentz
covariant. Furthermore, the Ward identity for the gluon self-energy
requires that it be transverse to the external momentum, 
\begin{equation}
p^{\mu} \Pi_{\mu\nu,{\rm R}}^{ab (1)} (p) = 0.\label{ward}
\end{equation}
Since the leading part of the retarded self-energy is independent of
the gauge fixing parameter $\xi$, in evaluating \eqref{YM} in this regime, 
we can choose it to have the value
$\xi=1$ for simplicity (Feynman gauge). In this case, the numerator of
the gauge propagator in \eqref{YMprop}  is simply the metric tensor independent of
any momentum. The momentum dependence in the numerator in \eqref{YM} 
comes from the
vertices and is, therefore, at most quadratic in the momenta. (Note
that the three point vertex depends on the momentum, but the quartic
vertex is independent of the momentum.) Since we are interested in the
``hard" internal momentum region where $k_{\mu}\gg p_{\mu}$, we can
expand the numerator in powers of the internal momentum. Similarly, we
can also expand the denominator in \eqref{YM} in this region as (recall that
$k^{2}=0$ because of the delta function) 
\begin{equation}
\frac{1}{(p+k)^{2}} = \frac{1}{2p\cdot k} - \frac{p^{2}}{(2k\cdot
  p)^{2}} + \cdots. 
\end{equation}
Using these in \eqref{YM}, one explicitly finds that the leading contribution
in this region comes from terms in the integrand (multiplying the
delta function) which are of degree zero in both the external as well
as the internal momentum. There is only one such tensor structure
available at zero temperature which is consistent with the Ward
identity \eqref{ward}, namely, 
\begin{equation}
\eta_{\mu\nu} - \frac{p_{\mu}k_{\nu} + p_{\nu}k_{\mu}}{p\cdot k} +
\frac{p^{2}k_{\mu}k_{\nu}}{(p\cdot k)^{2}}, 
\end{equation} 
and the integrand has to be proportional to this structure. Explicit
calculation indeed shows that in this region, \eqref{YM} takes the
form  
\begin{eqnarray}
\lefteqn{\Pi_{\mu\nu,{\rm R}}^{ab (1)} (p)
 =  -g^{2}N \delta^{ab}\int \frac{d^{4}k}{(2\pi)^{3}} \delta^{+}
 (k^{2})}\nonumber\\ 
 & & \quad\times \left(\eta_{\mu\nu} - \frac{p_{\mu}k_{\nu} +
 p_{\nu}k_{\mu}}{p\cdot k} + \frac{p^{2}k_{\mu}k_{\nu}}{(p\cdot
 k)^{2}}\right).\label{selfenergy} 
\end{eqnarray}
This is the leading contribution in the ``hard" internal momentum
region and as we will show shortly, this leads to the hard thermal
loop results at high temperature. 

There are several things to note from the structure in
\eqref{selfenergy}. First, the quantity in the parenthesis is
manifestly Lorentz covariant of degree zero in both the internal and
the external momenta. Furthermore, it is manifestly transverse (gauge
invariant). We note that the integral in \eqref{selfenergy} is
quadratically divergent and is usually set to zero in the dimensional
regularization. However, if we only carry out the $k_{0}$ integration it does not
vanish and as we will show shortly,  it is this leading
contribution that leads to the hard thermal loop results at high
temperature. (We remark here parenthetically that the conventional
ultraviolet divergent terms in the self-energy are, in contrast, gauge
dependent and are only logarithmically divergent, yielding at high
temperature to only $\ln T$ contributions \cite{frenkel5}.) We want to 
emphasize that the reason for carrying out only the 
$k_{0}$ integration is that the thermal operator acts on the integrand after the energy
integrations have been carried out and  before evaluating  the integrations 
over the spatial momenta \cite{silvana1}.  To apply
the thermal operator representation, we need to integrate over the
$k_{0}$ variable in \eqref{selfenergy} and this leads to 
\begin{eqnarray}
\lefteqn{\Pi_{\mu\nu,{\rm R}}^{ab (1)} (p)
 = -g^{2}N \delta^{ab} \int \frac{d^{3}k}{(2\pi)^{3}}}\nonumber\\
 & & \quad \times \frac{1}{2E_{k}} \left(\eta_{\mu\nu} -
 \frac{p_{\mu}\hat{k}_{\nu} + p_{\nu}\hat{k}_{\mu}}{p\cdot \hat{k}} +
 \frac{p^{2}\hat{k}_{\mu}\hat{k}_{\nu}}{(p\cdot
 \hat{k})^{2}}\right),\label{selfenergy1} 
\end{eqnarray}
where $E_{k} = |\vec{k}|$ and we have defined
$\hat{k}_{\mu} = (1, - \hat{\vec{k}})$.
Applying now the thermal operator to the integrand, we obtain, (It is worth clarifying
here once again that although some of the diagrams in \eqref{YM} involve two 
propagators, only the thermal operator corresponding to the on-shell propagator 
leads to a nontrivial result. The retarded propagators are unchanged by the application 
of the thermal operator as discussed in \eqref{tor3}.)
\begin{widetext}
\begin{eqnarray}
\Pi_{\mu\nu,{\rm R}}^{ab (1) (T)} & = & -g^{2}N\delta^{ab} \int
\frac{d^{3}k}{(2\pi)^{3}} {\cal O}^{(T)} (E_{k})
\frac{1}{2E_{k}}\left(\eta_{\mu\nu} - \frac{p_{\mu}\hat{k}_{\nu} +
  p_{\nu}\hat{k}_{\mu}}{p\cdot \hat{k}} +
\frac{p^{2}\hat{k}_{\mu}\hat{k}_{\nu}}{(p\cdot
  \hat{k})^{2}}\right)\nonumber\\ 
& = & \Pi_{\mu\nu,{\rm R}}^{ab (1)} (p) -g^{2}N \delta^{ab} \int
\frac{d^{3}k}{(2\pi)^{3}} \frac{n_{\rm B} (E_{k})}{E_{k}}
\left(\eta_{\mu\nu} - \frac{p_{\mu}\hat{k}_{\nu} +
  p_{\nu}\hat{k}_{\mu}}{p\cdot \hat{k}} +
\frac{p^{2}\hat{k}_{\mu}\hat{k}_{\nu}}{(p\cdot \hat{k})^{2}}\right) =
\Pi_{\mu\nu,{\rm R}}^{ab (1)} (p) + \Pi_{\mu\nu,{\rm R}}^{ab (1) 
  \beta}.\label{selfenergy2} 
\end{eqnarray}
\end{widetext}
The temperature independent part $\Pi_{\mu\nu,{\rm R}}^{ab (1)}$ can now be 
set to zero using dimensional regularization for the spatial momentum. 

We note that since $E_{k}=|\vec{k}|$, the radial momentum integration
in the second term in \eqref{selfenergy2} can be done to yield 
\begin{equation}
\int_{0}^{\infty}  dk\ k n_{\rm B} (k) = \frac{\pi^{2}T^{2}}{6},
\end{equation}
which leads us to the temperature dependent part of the self-energy
\begin{eqnarray}
\lefteqn{\Pi_{\mu\nu,{\rm R}}^{ab (1) \beta} (p) =
  -\frac{g^{2}NT^{2}\delta^{ab}}{48\pi}}\nonumber\\ 
& & \times \int d\Omega \left(\eta_{\mu\nu} -
\frac{p_{\mu}\hat{k}_{\nu} + p_{\nu}\hat{k}_{\mu}}{p\cdot \hat{k}} +
\frac{p^{2}\hat{k}_{\mu}\hat{k}_{\nu}}{(p\cdot \hat{k})^{2}}\right). 
\end{eqnarray}
This is the well known result for the temperature dependent part of the
retarded self-energy in the hard thermal loop approximation in the
forward scattering description. The angular integrations break the
manifest Lorentz invariance. (At finite temperature, Lorentz invariance is
broken because the rest frame of the heat bath defines a preferred reference frame.) 
However, the beautiful properties of the
integrand follow simply from the properties of the zero temperature
amplitude. 

This example demonstrates that,  in addition to establishing a graphical
correspondence, the thermal operator representation can
also be conveniently used  to calculate the thermal amplitudes in the hard
thermal loop approximation from the ``hard" internal momentum
amplitudes at zero temperature. Such a calculation also clarifies
various important features of the thermal amplitudes at high
temperature. 

\section{Conclusion}

In this paper, we have derived systematically the forward scattering
description for retarded amplitudes to all orders at zero
temperature. This graphical derivation then allows us to obtain the
forward scattering description for such amplitudes to all
orders at finite temperature through the thermal operator
representation. Although our derivation has been within the context of
a scalar field theory, the derivation can be generalized easily to
other theories with or without a chemical potential. Furthermore, although 
we have used the real time formalism of closed time path for our discussions 
for simplicity, the results also hold in the imaginary time formalism (which we 
do not go into). Besides giving a
graphical derivation of the forward scattering description at finite
temperature, such a relation can be used as a powerful tool for
calculations at high temperature and clarifies various properties of
thermal amplitudes. As an example, we have calculated the one loop
retarded self-energy for gluons in the Yang-Mills theory at finite
temperature starting from the forward scattering description at zero
temperature. This derivation emphasizes that various nice features of
these amplitudes such as gauge invariance, transversality, manifest
Lorentz covariance etc  arise simply because the zero temperature
amplitude already possesses such properties. This description of the 
forward scattering amplitudes at finite temperature provides yet another 
example of the usefulness of the thermal operator representation.
\vspace{.3in}

\noindent{\bf Acknowledgment}

We would like to thank Gerald Dunne for making us aware of reference
\cite{feynman}. 
This work was supported in part by the US DOE Grant number DE-FG 02-91ER40685,
by MCT/CNPq as well as by FAPESP, Brazil.

\end{document}